\documentclass[aps,rmp,floatfix,twocolumn,showpacs]{revtex4-1}
\usepackage[usenames]{color}
\usepackage{graphicx}
\usepackage{multirow}
\usepackage{subfigure}
\usepackage{epstopdf}
\usepackage{bm}
\usepackage{amssymb}
\usepackage{amsmath}
\usepackage{times}
\include{macros}

\renewcommand{\c}[1]{\mathcal{#1}}

\def\be{\begin{equation}} \def\ee{\end{equation}}
\def\bea{\begin{eqnarray}} \def\eea{\end{eqnarray}}

\begin{document}
\title{Topological insulators and superconductors}
\author{Xiao-Liang Qi$^{1,2}$ and Shou-Cheng Zhang$^2$}
\affiliation{$^1$Microsoft Research, Station Q, Elings
Hall, University of California, Santa Barbara, CA 93106, USA\\
$^2$Department of Physics, Stanford University, Stanford, CA 94305}
\begin{abstract}
Topological insulators are new states of quantum matter which can not be adiabatically connected to conventional insulators and semiconductors. They are characterized by a full insulating gap in the bulk and gapless edge or surface states which are protected by time-reversal symmetry. These topological materials have been theoretically predicted and experimentally observed in a variety of systems, including HgTe quantum wells, BiSb alloys, and Bi$_2$Te$_3$ and Bi$_2$Se$_3$ crystals. We review theoretical models, materials properties and experimental results on two-dimensional and three-dimensional topological insulators, and discuss both the topological band theory and the topological field theory. Topological superconductors have a full pairing gap in the bulk and gapless surface states consisting of Majorana fermions. We review the theory of topological superconductors in close analogy to the theory of topological insulators. \end{abstract}

\pacs{73.20.-r, 73.43.-f, 85.75.-d, 74.90.+n}

\maketitle

\tableofcontents

\section{Introduction}
\label{sec:intro}

Ever since the Greeks invented the concept of the atom, fundamental science has focused on finding ever smaller building blocks of matter. In the 19th century, the discovery of elements defined the golden age of chemistry. Throughout most of the 20th
century, fundamental science was dominated by the search for elementary particles. In condensed matter physics, there are no new building blocks of matter to be discovered: one is dealing with the same atoms and electrons as those discovered centuries ago. Rather, one is interested in how these basic building blocks are put together to form new states of matter. Electrons and atoms in the quantum world can form many different states of matter: for example, they can form crystalline solids, magnets and superconductors. The greatest triumph of condensed matter physics in the last century is the classification of these quantum states by the principle of spontaneous symmetry breaking~\cite{anderson1997}. For example, a crystalline solid breaks translation symmetry, even though the interaction among its atomic building blocks is translationally invariant. A magnet breaks rotation symmetry, even though the fundamental interactions are isotropic. A superconductor breaks the more subtle gauge symmetry, leading to novel phenomena such as flux quantization and Josephson effects. The pattern of symmetry breaking leads to a unique order parameter, which assumes a nonvanishing expectation value only in the ordered state, and a general effective field theory can be formulated based on the order parameter. The effective field theory, generally called Landau-Ginzburg theory~\cite{LLStatPhys}, is determined by general properties such as dimensionality and symmetry of the order parameter, and gives a universal description of quantum states of matter.

In 1980, a new quantum state was discovered which does not fit into this simple paradigm~\cite{klitzing1980}. In the quantum Hall (QH) state, the bulk of the two-dimensional (2D) sample is insulating, and the electric current is carried only along the edge of the sample. The flow of this unidirectional current avoids dissipation and gives rise to a quantized Hall effect. The QH state provided the first example of a quantum state which is topologically distinct from all states of matter known before. The precise quantization of the Hall conductance is explained by the fact that it is a topological invariant, which can only take integer values in units of $e^2/h$, independent of material details~\cite{laughlin1981,thouless1982}. Mathematicians have introduced the concept of topological invariance to classify different geometrical objects into broad classes. For example, 2D surfaces are classified by the number of holes in them, or genus. The surface of a perfect sphere is topologically equivalent to the surface of an ellipsoid, since these two surfaces can be smoothly deformed into each other without creating any holes. Similarly, a coffee cup is topologically equivalent to a donut, since both of them contain a single hole. In mathematics, topological classification discards small details and focuses on the fundamental distinction of shapes. In physics, precisely quantized physical quantities such as the Hall conductance also have a topological origin, and remain unchanged by small changes in the sample.

It is obvious that the link between physics and topology should be more general than the specific case of QH states. The key concept is that of a ``smooth deformation". In mathematics, one considers smooth deformations of shapes without the violent action of creating a hole in the deformation process. The operation of smooth deformation groups shapes into topological equivalence classes. In physics, one can consider general Hamiltonians of many-particle systems with an energy gap separating the ground state from the excited states. In this case, one can define a smooth deformation as a change in the Hamiltonian which does not close the bulk gap. This topological concept can be applied to both insulators and superconductors with a full energy gap, which are the focus of this review article. It cannot be applied to gapless states such as metals, doped semiconductors, or nodal superconductors. According to this general definition, if we put in contact two quantum states belonging to the same topological class, the interface between them does not need to support gapless states. On the other hand, if we put in contact two quantum states belonging to different topological classes, or put a topologically nontrivial state in contact with the vacuum, the interface must support gapless states.

From these simple arguments, we immediately see that the abstract concept of topological classification can be applied to condensed matter system with an energy gap, where the notion of a smooth deformation can be defined~\cite{zhang2008}. Further progress can be made through the concepts of topological order parameter and topological field theory (TFT), which are powerful tools describing topological states of quantum matter. Mathematicians have expressed the intuitive concept of genus in terms of an integral, called topological invariant, over the local curvature of the surface~\cite{nakahara1990}. Whereas the integrand depends on details of the surface geometry, the value of the integral is independent of such details and depends only on the global topology. In physics, topologically quantized physical quantities can be similarly expressed as invariant integrals over the frequency-momentum space~\cite{thouless1982,Thouless1998}. Such quantities can serve as a topological order parameter which uniquely determines the nature of the quantum state. Furthermore, the long-wavelength and low-energy physics can be completely described by a TFT, leading to powerful predictions of experimentally measurable topological effects~\cite{zhang1992}. Topological order parameters and TFTs for topological quantum states play the role of conventional symmetry-breaking order parameters and effective field theories for broken-symmetry states.

The QH states belong to a topological class which explicitly breaks time-reversal (TR) symmetry, for example, by the presence of a magnetic field. In recent years, a new topological class of materials has been theoretically predicted and experimentally observed~\cite{bernevig2006c,koenig2007,fu2007b,hsieh2008,zhang2009,xia2009,chen2009}. These new quantum states belong to a class which is invariant under TR, and where spin-orbit coupling (SOC) plays an essential role. Some important concepts were developed in earlier works~\cite{haldane1988,zhang2001,murakami2003,sinova2004,murakami2004a,kane2005a,bernevig2006a}, culminating in the construction of the topological band theory (TBT) and the TFT of 2D and 3D topological insulators~\cite{kane2005b,moore2007,roy2009,fu2007b,fu2007a,qi2008b}.
All TR invariant insulators in nature (without ground state degeneracy) fall into two distinct classes, classified by a $\mathbb{Z}_2$ topological order parameter. The topologically nontrivial state has a full insulating gap in the bulk, but has gapless edge or surface states consisting of an odd number of Dirac fermions. The topological property manifests itself more dramatically when TR symmetry is preserved in the bulk but broken on the surface, in which case the material is fully insulating both inside the bulk and on the surface. In this case, Maxwell's laws of electrodynamics are dramatically altered by a topological term with a precisely quantized coefficient, similar to the case of the QH effect. The 2D topological insulator, synonymously called the quantum spin Hall (QSH) insulator, was first theoretically predicted in 2006~\cite{bernevig2006c} and experimentally observed~\cite{koenig2007,roth2009} in HgTe/CdTe quantum wells (QW). A topologically trivial insulator state is realized when the thickness of the QW is less than a critical value, and the topologically nontrivial state is obtained when that thickness exceeds the critical value. In the topologically nontrivial state, there is a pair of edge states with opposite spins propagating in opposite directions. Four-terminal measurements~\cite{koenig2007} show that the longitudinal conductance in the QSH regime is quantized to $2e^2/h$, independently of the width of the sample. Subsequent nonlocal transport measurements~\cite{roth2009} confirm the edge state transport as predicted by theory. The first discovery of the QSH topological insulator in HgTe was ranked by Science Magazine as one of the top ten breakthroughs among all sciences in year 2007, and the subject quickly became mainstream in condensed matter physics~\cite{day2008}. The 3D topological insulator was predicted in the Bi$_{1-x}$Sb$_x$ alloy within a certain range of compositions $x$~\cite{fu2007a}, and angle-resolved photoemission spectroscopy (ARPES) measurements soon observed an odd number of topologically nontrivial surface states~\cite{hsieh2008}. Simpler versions of the 3D topological insulator were theoretically predicted in Bi$_2$Te$_3$, Sb$_2$Te$_3$~\cite{zhang2009} and Bi$_2$Se$_3$~\cite{zhang2009,xia2009} compounds with a large bulk gap and a gapless surface state consisting of a single Dirac cone. ARPES experiments indeed observed the linear dispersion relation of these surface states~\cite{xia2009,chen2009}. These pioneering theoretical and experimental works opened up the exciting field of topological insulators, and the field is now expanding at a rapid pace~\cite{zhang2008,kane2008,koenig2008,moore2009,qi2010phystoday,hasan2010}. Beyond the topological materials mentioned above, more than fifty new compounds have been predicted to be topological insulators~\cite{chadov2010,lin2010a,franz2010,yan2010a}, and two of them have been experimentally observed recently~\cite{sato2010,chen2010}. This collective body of work establishes beyond any reasonable doubt the ubiquitous existence in nature of this new topological state of quantum matter. It is remarkable that such topological effects can be realized in common materials, previously used for infrared detection or thermoelectric applications, without requiring extreme conditions such as high magnetic fields or low temperatures. The discovery of topological insulators has undoubtedly had a dramatic impact on the field of condensed matter physics.

\begin{figure}[t]
\begin{center}
\includegraphics[width=9cm,angle=0]{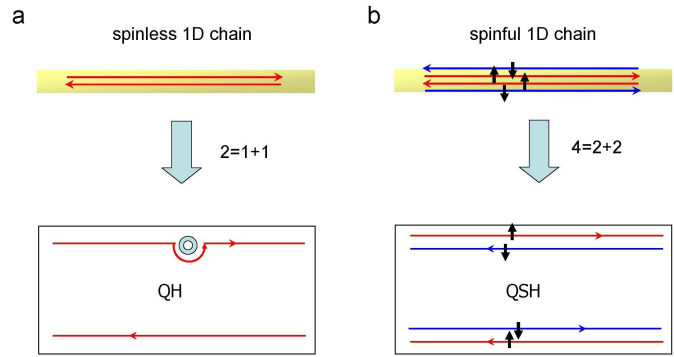}
\end{center}
\caption{Analogy between QH and QSH effects: (a) A spinless 1D system has both forward and backward movers. These two basic degrees of freedom are spatially separated in a QH bar, as expressed by the symbolic equation
``$2 = 1 + 1$". The upper edge supports only a forward mover and the lower edge supports only a backward mover. The states are robust and go around an impurity without scattering. (b) A spinful 1D system has four basic degrees of freedom, which are spatially separated in a QSH bar. The upper edge supports a forward mover with spin up and a backward mover with spin down, and conversely for the
lower edge. That spatial separation is expressed by the symbolic equation ``$4 = 2 + 2$". Adapted from~\onlinecite{qi2010phystoday}.} \label{qh_qsh}
\end{figure}

After briefly reviewing the history of the theoretical prediction and the experimental observation of the topological materials in nature, we now turn to the history of the conceptual developments, and retrace the intertwined paths taken by theorists. An important step was taken in 1988 by Haldane~\cite{haldane1988}, who borrowed the concept of the parity anomaly~\cite{redlich1984a,semenoff1984} in quantum electrodynamics to construct a theoretical model of the QH state on the 2D honeycomb lattice. This model does not require an external magnetic field nor the associated orbital quantization and Landau levels (LLs). However, it is in the same topological class as the ordinary QH states, and requires both two dimensionality and the breaking of the TR symmetry. There was a misconception at the time that topological quantum states could only exist under these conditions. Another important step was the construction in 1989 of a TFT of the QH effect based on the Chern-Simons (CS) term~\cite{zhang1992}. This theory captures the most important topological aspects of the QH effect in a single and unified effective field theory. At this point, the path towards generalizing the QH states became clear: since the CS term can exist in all even spatial dimensions, the topological physics of the QH states can be generalized to such dimensions. However, it was unclear at the time what kind of microscopic interactions could be responsible for these topological states. In 2001, Zhang and Hu~\cite{zhang2001} explicitly constructed a microscopic model for the generalization of the QH state in 4D. A crucial ingredient of this model is its invariance under TR symmetry, in sharp contrast to the QH state in 2D which explicitly breaks TR symmetry. This fact can also be seen directly from the CS effective action in $4+1$ spacetime dimensions, which is invariant under TR symmetry. With this generalization of the QH state, two basic obstacles, the breaking of TR symmetry and the restriction to 2D, were removed. Partly because of the mathematical complexity involved in this work, it was not appreciated by the general community at the time --- but is clear now --- that this state is the root state from which all TR invariant topological insulators in 3D and 2D are derived~\cite{qi2008b}. TR invariant topological insulators can be classified in the form of a family tree, where the 4D state is the ``grandfather" state and begets exactly two generations of descendants, the 3D and 2D topological insulators, by the procedure of dimensional reduction~\cite{qi2008b,schnyder2008,kitaev2009,ryu2010}.

Motivated by the construction of a TR invariant topological state, theorists started to look for a physical realization of this new topological class, and discovered the intrinsic spin Hall effect~\cite{murakami2003,sinova2004,murakami2004a}. Murakami~\emph{et al.}~\cite{murakami2003} state their motivation clearly in the introduction: ``{\it Recently, the QH effect has been generalized to four spatial dimensions [...]. The QH response in that system is physically realized through the SOC in a TR symmetric system}". Soon after, it was realized in 2004 that the two key ideas, TR symmetry and SOC, can also be applied to insulators as well, leading to the concept of spin Hall insulator~\cite{murakami2004a}. The spin Hall effect in insulators is dissipationless, similarly to the QH effect. The concept of spin Hall insulator motivated Kane and Mele in 2005 to investigate the QSH effect in graphene~\cite{kane2005a}, a material first discovered experimentally that same year. Working independently, Bernevig and Zhang studied the QSH effect in strained semiconductors, where SOC generates LLs without the breaking of TR symmetry~\cite{bernevig2006a}. Unfortunately the energy gap in graphene caused by the intrinsic SOC is insignificantly small~\cite{yao2007,min2006}. Even though neither models have been experimentally realized, they played important roles for the conceptual developments. In 2006, Bernevig, Hughes and Zhang~\cite{bernevig2006c} successfully predicted the first topological insulator to be realized in HgTe/CdTe QWs.

The QSH state in 2D can be roughly understood as two copies of the QH state, where states with opposite spin counter-propagate at the edge. A natural question arises as to whether the edge states of the QSH state are stable. In a deeply insightful paper~\cite{kane2005b}, Kane and Mele showed in 2005 that the stability depends on the number of pairs of edge states. An odd number of pairs is stable, whereas an even number of pairs is not. This observation led Kane and Mele to propose a $\mathbb{Z}_2$ classification of TR invariant 2D insulators. In addition, they devised a precise algorithm for the computation of a $\mathbb{Z}_2$ topological invariant within TBT. TBT was soon extended to 3D by Fu, Kane and Mele, Moore and Balents and Roy~\cite{moore2007,fu2007b,fu2007a,roy2009a}, where sixteen topologically distinct states are possible. Most of these states can be viewed as stacked 2D QSH insulator planes, but one of them, the strong topological insulator, is genuinely 3D. The topological classification according to TBT is only valid for noninteracting systems, and it was not clear at the time whether these states are stable under more general topological deformations including interactions. Qi, Hughes and Zhang introduced the TFT of topological insulators~\cite{qi2008b}, and demonstrated that these states are indeed generally stable in the presence of interactions. Furthermore, a topologically invariant topological order parameter can be defined within the TFT as a experimentally measurable, quantized topological magnetoelectric effect. The standard Maxwell's equations are modified by the topological terms, leading to the axion electrodynamics of the topological insulators. This work also showed that the 2D and 3D topological insulators are descendants of the 4D topological insulator state discovered in 2001~\cite{zhang2001}, and motivated this series of recent developments. At this point, the two different paths based on the TBT and TFT converged, and an unified theoretical framework emerged.

There are a number of excellent reviews on this subject~\cite{koenig2008,moore2009,qi2010phystoday,hasan2010}. This article attempts to give a simple pedagogical introduction to the subject and reviews the current status of the field. In Sec.~\ref{sec:2dti} and Sec.~\ref{sec:3dti}, we review the standard models, materials and experiments for the 2D and the 3D topological insulators. These two sections can be understood without any prior knowledge of topology. In Sec.~\ref{sec:general}, we review the general theory of topological insulators, presenting both the TFT and the TBT. In Sec.~\ref{sec:TSC}, we discuss an important generalization of topological insulators--topological superconductors.

\section{Two-Dimensional Topological Insulators}
\label{sec:2dti}
The QSH state, or the 2D topological insulator was first discovered in the HgTe/CdTe quantum wells. Bernevig, Hughes and Zhang~\cite{bernevig2006c} initiated the
search for the QSH state in semiconductors with an ``inverted"
electronic gap, and predicted a quantum phase transition in
HgTe/CdTe quantum wells as a function of the thickness $d_\mathrm{QW}$ of the quantum well. The quantum well system is predicted to be a
conventional insulator for $d_\mathrm{QW}<d_c$, and a QSH insulator with a single pair of helical edge states for
$d_\mathrm{QW}>d_c$, where $d_c$ is a critical thickness. The first experimental confirmation of the existence of the QSH state in
HgTe/CdTe quantum wells was carried out by K\"{o}nig \emph{et al.}~\cite{koenig2007}.
This work reports the observation of a nominally insulating
state which conducts only through 1D edge channels, and is
strongly influenced by a TR symmetry-breaking magnetic
field. Further transport
measurements~\cite{roth2009} reported unique nonlocal conduction
properties due to the helical edge states.

The QSH insulator state is invariant under TR, has a
charge excitation gap in the 2D bulk, but has topologically
protected 1D gapless edge states that lie inside the bulk insulating
gap. The edge states have a distinct helical property: two states
with opposite spin polarization counter-propagate at a given
edge~\cite{kane2005a,wu2006,xu2006}. For this reason, they are
also called helical edge states, i.e. the spin is correlated
with the direction of motion\cite{wu2006}. The edge states come in Kramers
doublets, and TR symmetry ensures the crossing of their energy levels at special points in the Brillouin zone.
Because of this level crossing, the spectrum of a QSH
insulator cannot be adiabatically deformed into that of a
topologically trivial insulator without helical edge states.
Therefore, in this precise sense, the QSH insulator represents a
new topologically distinct state of matter. In the special case that SOC preserves a $U(1)_s$ subgroup of the full $SU(2)$ spin rotation group, the topological properties of the QSH state can be characterized by the spin Chern number~\cite{sheng2006}. More generally, the topological properties of the QSH state are mathematically characterized by a $\mathbb{Z}_2$ topological invariant~\cite{kane2005b}. States with an even number of Kramers pairs of edge states at a given edge are topologically trivial, while those with an odd number are topologically nontrivial. The $\mathbb{Z}_2$ topological quantum number can also be defined for generally interacting systems and experimentally measured in terms of the fractional charge and quantized current on the edge~\cite{qi2008}, and spin-charge separation in the bulk~\cite{qi2008a,ran2008}.

In this section, we shall focus on the basic theory of the QSH state in the HgTe/CdTe system because of its simplicity and experimental relevance, and provide an explicit and pedagogical discussion of the helical edge states and their transport properties. There are several other theoretical proposals for the QSH state,
including bilayer bismuth~\cite{murakami2006}, and the ``broken-gap" type-II AlSb/InAs/GaSb quantum wells~\cite{liu2008}. Initial experiments in the AlSb/InAs/GaSb system already show encouraging signatures~\cite{knez2010}. The QSH system has also been proposed for the transition metal oxide Na$_2$IrO$_3$~\cite{shitade2009}. The concept of fractional QSH state was proposed at the same time as the QSH~\cite{bernevig2006a}, and has been investigated theoretically in more details recently~\cite{young2008,levin2009}.

\subsection{Effective model of the two-dimensional time-reversal invariant topological insulator in HgTe/CdTe quantum wells}
\label{sec:bhz}

In this section we review the basic electronic structure of bulk HgTe and CdTe, and present a simple model first introduced by Bernevig, Hughes and Zhang~\cite{bernevig2006c} (BHZ) to describe the physics of those subbands of HgTe/CdTe quantum wells that are relevant for the QSH effect. HgTe and CdTe crystallize in the
zincblende lattice structure. This structure has the same geometry
as the diamond lattice, i.e. two interpenetrating face-centered-cubic
lattices shifted along the body diagonal, but with a different
atom on each sublattice. The presence of two different atoms per lattice site breaks inversion symmetry, and thus reduces the
point group symmetry from $O_h$ (cubic) to $T_d$ (tetrahedral).
However, even though inversion symmetry is explicitly broken,
this only has a small effect on the physics of the QSH effect. To simplify the discussion, we shall first ignore this bulk inversion asymmetry (BIA).

\begin{figure}[t]
\centering
\includegraphics[scale=0.32]{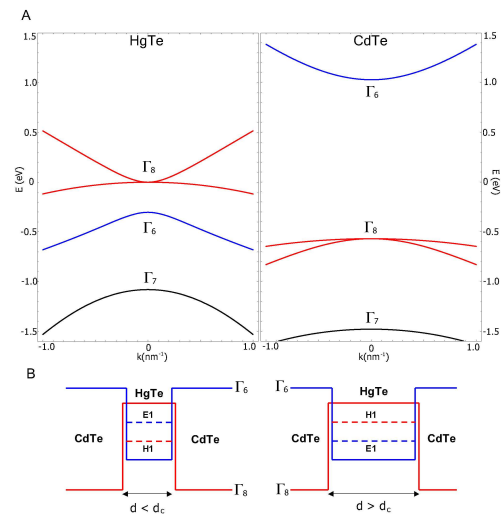}
\caption{(a) Bulk band structure of HgTe and CdTe; (b)
schematic picture of quantum well geometry and lowest subbands for
two different thicknesses. From~\onlinecite{bernevig2006c}. } \label{bandstructure}
\end{figure}

For both HgTe and CdTe, the important bands near the Fermi
level are close to the $\Gamma$ point in the Brillouin zone~[Fig.~\ref{bandstructure}(a)]. They are a $s$-type band ($\Gamma_6$), and a $p$-type band split by SOC into a $J=3/2$ band ($\Gamma_8$) and a $J=1/2$ band
($\Gamma_7$). CdTe has a band ordering similar to GaAs
with a $s$-type ($\Gamma_6$) conduction band, and $p$-type valence
bands ($\Gamma_8,\Gamma_7$) which are separated from the
conduction band by a large energy gap ($\sim1.6~$eV). Because of the large SOC present in the heavy element Hg, the usual band ordering is {\it inverted}: the negative energy gap of $-300$~meV indicates that the $\Gamma_8$ band, which usually forms the valence band, is above
the $\Gamma_6$ band. The light-hole $\Gamma_8$
band becomes the conduction band, the heavy-hole band
becomes the first valence band, and the $s$-type band ($\Gamma_6$)
is pushed below the Fermi level to lie between the heavy-hole
band and the spin-orbit split-off band ($\Gamma_7$) [Fig.~\ref{bandstructure}(a)]. Due to the degeneracy between heavy-hole and light-hole bands at the $\Gamma$ point, HgTe is a zero-gap
semiconductor.

\begin{figure}[t]
\centering
 \includegraphics[width=0.8\linewidth]{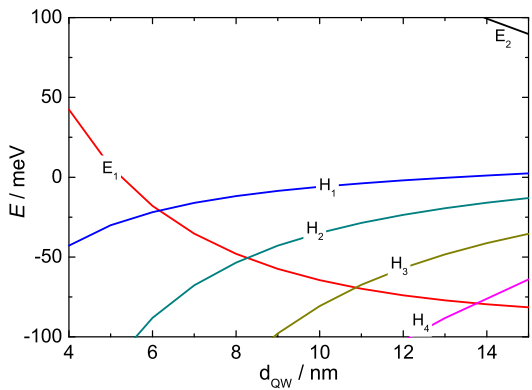}\\
 \caption{Energy levels of the QW as
 a function of QW width. From~\onlinecite{koenig2008}.}
 \label{FigEd}
\end{figure}

When HgTe-based quantum well structures are grown, the peculiar
properties of the well material can be utilized to tune the electronic structure. For wide QW layers, quantum confinement is
weak and the band structure remains ``inverted". However, the
confinement energy increases when the well width is reduced.
Thus, the energy levels will be shifted and, eventually, the
energy bands will be aligned in a ``normal" way, if the QW thickness $d_\mathrm{QW}$ falls below a critical thickness $d_{c}$. We can understand this
heuristically as follows: for thin QWs the
heterostructure should behave similarly to CdTe and have a normal
band ordering, i.e. the bands with primarily $\Gamma_6$
symmetry are the conduction subbands and the $\Gamma_8$ bands
contribute to the valence subbands. On the other hand, as $d_\mathrm{QW}$ is increased, we expect the material
to behave more like HgTe which has inverted bands. As $d_\mathrm{QW}$ increases, we expect to reach a critical thickness
where the $\Gamma_8$ and $\Gamma_6$ subbands cross and become
inverted, with the $\Gamma_8$ bands becoming conduction subbands
and the $\Gamma_6$ bands becoming valence subbands [Fig.~\ref{bandstructure}(b)]~\cite{novik05,bernevig2006c}. The shift of energy levels with $d_\mathrm{QW}$ is depicted in Fig.~\ref{FigEd}.
The QW states derived from the heavy-hole $\Gamma_8$ band are denoted by $H_n$, where the subscript $n=1,2,3,\ldots$ describes well states with increasing number of nodes in the $z$ direction. Similarly, the QW states derived from the electron $\Gamma_6$ band are denoted by $E_n$. The inversion between $E_1$ and $H_1$ bands occurs at a critical thickness $d_\mathrm{QW}=d_{c}\sim 6.3$~ nm [Fig.~\ref{FigEd}]. In the following, we develop a simple model and discuss why we
expect QWs with $d_\mathrm{QW}>d_c$ to form TR invariant 2D topological insulators with protected edge states.

Under our assumption of inversion symmetry, the relevant subbands, $E_1$ and $H_1$, must be doubly degenerate since TR
symmetry is present. We express states in the basis
$\{\vert
E_1 +\rangle,\vert H_1 +\rangle,\vert E_1 -\rangle,\vert H_1
-\rangle\}$, where $\vert E_1 \pm\rangle$ and $\vert H_1
\pm\rangle$ are two sets of Kramers partners. The states
$\vert E_1 \pm\rangle$ and $\vert H_1 \pm\rangle$ have opposite parity, hence a Hamiltonian matrix element that
connects them must be odd under parity. Thus, to lowest order in
$k$, $(\vert E_1 +\rangle,\vert H_1 +\rangle)$ and $(\vert E_1
-\rangle,\vert H_1 -\rangle)$ will each be coupled generically via
a term linear in $k$. The $\vert H_1 +\rangle$ heavy-hole state
is formed from the spin-orbit coupled $p$-orbitals $\vert p_x+ip_y,
\uparrow\rangle$, while the $\vert H_1 -\rangle$ heavy-hole
state is formed from the spin-orbit coupled $p$-orbitals $\vert
-(p_x-ip_y), \downarrow\rangle$. Therefore, to preserve
rotation symmetry around the growth axis $z$, the matrix
elements must by proportional to $k_{\pm}=k_x\pm ik_y.$ The only
terms allowed in the diagonal elements are terms that have even
powers of $k$ including $k$-independent terms. The subbands must
come in degenerate pairs at each $\mathbf{k}$, so there can be no matrix elements between the $+$ state and the $-$ state of the same band. Finally, if there are nonzero matrix elements between $\vert E_1
+\rangle,\vert H_1 -\rangle$ or $\vert E_1 -\rangle,\vert H_1
+\rangle$, this would induce a higher-order process coupling the $\pm$ states of the same band and splitting the
degeneracy. Therefore, these matrix elements are forbidden as well. These simple arguments led to the following model,
\begin{eqnarray}
{\cal{H}}&=&\left(\begin{array}{cc} h(\mathbf{k})& 0\\
0&h^{*}(-\mathbf{k})\end{array}\right),\label{contH}\\
h(\mathbf{k})&=&\epsilon (\mathbf{k}) \mathbb{I}_{2\times 2}+d_a(\mathbf{k})\sigma^a,
\label{BHZ}
\end{eqnarray}\noindent
where $\mathbb{I}_{2\times 2}$ is the $2\times 2$ identity matrix, and
\begin{eqnarray}
\epsilon (\mathbf{k})&=&C-D(k_{x}^2+k_{y}^2), \nonumber\\
d_a (\mathbf{k})&=&\left(A k_x,-Ak_y,M(\mathbf{k})\right),\nonumber\\
M(\mathbf{k})&=& M-B(k_{x}^2+k_{y}^2),\label{BHZ1} \end{eqnarray}\noindent
where $A,B,C,D,M$ are material parameters that depend on the QW geometry, and we choose the zero of energy to be the
valence band edge of HgTe at $\mathbf{k}=0$ [Fig.~\ref{bandstructure}].

The bulk energy spectrum of the BHZ model is given by
\begin{eqnarray}
E_{\pm}&=&\epsilon (k)\pm\sqrt{d_a d_a}\\
&=&\epsilon (k)\pm \sqrt{A^2(k^2_x +  k^2_y)+M^2 (k)}.
\end{eqnarray}
For $B=0$, the model reduces to two copies of the massive Dirac Hamiltonian in ($2+1$)D. The mass $M$ corresponds to the energy difference between the $E_1$ and $H_1$ levels at the $\Gamma$ point. The mass $M$ changes sign at the critical thickness $d_c$, where $E_1$ and $H_1$ become degenerate. At the critical point, the system is described by two copies of the massless Dirac Hamiltonian, one for each spin, and at a single valley $\mathbf{k}=0$. This situation is similar to graphene~\cite{castroneto2009}, which is also described by the massless Dirac Hamiltonian in ($2+1$)D. However, the crucial difference lies in the fact that graphene has four Dirac cones, consisting of two valleys and two spins, whereas we have two Dirac cones, one for each spin, and at a single valley. For $d_\mathrm{QW}>d_c$, the $E_1$ level falls below the $H_1$ level at the $\Gamma$ point, and the mass $M$ becomes negative. A pure massive Dirac model does not differentiate between a positive or negative mass $M$. Since we are dealing with a nonrelativistic system, the $B$ term is generally allowed. In order to make the distinction clear, we call $M$ the Dirac mass, and $B$ the Newtonian mass, since it describes the usual nonrelativistic mass term with quadratic dispersion relation. We shall show later that the relative sign between the Dirac mass $M$ and the Newtonian mass $B$ is crucial to determine whether the model describes a topological insulator state with protected edge states or not.

HgTe has a crystal structure of the zincblende type which lacks inversion symmetry, leading to a BIA term in the Hamiltonian, given to leading order by~\cite{koenig2008}
\begin{eqnarray}
H_\mathrm{BIA}=\left(\begin{array}{cccc}0&0&0&-\Delta_z\\0&0&\Delta_z&0\\
0&\Delta_z&0&0\\-\Delta_z&0&0&0\end{array}
\right).
\end{eqnarray}
This term plays an important role in determining the spin orientation of the helical edge state. The topological phase transition in the presence of BIA has been investigated recently~\cite{koenig2008,murakami2007}. In addition, in an asymmetric QW structural inversion symmetry can be broken by a build-in electric field, leading to a SOC term of Rashba type in the effective Hamiltonian~\cite{rothe2010,strom2010}. For simplicity, we will focus on symmetric QW witout SIA. In Table~\ref{tab:HgTeParameters}, we give the parameters of the BHZ model for various values of $d_\mathrm{QW}$.

\begin{center}
\begin{table}[t]
\begin{tabular}{|c|c|c|c|c|c|c|c|c|}
\hline
$d$~(\AA)&
$A$~(eV$\cdot$\AA)&$B$~(eV$\cdot$\AA$^2$)&$D$~(eV)&$M$~(eV)&$\Delta_z$~(eV)\\
\hline
$55$&$3.87$&$-48.0$&$-30.6$&$0.009$&$0.0018$\\
$61$&$3.78$&$-55.3$&$-37.8$&$-0.00015$&$0.0017$\\
$70$&$3.65$&$-68.6$&$-51.2$&$-0.010$&$0.0016$\\
\hline
\end{tabular}\caption{Material parameters for HgTe/CdTe quantum wells with different
well thicknesses $d$.}\label{tab:HgTeParameters}
\end{table}
\end{center}

For the purposes of studying the topological properties of this
system, as well as the edge states, it is sometimes convenient to work with a lattice regularization of the continuum model (\ref{contH}) which gives the energy spectrum over the entire Brillouin zone, i.e. a tight-binding representation. Since all the interesting physics at low energy occurs near the $\Gamma$ point, the behavior of the dispersion at energies much larger than the bulk gap at the $\Gamma$ point is not important. Thus, we can choose a regularization to simplify our calculations. This simplified lattice model consists of replacing (\ref{BHZ1}) by
\begin{eqnarray}
\epsilon (\mathbf{k})&=&C-2Da^{-2}(2-\cos k_xa-\cos k_ya),\nonumber\\
d_a (\mathbf{k})&=&\left(Aa^{-1} \sin k_xa,-Aa^{-1} \sin k_ya,M(\mathbf{k})\right),\nonumber\\
M(\mathbf{k})&=&M -2Ba^{-2} \left(2-\cos k_xa-\cos k_ya\right).
\label{BHZ2}
\end{eqnarray}
It is clear that near the
$\Gamma$ point, the lattice Hamiltonian reduces to the continuum BHZ model in Eq.~(\ref{contH}). For simplicity, below we work in units where the lattice constant $a=1$.

\subsection{Explicit solution of the helical edge states}\label{sec:edge2d}

The existence of topologically protected edge states is an important property of the QSH insulator. The edge states can be obtained by solving the BHZ model (\ref{BHZ}) with an open boundary condition. Consider the model Hamiltonian (\ref{BHZ}) defined on the half-space $x>0$ in the $xy$ plane. We can divide the model Hamiltonian into two
parts,
\begin{eqnarray}
    \hat{H}&=&\tilde{H}_0+\tilde{H}_1,\label{Hdecom2d}\\
    \tilde{H}_0&=&\tilde{\epsilon}(k_x)+\left(\begin{array}{cccc}
       \tilde{M}(k_x)&Ak_x&0&0\\
        Ak_x&-\tilde{M}(k_x)&0&0\\
        0&0&\tilde{M}(k_x)&-Ak_x\\
        0&0&-Ak_x&-\tilde{M}(k_x)
    \end{array}\right),\nonumber\\
    \tilde{H}_1&=&-Dk^2_{y}+\left(
    \begin{array}{cccc}
        -Bk_y^2&iAk_y&0&0\\
        -iAk_y&Bk_y^2&0&0\\
        0&0&-Bk_y^2&iAk_y\\
        0&0&-iAk_y&Bk_y^2
    \end{array}
    \right),
\end{eqnarray}
with $\tilde{\epsilon}(k_x)=C-Dk_x^2$ and $\tilde{M}(k_x)=M-Bk_x^2$.
All $k_x$-dependent terms are included in
$\tilde{H}_0$. For such a semi-infinite system, $k_x$ needs to be replaced by the operator $-i\partial_x$. On the other hand, translation symmetry along the $y$ direction is preserved, so that $k_y$ is a good quantum number. For $k_y=0$, we have $\tilde{H}_1=0$ and the wave equation is given by
\begin{eqnarray}
\tilde{H}_0 (k_x \rightarrow -i\partial_x) \Psi(x)=E\Psi(x).
\end{eqnarray}
Since $\tilde{H}_0$ is
block-diagonal, the eigenstates have the form
\begin{eqnarray}
        &&\Psi_\uparrow(x)=\left(
    \begin{array}{c}
        \psi_0\\
        \bold{0}
    \end{array}
    \right),\qquad
    \Psi_\downarrow(x)=\left(
    \begin{array}{c}
    \bold{0}\\
    \psi_0
    \end{array}
    \right),
    \label{eq:sur_wf}
\end{eqnarray}
where $\bold{0}$ is a two-component zero vector, and $\Psi_\uparrow(x)$ is related to $\Psi_\downarrow(x)$ by TR. For the edge states, the wave function $\psi_0(x)$ is localized at the edge and satisfies the wave equation
\begin{eqnarray}
    \left( \tilde\epsilon(-i\partial_x)+\left(\begin{array}{cc}\tilde{M}(-i\partial_x)&-iA_1\partial_x\\-iA_1\partial_x&-\tilde{M}(-i\partial_x)\end{array}\right) \right)\psi_0(x)=E\psi_0(x),
    \label{eq:sur_eqn}
\end{eqnarray}
which has been solved analytically for open boundary
conditions using different methods~\cite{koenig2008,zhou2008,linder2009,lu2010}. In order to show the existence of the edge states and
to find the region where the edge states exist,
we briefly review the derivation of the explicit form of the edge states by neglecting $\tilde\epsilon$ for simplicity~\cite{koenig2008}.

Neglecting $\tilde\epsilon$, the wave equation
(\ref{eq:sur_eqn}) has particle-hole symmetry. Therefore,
we expect that a special edge state with $E=0$
can exist. With the wave function ansatz $\psi_0=\phi e^{\lambda x}$,
the above equation can be simplified to
\begin{eqnarray}
    \left( M+B\lambda^2 \right)
    \tau_y\phi=A\lambda\phi,
    \label{eq:sur_eqn_cont1}
\end{eqnarray}
therefore the two-component wave function $\phi$ should be an eigenstate
of the Pauli matrix $\tau_y$. Defining a two-component spinor $\phi_\pm$ by $\tau_y\phi_\pm=\pm\phi_\pm$, Eq.~(\ref{eq:sur_eqn_cont1}) is simplified to a quadratic equation for $\lambda$. If $\lambda$ is a solution for $\phi_+$, then $-\lambda$ is a solution for $\phi_-$. Consequently, the general solution is given by
\begin{eqnarray}
    \psi_0(x)=(ae^{\lambda_1 x}+be^{\lambda_2 x})\phi_++(ce^{-\lambda_1 x}+de^{-\lambda_2 x})\phi_-,
    \label{eq:sur_wf1}
\end{eqnarray}
where $\lambda_{1,2}$ satisfy
\begin{eqnarray}
    \lambda_{1,2}=\frac{1}{2B}\left(A\pm\sqrt{A^2-4MB }\right).
    \label{eq:sur_lambda1}
\end{eqnarray}
The coefficients $a,b,c,d$ can be determined by imposing the open
boundary condition $\psi(0)=0$. Together with the normalizability of
the wave function in the region $x>0$, the open boundary condition
leads to an existence condition for the edge states:
$\Re\lambda_{1,2}<0$ ($c=d=0$) or $\Re\lambda_{1,2}>0$ ($a=b=0$), where $\Re$ stands for the real part. As
seen from Eq.~(\ref{eq:sur_lambda1}), these conditions can only
be satisfied in the inverted regime when $M/B>0$. Furthermore, one can show that when $A/B<0$, we have $\Re\lambda_{1,2}<0$, while when $A/B>0$, we have $\Re\lambda_{1,2}>0$. Therefore, the wave function for the edge states at the $\Gamma$ point is given by
\begin{eqnarray}
    \psi_0(x)=\left\{
    \begin{array}{cc}
        a\left( e^{\lambda_1 x}-e^{\lambda_2 x} \right)\phi_+,&A/B<0;\\
        c\left( e^{-\lambda_1 x}-e^{-\lambda_2 x} \right)\phi_-,&A/B>0.
    \end{array}
    \right.
    \label{eq:sur_wfpsi}
\end{eqnarray}
The sign of $A/B$ determines the spin polarization of the edge states, which is key to determine the helicity of the Dirac Hamiltonian for the topological edge states. Another important quantity characterizing the edge states is their decay length, which is defined as $l_c={\rm max}\left\{ |\Re\lambda_{1,2}|^{-1} \right\}$.

The effective edge model can be obtained by projecting the bulk Hamiltonian onto the edge states $\Psi_\uparrow$ and $\Psi_\downarrow$ defined in Eq. (\ref{eq:sur_wf}). This procedure leads to a $2\times 2$ effective Hamiltonian defined by $H_{\rm edge}^{\alpha\beta}(k_y)=\left\langle \Psi_\alpha\right|\left(\tilde{H}_0+\tilde{H}_1\right)\left|\Psi_\beta\right\rangle$. To leading order in $k_y$, we arrive at the effective Hamiltonian for the helical edge states:
\begin{eqnarray}
H_{\rm edge}= A k_y \sigma^z.
    \label{Effedge}
\end{eqnarray}
For HgTe QWs, we have $A\simeq 3.6$~eV$\cdot$\AA~\cite{koenig2008}, and the Dirac velocity of the edge states is given by $v=A/\hbar\simeq 5.5\times 10^5$~m/s.

\begin{figure}[t]
\begin{center}
\includegraphics[width=0.7\columnwidth]{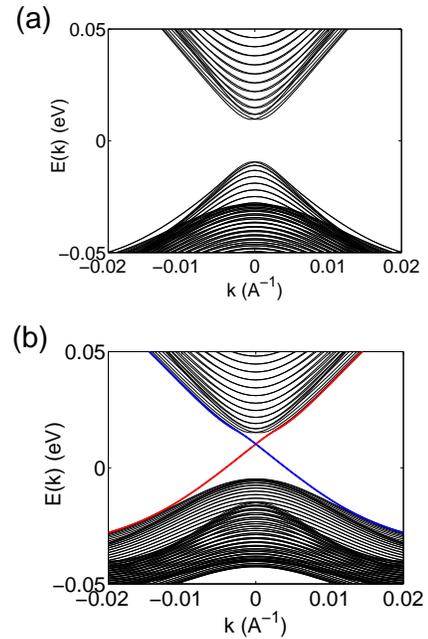}
\end{center}
\caption{Energy spectrum of the effective Hamiltonian (\ref{BHZ}) in a cylinder geometry. In a thin QW, (a) there is a gap between conduction band and valence band. In a thick QW, (b) there are gapless edge states on the left and right edge (red and blue lines, respectively). The dashed line stands for a typical value of the chemical potential within the bulk gap. Adapted from~\onlinecite{qi2010phystoday}.} \label{cylinderspec}
\end{figure}

The analytical calculation above can be confirmed by exact numerical diagonalization of the Hamiltonian (\ref{BHZ}) on a strip of finite width, which can also include the contribution of the $\epsilon(\mathbf{k})$ term~[Fig.~\ref{cylinderspec}]. The finite decay length of the helical edge states into the bulk determines the amplitude for interedge tunneling~\cite{zhou2008,hou2009,tanaka2009,teo2009,strom2009,zyuzin2010}.

\subsection{Physical properties of the helical edge states}\label{sec:edgeModel}

\subsubsection{Topological protection of the helical edge states}

From the explicit analytical solution of the BHZ model, there is a pair of helical edge states exponentially localized at the edge, and described by the effective helical edge theory (\ref{Effedge}). In this context, the concept of ``helical" edge state~\cite{wu2006} refers to the fact that states with opposite spin counter-propagate at a given edge, as we see from the edge state dispersion relation shown in Fig.~\ref{cylinderspec}(b), or the real space picture shown in Fig.~\ref{qh_qsh}(b). This is in sharp contrast to the ``chiral" edge states in the QH state, where the edge states propagate in one direction only, as shown in Fig.~\ref{qh_qsh}(a).

In the QH effect, the chiral edge states can not be backscattered for sample widths larger than the decay length of the edge states. In the QSH effect, one may naturally ask whether backscattering of the helical edge states is possible. It turns out that TR symmetry prevents the helical edge states from backscattering. The absence of backscattering relies on the destructive interference between all possible backscattering paths taken by the edge electrons.

\begin{figure}[t]
    \begin{center}
        \includegraphics[scale=0.4]{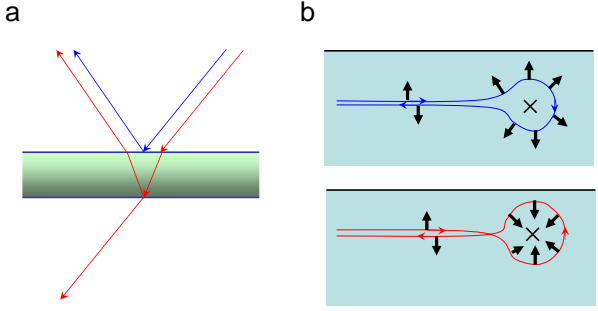}
    \end{center}
    \caption{(a) On a lens with antireflective coating, light reflected by top (blue line) and bottom (red line) surfaces interferes
destructively, leading to suppressed reflection. (b) Two
possible paths taken by an electron on a QSH edge when scattered by a
nonmagnetic impurity. The electron spin rotates by $180^\circ$ clockwise
along the blue curve, and counterclockwise along the red curve. A
geometrical phase factor associated with this rotation of the spin leads to destructive interference between the two paths. In other words, electron backscattering on the QSH edge is suppressed in a way similar to how the reflection of photons is suppressed by an antireflective coating. Adapted from~\onlinecite{qi2010phystoday}.}
    \label{antireflection}
\end{figure}

Before giving a semiclassical argument why this is so, we first consider an analogy from daily experience. Most eyeglasses and camera lenses have an antireflective coating [Fig.~\ref{antireflection}(a)], where light reflected from the top and bottom surfaces interfere destructively, leading to no net reflection and thus perfect
transmission. However, this effect is not robust, as it depends
on a precise matching between the wavelength of light and the thickness of the coating. Now we turn to the helical edge states.
If a nonmagnetic impurity is present near the edge, it can in principle cause backscattering of the helical edge states due to SOC. However, just as for the reflection of photons by a surface, an electron can be reflected by a nonmagnetic impurity, and different reflection paths interfere quantum-mechanically. A forward-moving electron with spin up on the QSH edge can make either a clockwise or a counterclockwise turn around the impurity [Fig.~\ref{antireflection}(b)]. Since only spin down electrons can propagate backwards, the electron spin has to rotate adiabatically, either by an angle of $\pi$ or $-\pi$, i.e. into the opposite direction. Consequently, the two paths differ by a full $\pi - (-\pi)=2\pi$ rotation of the electron spin. However, the wave function of a spin-1/2 particle picks up a negative sign under a full $2\pi$ rotation. Therefore, two backscattering paths related by TR always
interfere destructively, leading to perfect transmission. If the impurity carries a magnetic moment, TR symmetry is explicitly
broken, and the two reflected waves no longer interfere destructively. In this way, the robustness of the QSH edge state is protected by TR symmetry.

The physical picture described above applies only to the case of a single pair of QSH edge states~\cite{kane2005a,wu2006,xu2006}. If there are two
forward-movers and two backward-movers on a given
edge, an electron can be scattered from a forward-moving to a
backward-moving channel without reversing its spin. This spoils
the perfect destructive interference described above, and leads to dissipation. Consequently, for the QSH state to be robust, edge states must consist of an odd number of forward (backward) movers. This
even-odd effect is the key reason why the QSH insulator is
characterized by a $\mathbb{Z}_2$ topological quantum number~\cite{kane2005a,wu2006,xu2006}.

The general properties of TR symmetry are important for understanding the properties of the edge theory. The anti-unitary TR operator $T$ takes different forms depending on whether the degrees of freedom have integer or half-odd-integer spin. For half-odd-integer spin, we have $T^2=-1$
which implies, by Kramers' theorem, that any single-particle
eigenstate of the Hamiltonian must have a degenerate partner. From Fig.~\ref{cylinderspec}(b), we see that the two dispersion branches at one given edge cross each other at the TR invariant $k=0$ point. At this point, these two degenerate states exactly satisfy Kramers' theorem. If we add TR invariant perturbations to the Hamiltonian, we can move the degenerate point up and down in energy, but cannot remove the degeneracy. In this precise sense, the helical edge states are topologically protected by TR symmetry.

If TR symmetry is not present, a simple ``mass" term
can be added to the Hamiltonian so that the spectrum becomes
gapped:
\begin{eqnarray}
H_{\rm
mass}=m \int\frac{dk}{2\pi}\left(\psi^\dagger_{k+}\psi_{k-}+\mathrm{h.c.}\right),
\nonumber
\end{eqnarray}
where $\mathrm{h.c.}$ denotes Hermitian conjugation, and $\psi^\dag_{k\pm},\psi_{k\pm}$ are creation/annihilation operators for an edge electron of momentum $k$, with $\pm$ denoting the electron spin. The action of TR symmetry on the electron operators is given by
\begin{eqnarray}
T\psi_{k+}T^{-1}=\psi_{-k,-},~T\psi_{k-}T^{-1}=-\psi_{-k,+},
\end{eqnarray}
which implies
\begin{eqnarray}
TH_{\rm mass}T^{-1}=-H_{\rm mass}.\nonumber
\end{eqnarray}
Consequently, $H_{\rm mass}$ is a TR symmetry breaking
perturbation. More generally, if we define the ``chirality"
operator
\begin{eqnarray}
C=N_+-N_-=\int
\frac{dk}{2\pi}\left(\psi_{k+}^\dagger\psi_{k+}-\psi_{k-}^\dagger
\psi_{k-}\right)\nonumber,
\end{eqnarray}
any operator that changes $C$ by $2(2n-1),~n\in\mathbb{Z}$ is
odd under TR. In other words, TR symmetry only allows $2n$-particle backscattering, described by operators such as
$\psi_{k+}^\dagger\psi_{k'+}^\dagger\psi_{p-}\psi_{p'-}$ (for $n=1$).
Therefore, the most relevant perturbation
$\psi_{k+}^\dagger\psi_{k'-}$ is forbidden by TR symmetry, which is essential for the topological stability of the
edge states. This edge state effective theory is nonchiral, and
is qualitatively different from the usual spinless or spinful
Luttinger liquid theories. It can be considered as a new class of
1D critical theories, dubbed a ``helical
liquid"~\cite{wu2006}. Specifically, in the
noninteracting case no TR invariant perturbation is
available to induce backscattering, so that the edge state is
robust.

Consider now the case of two flavors of helical edge states on
the boundary, i.e. a 1D system consisting of two left-movers and
two right-movers with Hamiltonian
\begin{eqnarray}
H=\int\frac{dk}{2\pi}\sum_{s=1,2}\left(\psi^\dagger_{ks+}vk\psi_{ks+}-\psi^\dagger_{ks-}vk\psi_{ks-}\right).\nonumber
\end{eqnarray}
A mass term such as $\tilde{m}\int
\frac{dk}{2\pi}\left(\psi^\dagger_{k1+}\psi_{k2-}-\psi^\dagger_{k1-}\psi_{k2+}+\mathrm{h.c.}\right)$ (with $\tilde{m}$ real)
can open a gap in the system while preserving time-reversal symmetry. In other words, two copies of the helical liquid form a a topologically trivial theory. More generally, an edge system with TR symmetry is a nontrivial helical liquid when there is an odd number of left- (right-) movers, and trivial when there is an even number of them. Thus the topology of QSH systems are characterized by a $\mathbb{Z}_2$ topological quantum number.

\subsubsection{Interactions and quenched disorder}

We now review the effect of interactions and quenched disorder on the QSH edge liquid~\cite{wu2006,xu2006}. Only two TR invariant nonchiral interactions can be added to Eq.~\ref{Effedge}, the forward and Umklapp scatterings
\begin{eqnarray} H_\mathrm{f}&=& g \int dx\,
\psi^\dagger_{+} \psi_{+}
\psi^\dagger_{-} \psi_{-} \label{eq:fw}\\
H_\mathrm{u}&=& g_\mathrm{u} \int dx\,e^{-i 4 k_F x}
\psi^\dagger_{+} (x) \psi^\dagger_{+} (x+a)
\nonumber \\
&&\times\psi_{-} (x+a)\psi_{-}(x) +\mathrm{h.c.}, \label{eq:umklapp} \end{eqnarray}
where the two-particle operators $\psi^\dag\psi^\dag,\psi\psi$ are point-split with the lattice constant $a$ which plays the role of a short-distance cutoff. The
chiral interaction terms only renormalize the Fermi velocity $v$,
and are thus ignored. It is well known that the forward scattering
term gives a nontrivial Luttinger parameter $K=\sqrt{(v-g)/(v+g)}$, but keeps the system gapless. Only the Umklapp term has the potential to open up a gap at the commensurate filling $k_F=\pi/2$. The bosonized form of the Hamiltonian reads
\begin{eqnarray}
H&=& \int d x \frac{\bar{v}}{2}\Big\{ \frac{1}{K} (\partial_x
\phi)^2 + K (\partial_x \theta)^2 \Big\}
+ \frac{g_\mathrm{u} \cos \sqrt {16 \pi} \phi }{2 (\pi a)^2}, \nonumber \\
\end{eqnarray}
where $\bar{v}=\sqrt{v^2-g^2}$ is the renormalized velocity, and we define nonchiral bosons $\phi=\phi_R+\phi_L$ and $\theta=\phi_R-\phi_L$, respectively, where $\phi_R$ and $\phi_L$ are chiral bosons describing the spin up (down) right-mover and the spin down (up) left-mover, respectively. $\phi$ contains both spin and charge degrees of freedom, and is equivalent to the combination $\phi_c-\theta_s$ in the spinful Luttinger liquid, with $\phi_c$ and $\theta_s$ the charge and spin bosons, respectively~\cite{giamarchi2003}. It is also a compact variable with period $\sqrt\pi$. A renormalization group analysis
shows that the Umklapp term is relevant for $K<1/2$ with a
pinned value of $\phi$. Consequently, a gap $\Delta\sim a^{-1}
(g_u)^{\frac{1}{2-4K}}$ opens and spin transport is blocked.
The mass order parameters $N_{x,y}$ the bosonized form of which is $N_x= \frac{i\eta_R \eta_L} {2\pi a} \sin \sqrt{4\pi} \phi$, $N_y=
\frac{i\eta_R \eta_L} {2\pi a} \cos \sqrt{4\pi} \phi$, are odd
under TR. For $g_\mathrm{u}<0$, $\phi$ is pinned at either
$0$ or $\sqrt\pi/2$, and the $N_y$ order is Ising-like. At
$T=0$, the system is in a Ising ordered phase, and TR symmetry is spontaneously broken. On the other hand, when $0<T\ll \Delta$,
$N_y$ is disordered, the gap remains, and TR symmetry is restored by thermal fluctuations. A similar reasoning applies to the case $g_\mathrm{u}>0$ where $N_x$ is the order parameter.

There is also the possibility of two-particle backscattering due
to quenched disorder, described by the term
\begin{eqnarray}
H_\mathrm{dis}&=& \int dx\,
\frac{ g_\mathrm{u}(x) }{2 (\pi a)^2}\cos \sqrt {16 \pi} (\phi
(x,\tau)+\alpha(x) ),
 \end{eqnarray}
where the scattering strength $g_\mathrm{u}(x)$ and phase $\alpha(x)$ are Gaussian random variables. The standard replica analysis shows that
disorder becomes relevant at $K<3/8$~\cite{giamarchi1988,wu2006,xu2006}.
At $T=0$, $N_{x,y}(x)$ exhibits glassy behavior, i.e. disordered in the spatial direction but static in the
time direction. Spin transport is thus blocked and TR symmetry
is again spontaneously broken at $T=0$. At low but
finite $T$, the system remains gapped with TR symmetry restored.

In the above, we have seen that the helical liquid can {\it in
principle} be destroyed. However, for a reasonably weak interacting system, i.e. $K\approx 1$, the one-component helical liquid remains gapless. In an Ising ordered phase, the low-energy excitations on
the edge are Ising domain walls which carry fractional $e/2$
charge~\cite{qi2008}. The properties of multi-component helical
liquids in the presence of disorder has also been studied~\cite{xu2006}.

A magnetic impurity on the edge of a QSH insulator is expected to act as a local mass term for the edge theory, and thus is expected to lead to a suppression of the edge conductance. While this is certainly true for a static magnetic impurity, a \emph{quantum} magnetic impurity, i.e. a Kondo impurity, leads to subtler behavior~\cite{wu2006,maciejko2009}. In the presence of a quantum magnetic impurity, due to the combined effects of interactions and SOC one must also generally consider local two-particle backscattering processes~\cite{meidan2005} similar to Eq.~(\ref{eq:umklapp}), but occurring only at the position of the impurity. At high temperatures, both weak Kondo and weak two-particle backscattering are expected to give rise to a logarithmic temperature dependence as in the usual Kondo effect~\cite{maciejko2009}, and their effect is not easily distinguishable. However, at low temperatures the physics depends drastically on the strength of Coulomb interactions on the edge, parameterized by the Luttinger parameter $K$. For weak Coulomb interactions $K>1/4$, the edge conductance is restored to the unitarity limit $2e^2/h$ with unusual power laws characteristic of a ``local helical liquid''~\cite{wu2006,maciejko2009}. For strong Coulomb interactions $K<1/4$, the conductance vanishes at $T=0$, but is restored at low $T$ by a fractionalized tunneling current of charge $e/2$ quasiparticles~\cite{maciejko2009}. The tunneling of a charge $e/2$ quasiparticle is described by an instanton process which is the time counterpart to the static $e/2$ charge on a spatial magnetic domain wall along the edge~\cite{qi2008}. In addition to the single-channel Kondo effect just described, the possibility of an even more exotic two-channel Kondo effect on the edge of the QSH insulator has also been studied recently~\cite{law2010}.

\subsubsection{Helical edge states and the holographic principle}

\begin{figure}[t]
\centering
\includegraphics[width=0.45\textwidth]{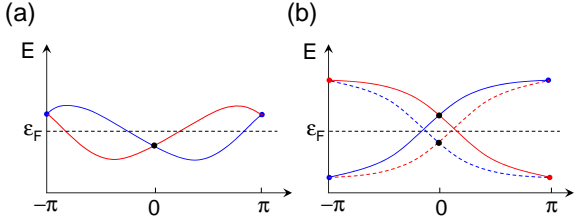}
\caption{(a) Energy dispersion of a 1D
TR invariant system. The Kramers degeneracy is required at
$k=0$ and $k=\pi$, so that the energy spectrum always crosses $4n$ times the Fermi level $\epsilon_F$. (b) Energy dispersion of
the helical edge states on one boundary of the QSH system (solid
lines). At $k=0$ the edge states are Kramers partners, while at $k=\pi$ they merge into the bulk and pair with the
edge states of the other boundary (dash lines). In both (a) and
(b), red and blue lines represent the two partners of a Kramers pair. From~\onlinecite{koenig2008}.}
\label{nogotheorem}
\end{figure}

There is an alternative way to understand the qualitative
difference between an even and odd number of edge states in terms of a ``fermion doubling" theorem~\cite{wu2006}. This theorem states that there is always an even number of Kramers pairs at the Fermi energy for a TR invariant, but otherwise arbitrary 1D band structure. A single pair of helical states can occur only ``holograhically", i.e. when the 1D system is the boundary of a 2D system. This fermion doubling theorem is a
TR invariant generalization of the Nielsen-Ninomiya no-go theorem for chiral fermions on a lattice~\cite{nielsen1981}.
For spinless fermions, there is always an equal number of left-movers and right-movers at the Fermi level, which leads to the
fermion doubling problem in odd spatial dimensions. A geometrical
way to understand this result is that for periodic functions (i.e. energy spectra of a lattice model), ``what goes up must eventually come down". Similarly, for a TR symmetric system with half-odd-integer spins, Kramers' theorem requires that each eigenstate of the Hamiltonian
is accompanied by its TR conjugate or Kramers partner, so that the number of low-energy channels is doubled. A Kramers pair of states at $k=0$ must recombine into pairs when $k$ goes from $0$ to $\pi$ and $2\pi$, which requires the bands to cross the Fermi level $4n$ times [Fig.~\ref{nogotheorem}(a)]. However, there is an exception to this
theorem, which is analogous to the reason why a chiral liquid can exist in the QH effect. A helical liquid with an odd number of fermion branches \emph{can} occur if it is
holographic, i.e. if it appears at the boundary (edge) of a 2D system. In this case, the edge states are Kramers partners at $k=0$, but merge into the bulk at some finite $k_c$, such that they do not
have to be combined at $k=\pi$. More accurately, the edge states on both left and right boundaries becomes bulk states for $k>k_c$ and form a Kramers pair [Fig.~\ref{nogotheorem}(b)]. This is exactly the behavior discussed in Sec.~\ref{sec:edge2d} in the context of the analytical solution
the edge state wave functions.

The fermion doubling theorem also provides a physical understanding of the topological stability of the helical liquid. Any local
perturbation on the boundary of a 2D QSH system is equivalent to the action of coupling a ``dirty surface layer" to the unperturbed helical
edge states. Whatever perturbation is considered, the ``dirty
surface layer" is always 1D, such that there
is always an even number of Kramers pairs of low-energy
channels. Since the helical liquid has only an odd number of
Kramers pairs, the coupling between them can only annihilate an
{\em even} number of Kramers pairs if TR is preserved. As a result, at least one pair of gapless edge states can survive.

This fermion doubling theorem can be generalized to 3D in a straightforward way. In the 2D QSH state, the simplest helical edge state consists of a single massless Dirac fermion in ($1+1$)D. The simplest 3D topological insulator contains a surface state consisting of a single massless Dirac fermion in ($2+1$)D. A single massless Dirac fermion would also violate the fermion doubling theorem and cannot exist in a purely 2D system with TR symmetry. However, it can exist holographically, as the boundary of a 3D topological insulator. More generically, there is a one-to-one correspondence between topological insulators and robust gapless theories in one lower dimension\cite{kitaev2009,freedman2010,teo2010}

\subsubsection{Transport theory of the helical edge states}
\label{sec:transport}

In conventional diffusive electronics, bulk transport satisfies Ohm's law. Resistance is
proportional to the length and inversely proportional to the
cross-sectional area, implying the existence of a local
resistivity or conductivity tensor. However, in systems such as
the QH and QSH states, the existence of edge states necessarily leads to nonlocal transport which invalidates the concept of local resistivity. Such nonlocal transport has been experimentally observed in the QH regime in the presence of a large magnetic field~\cite{beenakker1991}, and the nonlocal transport is well described by a quantum transport theory based on the Landauer-B\"{u}ttiker formalism~\cite{buttiker1988}.
A similar transport theory has been developed for the helical edge states of the QSH states, and the nonlocal transport experiments are in excellent agreement with the theory~\cite{roth2009}.
These measurements are now widely acknowledged as constituting
definitive experimental evidence for the existence of edge states
in the QSH regime~\cite{buttiker2009}.

Within the general Landauer-B\"{u}ttiker formalism~\cite{buttiker1986}, the current-voltage relationship is expressed as
\begin{equation}\label{Landauer}
I_i=\frac{e^2}{h} \sum_j (T_{ji} V_i - T_{ij} V_j),
\end{equation}
where $I_i$ is the current flowing out of the $i$th electrode
into the sample region, $V_i$ is the voltage on the $i$th
electrode, and $T_{ji}$ is the transmission probability from the
$i$th to the $j$th electrode. The total current is conserved in
the sense that $\sum_i I_i=0$. A voltage lead $j$ is defined by
the condition that it draws no net current, i.e. $I_j=0$.
The physical currents remain unchanged if the voltages on all
electrodes are shifted by a constant amount $\mu$, implying that
$\sum_i T_{ij}=\sum_i T_{ji}$. In a TR invariant
system, the transmission coefficients satisfy the condition
$T_{ij}=T_{ji}$.

For a general 2D sample, the number of transmission
channels scales with the width of the sample, so that the
transmission matrix $T_{ij}$ is complicated and nonuniversal.
However, a tremendous simplification arises if the quantum
transport is entirely dominated by the edge states. In the QH
regime, chiral edge states are responsible for the transport. For
a standard Hall bar with $N$ current and voltage leads attached, the transmission matrix
elements for the $\nu=1$ QH state are given by $T({\rm
QH})_{i+1,i}=1$, for $i=1,\ldots,N$, and all other matrix elements
vanish identically. Here we periodically identify the $i=N+1$
electrode with $i=1$. Chiral edge states are protected from
backscattering, therefore, the $i$th electrode transmits
perfectly to the neighboring ($i+1$)th electrode on one side only.
In the example of current leads on the electrodes $1$ and $4$, and
voltage leads on the electrodes $2$, $3$, $5$ and $6$, (see the inset of Fig.~\ref{FigexpBcross} for the labeling), one finds
that $I_1=-I_4\equiv I_{14}$, $V_2-V_3=0$ and
$V_1-V_4=\frac{h}{e^2} I_{14}$, giving a four-terminal resistance
of $R_{14,23}=0$ and a two-terminal resistance of
$R_{14,14}=\frac{h}{e^2}$.

The helical edge states can be viewed as two copies of chiral edge
states related by TR symmetry. Therefore, the
transmission matrix is given by $T({\rm QSH})=T({\rm
QH})+T^\dagger({\rm QH})$, implying that the only nonvanishing
matrix elements are given by
\begin{equation}\label{Transmission}
T({\rm QSH})_{i+1,i}=T({\rm QSH})_{i,i+1}=1.
\end{equation}
Considering again the example of current leads on the electrodes
$1$ and $4$, and voltage leads on the electrodes $2$, $3$, $5$ and
$6$, one finds that $I_1=-I_4\equiv I_{14}$,
$V_2-V_3=\frac{h}{2e^2} I_{14}$ and $V_1-V_4=\frac{3h}{e^2}
I_{14}$, giving a four-terminal resistance of
$R_{14,23}=\frac{h}{2e^2}$ and a two-terminal resistance of
$R_{14,14}=\frac{3h}{2e^2}$. Four terminal resistance with different configurations of voltage and current probes can be predicted in the same way, which are all rational fractions of $h/e^2$.
The experimental data [Fig.~\ref{nonlocalFig3}] neatly confirms all these highly nontrivial theoretical predictions~\cite{roth2009}. For
two micro Hall bar structures that differ only in the dimensions
of the area between the voltage contacts 3 and 4, the expected
resistance values $R_{14,23}=\frac{h}{2e^2}$ and
$R_{14,14}=\frac{3h}{2e^2}$ are indeed observed for gate voltages for which the samples are in the QSH regime.

As mentioned earlier, one might sense a paradox between the
dissipationless nature of the QSH edge states and the finite
four-terminal longitudinal resistance $R_{14,23}$, which vanishes
in the QH state. We can generally assume that the microscopic
Hamiltonian governing the voltage leads is invariant under TR symmetry. Therefore, one would naturally ask how such
leads could cause the dissipation of the helical edge states,
which are protected form backscattering by TR symmetry? In nature, TR symmetry can be broken in two ways, either at the level
of the microscopic Hamiltonian, or at the level of the macroscopic
irreversibility in systems whose microscopic Hamiltonian respects
TR symmetry. When the helical edge states propagate
without dissipation inside the QSH insulator between the
electrodes, neither forms of TR symmetry breaking are
present. As a result, the two counter-propagating channels can be
maintained at two different quasi-chemical potentials, leading to
a net current flow. However, once they enter the voltage leads,
they interact with a reservoir containing a large number of
low-energy degrees of freedom, and TR symmetry is
effectively broken by the macroscopic irreversibility. As a
result, the two counter-propagating channels equilibrate at the
same chemical potential, determined by the voltage of the lead.
Dissipation occurs with the equilibration process. The transport
equation (\ref{Landauer}) breaks the macroscopic TR
symmetry, even though the microscopic TR symmetry is
ensured by the relationship $T_{ij}=T_{ji}$. In contrast to the
case of the QH state, the absence of dissipation in the QSH helical
edge states is protected by Kramers' theorem, which relies on the
quantum phase coherence of wave functions. Thus, dissipation can
occur once phase coherence is destroyed in the metallic leads.
On the contrary, the robustness of QH chiral edge states does not
require phase coherence. A more rigorous and microscopic analysis
of the different role played by a metallic lead in QH and QSH
states has been performed~\cite{roth2009}, the result of
which agrees with the simple transport equations (\ref{Landauer})
and (\ref{Transmission}). These two equations correctly describe
the dissipationless quantum transport inside the QSH insulator,
and the dissipation inside the electrodes. As shown in Sec.~\ref{sec:nonlocaltransport}, these equations can be put to more
stringent experimental tests.

The unique helical edge states of the QSH state can be used to construct devices with interesting transport properties~\cite{akhmerov2009b,lbzhang2009,kharitonov2010}. Besides the edge state transport, the QSH state also leads to interesting bulk transport properties~\cite{novik2010}.

\subsection{Topological excitations}
\label{sec:fractional}

In the previous sections, we discussed the transport properties of the helical edge states in the QSH state. Unlike the case of the QH state, these transport properties are not expected to be precisely quantized, since they are not directly related to the $\mathbb{Z}_2$ topological invariant which characterizes the topological state. In this section, we show that it is possible to measure the $\mathbb{Z}_2$ topological quantum number directly in experiments. We shall discuss two examples. The first is the fractional charge and quantized current experiments at the edge of a QSH system~\cite{qi2008}. Second, we discuss the spin-charge separation effect occurring in the
bulk of the sample~\cite{qi2008a,ran2008}.

\subsubsection{Fractional charge on the edge}

The first theoretical proposal we discuss is that of a localized fractional
charge at the edge of a QSH sample when a magnetic domain wall is
present. The concept of fractional charge in a condensed matter
system induced at a mass domain wall goes back to the
Su-Schrieffer-Heeger (SSH) model~\cite{su1979}. For spinless
fermions, a mass domain wall induces a localized state with
one-half of the electron charge. However, for a real material such as
polyacetylene, two spin orientations are present for each
electron, and because of this doubling, a domain wall in
polyacetylene only carries integer charge. The beautiful proposal
of SSH, and its counterpart in field theory, the Jackiw-Rebbi
model~\cite{jackiw1976}, have never been
experimentally realized. As mentioned earlier, conventional 1D electronic
systems have four basic degrees of freedom, i.e. forward- and
backward-movers with two spins. However, a helical liquid
at a given edge of the QSH insulator has only two: a spin up (down) forward-mover and a spin down (up) backward-mover. Therefore,
the helical liquid has \emph{half} the degrees of freedom of a
conventional 1D system, and thus avoids the doubling
problem. Because of this fundamental topological property of the
helical liquid, a domain wall carries charge $e/2$. In addition,
if the magnetization is rotated periodically, a quantized charge
current will flow. This provides a direct
realization of the Thouless topological pump~\cite{thouless1983}.

\begin{figure}[t]
\includegraphics[width=3in] {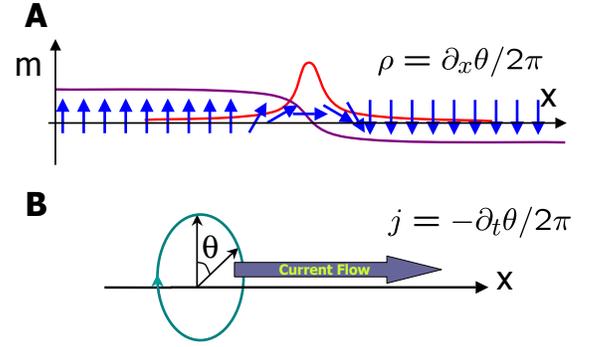}
\caption{ (a) Schematic picture of the half-charge on a domain wall. The blue arrows show a magnetic domain wall configuration and the purple line shows the mass kink. The red curve shows the charge density distribution. (b) Schematic picture of the pumping induced by the rotation of
magnetic field. The blue circle with arrow shows the rotation of the magnetic field vector. Adapted from~\onlinecite{qi2008}.}\label{QSHdomainAndChargePump}
\end{figure}

We begin with the edge Hamiltonian given in Eq.~(\ref{Effedge}).
These helical fermion states only have two degrees of freedom; the spin polarization is correlated with the direction of motion. A
mass term, being proportional to the Pauli matrices $\sigma^{1,2,3}$, can only be introduced in the Hamiltonian by
coupling to a TR symmetry breaking external field such as a magnetic field, aligned magnetic impurities~\cite{gao2009}, or interaction-driven ferromagnetic order on the edge~\cite{kharitonov2010}. To leading order in perturbation
theory, a magnetic field generates the mass terms
\begin{eqnarray}
H_M&=&\int dx \,\Psi^{\dagger}\sum_{a=1,2,3}m_a(x,t)\sigma^a\Psi\nonumber\\&=&\int
dx \Psi^{\dagger}\sum_{a,i}{t}_{ai}B_i(x,t)\sigma^a\Psi,\label{HM}
\end{eqnarray}\noindent
where $\Psi=\left(\psi_{+},\psi_{-}\right)^{T}$ and the model-dependent coefficient matrix $t_{ai}$ is determined by the
coupling of the edge states to the magnetic field. According to
the work of Goldstone and Wilczek~\cite{goldstone1981}, at zero
temperature the ground-state charge density $j_0\equiv\rho$ and current $j_1\equiv j$ in a
background field $m_a(x,t)$ is given by
\begin{eqnarray}
j_\mu=\frac1{2\pi}\frac{1}{\sqrt{m_\alpha
m^\alpha}}\epsilon^{\mu\nu}\epsilon^{\alpha\beta}m_\alpha\partial_\nu
m_\beta,~\alpha,\beta=1,2,\nonumber
\end{eqnarray}
with $\mu,\nu=0,1$ corresponding to the time and space components,
respectively, and $m_3$ does not enter the long-wavelength
charge-response equation. If we parameterize the mass terms in terms of an angular variable $\theta$, i.e.
$m_1=m\cos\theta$, $m_2=m\sin\theta$, the response equation is
simplified to
\begin{equation}
\rho=\frac{1}{2\pi}\partial_{x}\theta(x,t),\;\;\;
j=-\frac{1}{2\pi}\partial_{t}\theta(x,t).\label{GWformula}
\end{equation}
Such a response is topological in the sense that the net charge
$Q$ in a region $[x_1,x_2]$ at time $t$ depends only on the
boundary values of  $\theta(x,t)$ \emph{i.e.}
$Q=\left[\theta(x_2,t)-\theta(x_1,t)\right]/2\pi$. In particular,
a half-charge $\pm e/2$ is carried by an anti-phase domain wall of
$\theta$ [Fig.~\ref{QSHdomainAndChargePump}(a)] \cite{jackiw1976}. Similarly,
the charge pumped by a purely time-dependent $\theta(t)$ field in
a time interval $[t_1,t_2]$ is  $\Delta Q_{\rm
pump}|_{t_1}^{t_2}=\left[\theta(t_2)-\theta(t_1)\right]/2\pi$.
When $\theta$ is rotated from $0$ to $2\pi$ adiabatically, a
quantized charge $e$ is pumped through the 1D system [Fig.~\ref{QSHdomainAndChargePump}(b)].

From the linear relation $m_a=t_{ai}B_i$, the angle $\theta$ can
be determined for a given magnetic field ${\bf B}$. Independent from the details of $t_{ai}$, opposite magnetic fields ${\bf B}$ and ${\bf -B}$ always correspond to opposite mass, so that $\theta({\bf B})=\theta(-{\bf
B})+\pi$. Thus the charge localized on an anti-phase magnetic domain
wall of magnetization field is always $e/2$ mod $e$, which is a direct manifestation of the $\mathbb{Z}_2$ topological quantum number of the QSH state. Such a half charge is detectable in a specially designed single-electron transistor device\cite{qi2008}.

\subsubsection{Spin-charge separation in the bulk}
\begin{figure}[htpb]
    \begin{center}
        \includegraphics[width=3in]{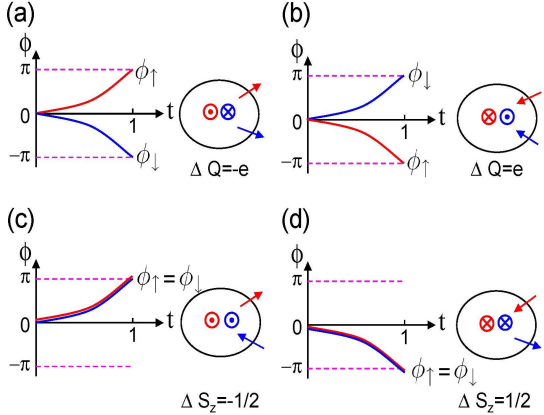}
    \end{center}
    \caption{Four different adiabatic processes from
    $\phi_\uparrow=\phi_\downarrow=0$ to $\phi_\uparrow=\phi_\downarrow=\pm\pi$. The red (blue) curve stands for
    the flux $\phi_{\uparrow(\downarrow)}(t)$, respectively. The symbol ``$\odot$" (``$\otimes$") represents
    increasing (decreasing) fluxes, and the arrows show the current
    into and out of the Gaussian loop, induced
    by the changing flux. Charge is pumped in the processes with
    $\phi_\uparrow(t)=-\phi_\downarrow(t)$, while spin is pumped in those with
    $\phi_\uparrow(t)=\phi_\downarrow(t)$. From~\onlinecite{qi2008a}.}
    \label{fig:spinonholon}
\end{figure}

In addition to the fractional charge on the edge, there have been theoretical proposals for a bulk spin-charge separation effect~\cite{qi2008a,ran2008}. These ideas are similar to the $\mathbb{Z}_2$ spin pump proposed in~\cite{fu2006}. We first present an argument which is physically intuitive, but only valid when there is at least a $U(1)_s$ spin rotation symmetry, e.g. when $S_z$ is conserved. In this case, the QSH effect is simply
defined as two copies of the QH effect, with opposite Hall conductances of
$\pm e^2/h$ for opposite spin orientations. Without loss of
generality, we first consider a disk geometry with an
electromagnetic gauge flux of
$\phi_\uparrow=\phi_\downarrow=hc/2e$, or simply $\pi$ in units of
$\hbar=c=e=1$, through a hole at the center [Fig.~\ref{fig:spinonholon}]. The gauge flux acts on both spin
orientations, and the $\pi$ flux preserves TR symmetry.
We consider adiabatic processes $\phi_\uparrow(t)$ and
$\phi_\downarrow(t)$, where
$\phi_\uparrow(t)=\phi_\downarrow(t)=0$ at $t=0$, and
$\phi_\uparrow(t)=\phi_\downarrow(t)=\pm\pi$ at $t=1$. Since the
flux of $\pi$ is equivalent to the flux of $-\pi$, there are four
different adiabatic processes all reaching the same final flux
configuration. In process (a),
$\phi_\uparrow(t)=-\phi_\downarrow(t)$ and
$\phi_\uparrow(t=1)=\pi$. In process (b),
$\phi_\uparrow(t)=-\phi_\downarrow(t)$ and
$\phi_\uparrow(t=1)=-\pi$. In process (c),
$\phi_\uparrow(t)=\phi_\downarrow(t)$ and
$\phi_\uparrow(t=1)=\pi$. In process (d),
$\phi_\uparrow(t)=\phi_\downarrow(t)$ and
$\phi_\uparrow(t=1)=-\pi$. These four processes are illustrated in
Fig.~\ref{fig:spinonholon}. Processes (a) and (b)
preserve TR symmetry at all intermediate stages, while processes (c) and
(d) only preserve TR symmetry at the final stage.

We consider a Gaussian loop surrounding the flux. As the flux
$\phi_\uparrow(t)$ is turned on adiabatically, Faraday's law of
induction states that a tangential electric field ${\bf
E}_\uparrow$ is induced along the Gaussian loop. The quantized
Hall conductance implies a radial current ${\bf
j}_\uparrow=\frac{e^2}{h} \hat{\mathbf{z}}\times {\bf E}_\uparrow$,
resulting in a net charge flow $\Delta Q_\uparrow$ through the
Gaussian loop:
\begin{eqnarray}
\Delta Q_\uparrow & = & - \int_0^1 dt \int d {\bf n}\cdot {\bf
j}_\uparrow = - \frac{e^2}{h} \int_0^1 dt \int d {\bf l}\cdot {\bf
E}_\uparrow \nonumber\\
& = & - \frac{e^2}{hc} \int_0^1 dt  \frac{\partial \phi}{\partial
t} = - \frac{e^2}{hc} \frac{hc}{2e}=-\frac{e}{2}.
\end{eqnarray}
An identical argument applied to the spin down component shows
that $\Delta Q_\downarrow=-e/2$. Therefore, this adiabatic process
creates the holon state with $\Delta Q=\Delta Q_\uparrow+\Delta
Q_\downarrow=-e$ and $\Delta S_z=\Delta Q_\uparrow-\Delta
Q_\downarrow=0$.

Applying similar arguments to process (b) gives $\Delta
Q_\uparrow=\Delta Q_\downarrow=e/2$, which leads to a chargeon
state with $\Delta Q=e$ and $\Delta S_z=0$. Processes (c) and (d)
give $\Delta Q_\uparrow=-\Delta Q_\downarrow=e/2$ and $\Delta
Q_\uparrow=-\Delta Q_\downarrow=-e/2$ respectively, which yield
the spinon states with $\Delta Q=0$ and $\Delta S_z=\pm 1/2$. The
Hamiltonians $H(t)$ in the presence of the gauge flux are the same
at $t=0$ and $t=1$, but differ in the intermediate stages of the
four adiabatic processes. Assuming that the ground state is unique
at $t=0$, we obtain four final states at $t=1$, which are the
holon, chargeon and the two spinon states. Both the spin and the
charge quantum numbers are sharply defined quantum
numbers~\cite{kivelson1982}. The insulating state has a bulk gap
$\Delta$, and an associated coherence length $\xi\sim A/\Delta$
where $A$ is the Dirac parameter in Eq.~\ref{contH}. As long as
the radius of the Gaussian loop $r_G$ far exceeds the coherence
length, {\it i.e.}, $r_G \gg \xi$, the spin and the charge quantum numbers are sharply defined with exponential accuracy.

When the spin rotation symmetry is broken but TR symmetry is still present, the concept of spin-charge separation is still well defined~\cite{qi2008a}.
A spinon state can be defined as a Kramers doublet without any charge, and a holon or a chargeon
is a Kramers singlet carrying charge $\pm e$. By combining the spin and charge flux threading\cite{essin2007}, it can be shown generally that these spin-charge separated quantum numbers are localized near a $\phi=\pi$ flux~\cite{qi2008a,ran2008}.

\subsection{Quantum anomalous Hall insulator}
\label{sec:qah}

Although TR invariance is essential in the QSH insulator, there is a TR symmetry breaking state of matter which is closely related to the QSH insulator: the quantum anomalous Hall (QAH) insulator. The QAH insulator is a band insulator with quantized Hall conductance but without orbital magnetic field. Nearly two decades ago, Haldane~\cite{haldane1988} proposed a model on a honeycomb lattice where the QH is realized without any external magnetic field, or the breaking of translational symmetry. However, the microscopic mechanism of circulating current loops within one unit cell has not been realized in any materials. Qi, Wu and Zhang~\cite{qi2005} proposed a simple model based on the concept of the QAH insulator with ferromagnetic moments interacting with band electrons via the SOC. This simple model can be realized in real materials. Two recent proposals~\cite{liu2009c,yu2010} make use of the properties of TR invariant topological insulators to realize the QAH state by magnetic doping. This is not accidental, but shows the deep relationship between these two states of matter. Thus we give a brief review of the QAH state in this subsection.

As a starting point, consider the upper $2\times 2$ block of the QSH Hamiltonian (\ref{BHZ}):
\begin{eqnarray}
h({\bf k})=\epsilon({\bf k})\mathbb{I}_{2\times 2}+d_a({\bf k})\sigma^a.\label{QAH}
\end{eqnarray}
If we consider only these two bands, this model describes a TR symmetry breaking system~\cite{qi2005}. As long as there is a gap between the two bands, the Hall conductance of the system is quantized~\cite{thouless1982}. The quantized Hall conductance is determined by the first Chern number of the Berry phase gauge field in the Brillouin zone, which, for the generic two-band model (\ref{QAH}), reduces to the following formula:
\begin{eqnarray}
\sigma_H=\frac{e^2}{h}\frac1{4\pi}\int dk_x\int dk_y\,\hat{\bf d}\cdot
\left(\frac{\partial
\hat{\bf d}}{\partial k_x}\times\frac{\partial \hat{\bf d}}{\partial
k_y}\right),\label{winding2band}
\end{eqnarray}
which is $e^2/h$ times the winding number of the unit vector $\hat{\bf d}({\bf k})={\bf d}({\bf k})/|{\bf d}({\bf k})|$ around the unit sphere. The ${\bf d}({\bf k})$ vector defined in Eq.~(\ref{BHZ1}) has a skyrmion structure for $M/B>0$ with winding number $1$, while the winding number is $0$ for $M/B<0$. Just as in an ordinary QH insulator, the system with nontrivial Hall conductance $e^2/h$ has one chiral edge state propagating on the edge. For the QSH system described by Eq.~(\ref{BHZ}), the lower $2\times 2$ block has the opposite Hall conductance, so that the total Hall conductance is zero, as guaranteed by TR symmetry. The chiral edge state of the QAH and its TR partner form the helical edge states of the QSH insulator.

\begin{figure}[hbt]
\centering
 \includegraphics[width=0.8\linewidth]{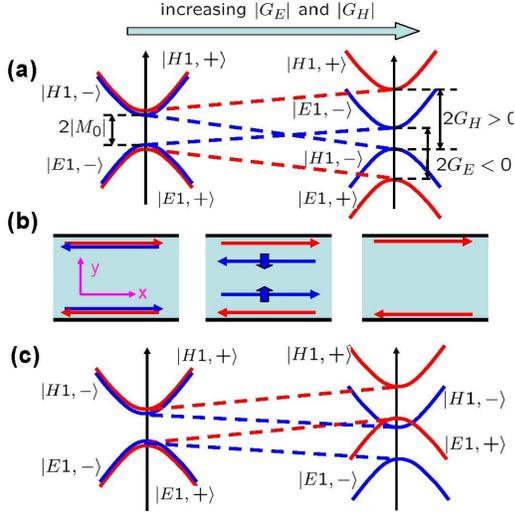}\\
 \caption{Evolution of band structure and edge states upon increasing the
    spin splitting. For (a) $G_E<0$ and $G_H>0$, the spin down states
    $|E1,-\rangle$ andn $|H1,-\rangle$ in the same block of the Hamiltonian
    (\ref{BHZ}) first touch each other, and then enter the normal regime. For (c) $G_E>0$ and
    $G_H>0$, gap closing occurs between $|E1,+\rangle$ and $|H1,-\rangle$, which belong to different blocks
    of the Hamiltonian, and thus will cross each other without opening a gap. (b) Behavior of the
    edge states during the level crossing. From~\onlinecite{liu2009c}.}\label{fig:QAH}
\end{figure}

When TR symmetry is broken, the two spin blocks are
no longer related, and their charge Hall conductances no longer
cancel exactly. For example, we can consider a different mass $M$ for the two blocks, which breaks TR symmetry. If one block is in the trivial insulator phase ($M/B<0$) and the other block is in the QAH phase ($M/B>0$), the whole system becomes a QAH state with Hall conductance $\pm e^2/h$. Physically, this can be realized by exchange coupling with magnetic impurities. In a system doped with magnetic impurities,
the spin splitting term induced by the magnetization is generically written as
\begin{eqnarray}
    H_{s}=\left(\begin{array}{cccc}G_E&0&0&0\\0&
        G_H&0&0\\0&0&-G_E&0\\0&0&0&-G_H\end{array}\right),
    \label{H_ss}
\end{eqnarray}
where $G_E$ and $G_H$ describe the splitting of $E1$ and $H1$ bands respectively, which are generically different. Adding $H_s$ to the Hamiltonian (\ref{BHZ}), we see that the mass term $M$ for the upper block is replaced by $M+(G_E-G_H)/2$, while that for the lower block is replaced by $M-(G_E-G_H)/2$. Therefore, the two blocks do acquire a different mass, which makes it possible to reach the QAH phase. After considering the effect of the identity term $(G_E+G_H)/2$, the condition for the QAH phase is given by $G_EG_H<0$. When $G_EG_H>0$ and $G_E\neq G_H$, the two blocks still acquire a different mass, but the system becomes metallic before the two blocks develop an opposite Hall conductance. Physically, we can also understand the physics from the
edge state picture [Fig.~\ref{fig:QAH}(b)]. On the boundary of a QSH insulator there are counter-propagating edge states carrying opposite spin. When the spin splitting term increases, one of the two blocks, say the spin down block, experiences a topological phase transition at $M=(G_E-G_H)/2$. The spin down edge states
penetrate deeper into the bulk due to the decreasing gap and
eventually disappear, leaving only the spin up state bound more
strongly to the edge. Thus, the system has only spin up edge states
and evolves from the QSH state to the QAH state [Fig.~\ref{fig:QAH}(b)]. Although the discussion above is based on the specific model (\ref{BHZ}), the mechanism to generate a QAH insulator from a QSH insulator is generic. A QSH insulator can always evolve into a QAH insulator once a TR symmetry breaking perturbation is introduced.

Fortunately, in Mn-doped HgTe QWs the condition $G_EG_H<0$ is indeed satisfied, so that the QAH phase exists in this system as long as the Mn spins are polarized. The microscopic reason for the opposite sign of $G_E$ and $G_H$ is the opposite sign of the $s$-$d$ and $p$-$d$ exchange couplings in this system~\cite{liu2009c}. Interestingly, in another family of QSH insulators, Bi$_2$Se$_3$ and Bi$_2$Te$_3$ thin films~\cite{liu2010a}, the condition $G_EG_H<0$ is also satisfied when magnetic impurities such as Cr or Fe are introduced into the system, but for a different physical reason. In HgTe QWs, the two bands in the upper block of the Hamiltonian (\ref{BHZ}) have the same direction of spin, but couple with the impurity spin with an opposite sign of exchange coupling because one band originates from $s$-orbitals while the other originates from $p$-orbitals. In Bi$_2$Se$_3$ and Bi$_2$Te$_3$, both bands originate from $p$-orbitals, which have the same sign of exchange coupling with the impurity spin, but the sign of spin in the upper block is opposite~\cite{yu2010}. Consequently, the condition $G_EG_H<0$ is still satisfied. More details on the properties of the Bi$_2$Se$_3$ and Bi$_2$Te$_3$ family of materials can be found in the next section, since as bulk materials they are both 3D topological insulators.

\subsection{Experimental results}
\label{sec:experiment}

\subsubsection{Quantum well growth and the band inversion transition}

\begin{figure}[h!]
\centering
 \includegraphics[width=0.92\linewidth]{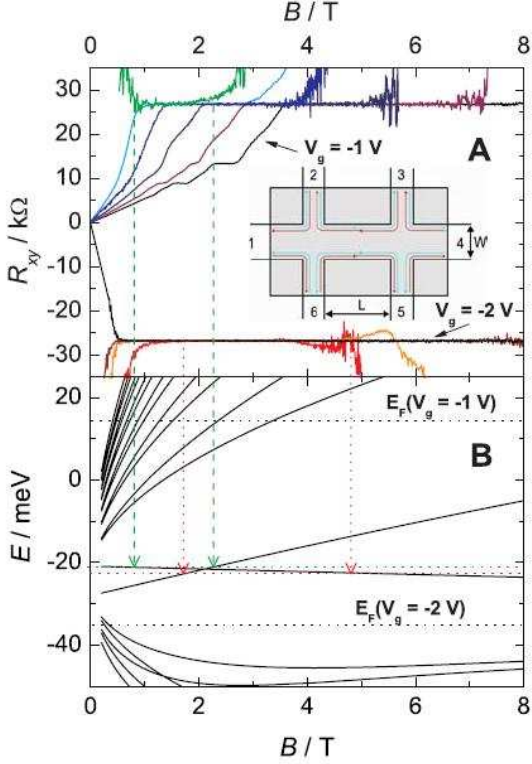}\\
 \caption{(a) Hall resistance $R_{xy}$ of a $(L\times W) =
(600\times 200)$~$\mu$m$^2$
 QW structure with $6.5$~nm well width for different carrier
concentrations obtained for
 different gate voltages $V_g$ in the range from $-1$~V to $-2$~V. For
decreasing $V_g$,
 the $n$-type carrier concentration decreases and a transition to $p$-type conduction
 is observed, passing through an insulating regime between $-1.4$~V and $-1.9$~V at zero field.
(b) Landau level fan chart of a $6.5$~nm quantum well obtained
from an eight-band $\mathbf{k}\cdot\mathbf{p}$ calculation. Black
dashed lines indicate the position of the Fermi energy,
$E_F$, for gate voltages $-1.0$~V and $-2.0$~V. Red and green dashed lines indicate the position of $E_F$ for the red and green
Hall resistance traces in (a). The crossing points of $E_F$ with the respective Landau levels are marked by arrows of the
same color. From~\onlinecite{koenig2007}.}\label{FigexpBcross}
\end{figure}

As shown above, the transition from a normal to an
inverted band structure coincides with the phase transition from a trivial insulator to the QSH insulator. In order to cover both the normal and the inverted band structure regime, HgTe QW samples with a QW
width in the range from $4.5$~nm to $12.0$~nm were grown~\cite{koenig2007,markusthesis,koenig2008} by molecular beam epitaxy (MBE). Samples with mobilities of several $10^5$~cm$^2$/(V$\cdot$s), even for low densities $n<5\times 10^{11}$~cm$^{-2}$, were
available for transport measurements. In such samples, the mean
free path is of the order of several microns. For the investigation of the QSH effect, devices in a Hall bar geometry [Fig.~\ref{FigexpBcross}, inset] of various dimensions were fabricated from QW structures with well widths of $4.5$~nm, $5.5$~nm,
$6.4$~nm, $6.5$~nm, $7.2$~nm, $7.3$~nm, $8.0$~nm and $12.0$~nm.

\begin{figure}[htb]
\centering
 \includegraphics[scale=0.75]{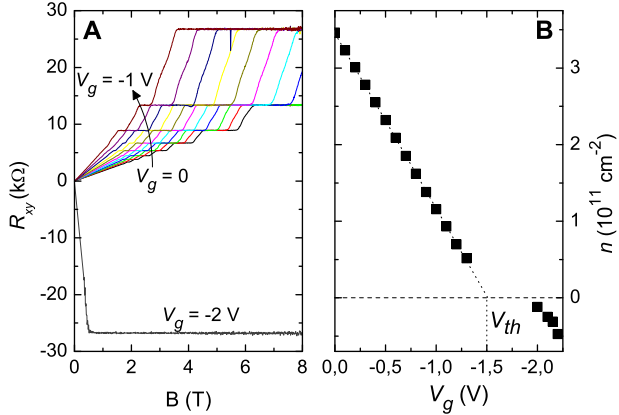}\\
 \caption{(a) Hall resistance $R_{xy}$ for various gate
voltages, indicating the transition from $n$- to $p$-conductance. (b)
Gate-voltage dependent carrier density deduced from Hall
measurements. From~\onlinecite{koenig2008}.}\label{Fignipdens}
\end{figure}

For the investigation of the QSH effect, samples with a low
intrinsic density $n(V_g=0)<5\times10^{11}$~cm$^{-2}$
were studied. When a negative gate voltage $V_g$ is applied to
the top gate electrode of the device, the usual decrease in
electron density is observed. In Fig.~\ref{Fignipdens}(a),
measurements of the Hall resistance $R_{xy}$ are presented for a
Hall bar with length $L=600~\mu$m and width $W=200~\mu$m. The decrease of the carrier density is
reflected in an increase of the Hall coefficient when the gate
voltage is lowered from $0$~V to $-1$~V. In this voltage range, the
density decreases linearly from $3.5\times10^{11}$~cm$^{-2}$
to $0.5\times10^{11}$~cm$^{-2}$ [Fig.~\ref{Fignipdens}(b)].
For even lower gate voltages, the sample becomes insulating, because the Fermi energy $E_F$ is shifted into the bulk gap. When a large negative voltage $V_g\leq-2$~V is
applied, the sample becomes conducting again. It can be inferred
from the change in sign of the Hall coefficient that the device is
$p$-conducting. Thus, $E_F$ has been
shifted into the valence band, passing through the entire bulk
gap.

\begin{figure}[hbt]
\centering
 \includegraphics[width=0.95\linewidth]{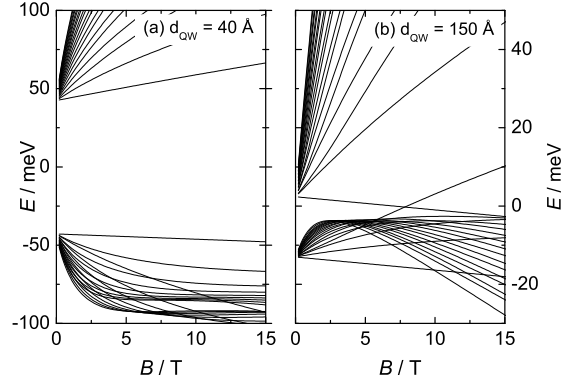}\\
 \caption{Landau level dispersion for quantum well thicknesses of (a) $4.0$~nm, (b) $15.0$~nm. The qualitative behavior is
 indicative for samples with (a) normal and (b) inverted band
 structure. From~\onlinecite{koenig2008}.}
 \label{FigLLs}
\end{figure}

The peculiar band structure of HgTe QWs gives rise to a
unique LL dispersion. For a normal band structure,
i.e., $d_\mathrm{QW}<d_{c}$, all LLs are shifted to higher
energies for increasing magnetic fields [Fig.~\ref{FigLLs}(a)].
This is the usual behavior and can be commonly observed in most materials.
When the band structure of the HgTe QW is inverted for
$d_{QW}>d_{c}$, however, a significant change is observed for the
LL dispersion [Fig.~\ref{FigLLs}~(b)]. Due to the inversion of electron-like and hole-like bands, states near the bottom of the conduction band have predominantly $p$ character. Consequently, the energy of the lowest LL decreases with increasing magnetic field. On the other hand, states near the top of the valence band have predominantly $s$ character, and the highest LL shifts to higher energies with increasing magnetic field. This leads to a crossing of these two peculiar LLs for a special value of the magnetic field. This behavior has been observed earlier by the W\"{u}rzburg group and can now be demonstrated analytically within the BHZ model~\cite{koenig2008}. The exact magnetic field ${\cal{B}}_{\rm cross}$ at which the crossing occurs depends on $d_\mathrm{QW}$. The existence of the LL crossing is a clear signature of an inverted band structure, which corresponds to a negative energy gap with $M/B<0$ in the BHZ model. The crossing of the LLs from the
conduction and valence bands can be
observed in experiments [Fig.~\ref{FigexpBcross}(a)]. For gate
voltages $V_g\geq-1.0$~V and $V_g\leq-2.0$~V, $E_F$ is
clearly in the conduction band and valence band, respectively.
When $E_F$ is shifted towards the bottom of the
conduction band, i.e. $V_g<-1.0$~V, a transition from a QH state with filling factor $\nu=1$, i.e.
$R_{xy}=h/e^2=25.8$~k$\Omega$, to an insulating state is
observed. Such behavior is expected independently of the details
of the band structure, when the lowest LL of the
conduction band crosses $E_F$ for a finite
magnetic field.
When $E_F$ is located within the gap, a nontrivial
behavior can be observed for devices with an inverted band
structure. Since the lowest LL of the conduction band lowers its energy with increasing magnetic field, it will cross $E_F$ for a certain magnetic field. Subsequently, one occupied LL is below $E_F$,
giving rise to the usual transport signatures of the quantum Hall
regime, i.e. $R_{xy}$ is quantized at $h/e^2$ and $R_{xx}$
vanishes. When the magnetic field is increased, the LLs from the
valence and conduction band cross. Upon crossing,
their "character" is exchanged, i.e. the level from the valence band
turns into a conduction band LL and vice versa. The lowest LL of the conduction band now rises in energy for larger magnetic fields.
Consequently, it will cross the $E_F$ for a certain
magnetic field. Since $E_F$ will be located within the fundamental
gap again afterwards, the sample will become insulating again.
Such a reentrant $n$-type QH state is shown in
Fig.~\ref{FigexpBcross}(a) for $V_g=-1.4$~V (green trace). For
lower gate voltages, a corresponding behavior is observed for a
$p$-type QH state (e.g. red trace for $V_g=-1.8$~V). As
Fig.~\ref{FigexpBcross}(b) shows, the experimental results are in
good agreement with the theoretically calculated LL dispersion.
The crossing point of the LLs in magnetic field,
${\cal{B}}_{\rm cross}$, can be determined accurately by tuning
$E_F$ through the energy gap. Thus, the width of the QW
layer can be verified experimentally~\cite{koenig2007}.

The observation of a reentrant QH state is a clear
indication of the nontrivial insulating behavior, which is a
prerequisite for the existence of the QSH state. In contrast,
trivial insulating behavior is obtained for devices with
$d_\mathrm{QW}<d_{c}$. For a normal band structure, the energy gap
between the lowest LLs of the conduction and valence bands increases in magnetic field [Fig.~\ref{FigLLs}(a)]. Thus, a sample remains insulating in magnetic field, if $E_F$ is located in the gap at zero field. The details of the physics of this reentrant
QH state can be understood within the BHZ model with an added orbital magnetic field~\cite{koenig2008}. This nontrivial LL crossing could also be detected optically~\cite{schmidt2009}.

\subsubsection{Longitudinal conductance in the quantum spin Hall state}

\begin{figure}[hbt]
\centering
 \includegraphics[width=0.92\linewidth]{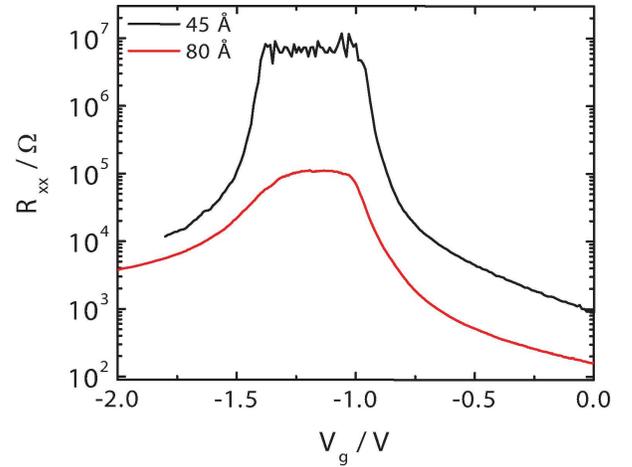}\\
 \caption{Longitudinal resistance of a $4.5$~nm QW [dashed (black)] and a $8.0$~nm QW [solid (red)] as a function of gate voltage. From~\onlinecite{koenig2008}.}
 \label{FigQSHEnip}
\end{figure}

Initial evidence for the QSH state was revealed when
Hall bars of dimensions $(L\times W) = (20.0\times13.3)~\mu$m$^2$ with different thickness $d_\mathrm{QW}$ are studied. For thin QW devices with $d_\mathrm{QW}<d_c$ and a normal band structure, the sample
shows trivial insulating behavior [Fig.~\ref{FigQSHEnip}]. A
resistance of several megaohms is measured when the Fermi level lies within the bulk insulating gap. This value can be
attributed to the noise level of the measurement setup,
and the intrinsic conductance is practically zero.
For a thicker device with $d_\mathrm{QW}>d_c$ and an inverted band structure, however, the resistance does not exceed $100$~k$\Omega$. This behavior is reproduced for various Hall bars with a QW width in the range from $4.5$~nm to $12.0$~nm. While devices with a normal band structure, i.e. $d_\mathrm{QW}<d_{c}\approx6.3$~nm, show trivial insulating behavior, a finite conductance in the insulating regime is observed for samples with an inverted band structure.

The obtained finite resistance $R\approx100$~k$\Omega$ is
significantly higher than the four-terminal resistance $h/(2e^2)
\approx12.9$~k$\Omega$ one anticipates for the geometry used
in the experiments. The enhanced resistance in these samples with
a length of $L=20~\mu$m can be understood as a consequence of inelastic
scattering. While, as discussed above, the helical
edge states are robust against single-particle elastic
backscattering, inelastic mechanisms can cause backscattering.
For $n$-doped HgTe quantum wells, the typical mobility of the
order of $10^5$~cm$^2$/(V$\cdot$s) implies an elastic mean free path of
the order of $1$~$\mu$m~\cite{daumer03}. Lower mobilities can be
anticipated for the QSH regime. The inelastic mean free path,
which determines the length scale of undisturbed transport by the
QSH edge states, can be estimated to be several times larger due
to the suppression of phonons and the reduced electron-electron
scattering at low temperatures. Thus, the inelastic scattering
length is of the order of a few microns.

For the observation of the QSH conductance, the sample dimensions
were reduced below the estimated inelastic mean free path. When
Hall bars with a length $L=1~\mu$m are studied, a four-terminal
resistance close to $h/(2e^2)$ is observed. The threshold voltage
$V_\mathrm{th}$ is defined such that the QSH regime is in the vicinity of $V_g=V_\mathrm{th}$.
\begin{figure}[htb]
\centering
 \includegraphics[width=0.95\linewidth]{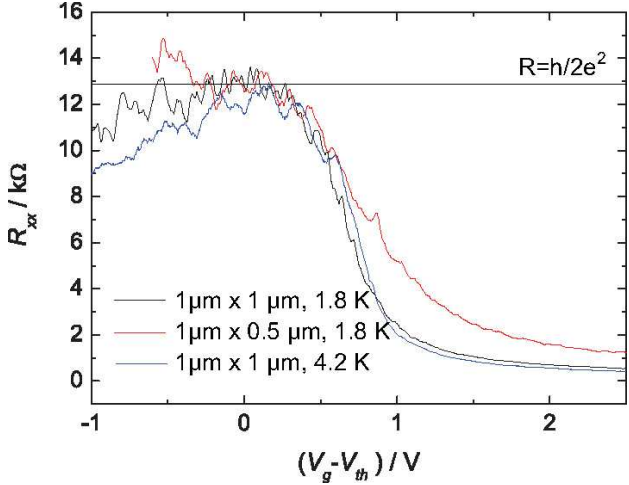}\\
 \caption{Longitudinal resistance as a function of gate voltage for two devices with $L=1~\mu$m. The width $W$ is
 $1~\mu$m [solid (black) and dotted (blue)] and $0.5~\mu$m [dashed (red)]. The solid and dashed traces were obtained at a temperature of $1.8$~K, and the dotted one at $4.2$~K. From~\onlinecite{koenig2008}.}
 \label{FigQSHEquant}
\end{figure}
The slight deviation of $R$ from the quantized value $h/(2e^2)$
can be attributed to some residual scattering. This is an
indication that the length of the edge states still exceeds the
inelastic mean free path. The results presented in Fig.~\ref{FigQSHEquant} provide evidence
that transport in the QSH regime indeed occurs due to edge states.
The two devices with $W=1.0~\mu$m and $W=0.5~\mu$m were fabricated from the same QW structure. The resistance of the two devices differ significantly in the $n$-conducting
regime, where transport is determined by bulk properties. In the QSH regime, however, both devices exhibit the same resistance, even though the width of the devices differs by a factor of two. This fact clearly shows that the conductance is due to the edge states, which are independent of the sample width.

\subsubsection{Magnetoconductance in the quantum spin Hall state}

Another indication that the observed nontrivial insulating
state is caused by the QSH effect is obtained by measurements in
a magnetic field. The following experimental results were
obtained on a Hall bar with dimensions $(L\times W) =
(20.0\times13.3)~\mu$m$^2$ in a vector magnet system at a temperature of
$1.4$~K~\cite{koenig2007,markusthesis}. When a magnetic field is applied
perpendicular to the QW layer, the QSH conductance decreases significantly already for small fields. A cusp-like
magnetoconductance peak is observed with a full width at half-maximum
${\cal{B}}_{\rm FWHM}$ of $28$~mT. Additional measurements show that
the width of the magnetoconductance peak decreases with decreasing temperature. For example, ${\cal{B}}_{\rm FWHM}=10$~mT is observed at 30~mK. For various devices of different sizes, a qualitatively similar
behavior in magnetic field is observed.

\begin{figure}[hbt]
\centering
 \includegraphics[width=0.95\linewidth]{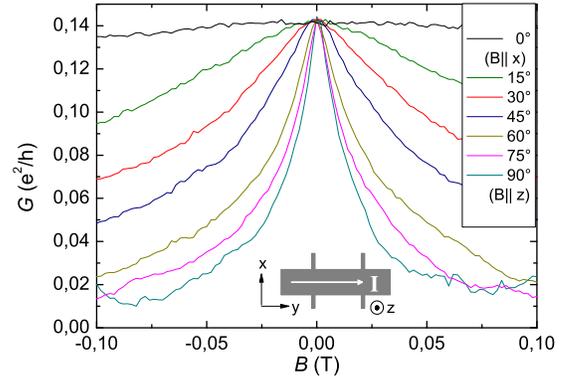}\\
 \caption{Four-terminal magnetoconductance $G_{14,23}$
 in the QSH regime as a function of tilt angle between
 the QW plane and the applied magnetic field for a
 $d=7.3$~nm QW structure with dimensions
 $(L\times W) = (20 \times 13.3)~\mu$m$^2$
 measured in a vector field cryostat at a temperature of $1.4$~K. From~\onlinecite{koenig2008}.}
 \label{FigQSHBanis}
\end{figure}

When the magnetic field is tilted towards the plane of the QW, the
magnetoconductance peak around ${\cal{B}}=0$ widens steadily
[Fig.~\ref{FigQSHBanis}]. For a tilt angle $\alpha=90^{\circ}$, i.e. when the magnetic field is in the QW plane, only a very small decrease in
the conductance is observed. The decrease of the conductance for an in-plane field can be described by ${\cal{B}}_{\rm
FWHM}\approx0.7$~T for any in-plane orientation.
From the results shown in Fig.~\ref{FigQSHBanis}, it is evident
that a perpendicular field has a much larger influence on the
QSH state than an in-plane field. The magnetoresistance in the
QSH regime has been investigated theoretically~\cite{koenig2008,maciejko2009c,chu2009,tkachov2010}. The large anisotropy can be understood by a slightly modified version of the BHZ model with the inclusion of BIA terms and anisotropy in the $g$-factor~\cite{koenig2008,maciejko2009c}. The cusp behavior in the magnetoconductance is possibly due to the presence of strong disorder; numerical simulations~\cite{maciejko2009c} are in good agreement with the experimental results.

\subsubsection{Nonlocal conductance}\label{sec:nonlocaltransport}

\begin{figure}[t]
\begin{center}
\includegraphics[scale=0.35]{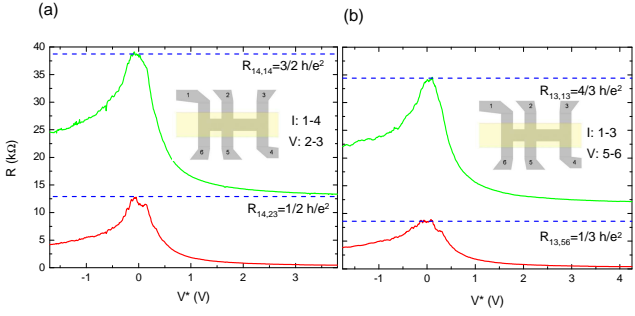}
\caption{Experimental measurements of the four- and two-terminal resistance: (a) $R_{14,23}$ (red line) and $R_{14,14}$ (green
line) and (b) $R_{13,56}$ (red line) and $R_{13,13}$ (green line).
The dotted blue lines indicate the expected resistance value from
the theory of the helical edge states. From~\onlinecite{roth2009}.}\label{nonlocalFig3}
\end{center}
\end{figure}

Further confidence in the helical edge state transport can be gained by performing more extended multi-terminal experiments~\cite{roth2009}. The longitudinal resistance of a device was measured by passing a current through contacts 1 and 4 [Fig.~\ref{nonlocalFig3}] and by detecting the voltage between contacts 2 and 3 ($R_{14,23}$). For this case, a result similar to the results found previously, i.e. a resistance $h/(2e^2)$ when the bulk of the device is gated into the insulating regime [Fig.~\ref{nonlocalFig3}(a)]. However, the longitudinal resistance is significantly different in a slightly modified
configuration, where the current is passed through contacts 1 and
3 and the voltage is measured between contacts 4 and 5
($R_{13,45}$) [Fig.~\ref{nonlocalFig3}(b)]. The result is
$R_{13,45} \approx 8.6$~k$\Omega$, which is markedly different
from what one would expect for either QH transport, or purely
diffusive transport, where this configuration would be equivalent
to the previous one. However, the application of the transport equations
(\ref{Landauer}) and (\ref{Transmission}) indeed predicts that the observed behavior is what one expects for helical edge
channels. One easily finds that this resistance value can be expressed as an integer fraction of the inverse conductance quanta $e^2/h$: $R_{13,45}=h/3e^2$. This result shows that the
current through the device is influenced by the number of ohmic
contacts in the current path. As discussed earlier, these ohmic
contacts lead to the equilibration inside
the contact of the chemical potentials of the two counter-propagating helical edge channels.

\begin{figure}[th]
\begin{center}
\includegraphics[width=0.92\linewidth]{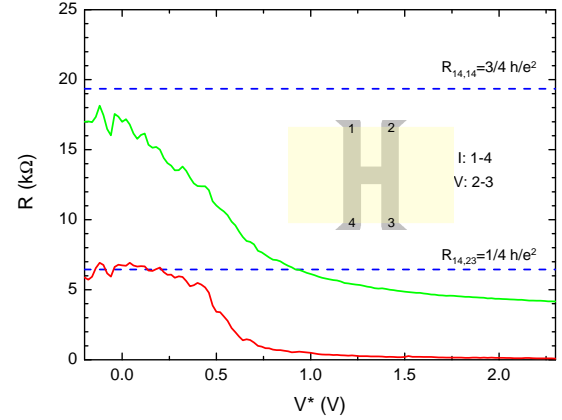}
\caption{Nonlocal four-terminal resistance and
two-terminal resistance measured on an H-bar device:
$R_{14,23}$ (red line) and $R_{14,14}$ (green line). The
dotted blue line represents the theoretically expected resistance. From~\onlinecite{roth2009}.}\label{nonlocalFig4}
\end{center}
\end{figure}

Another measurement that directly confirms the nonlocal character of the helical edge channel transport in the QSH regime is shown
in Fig.~\ref{nonlocalFig4}. This figure shows data obtained from a device in the shape of the letter ``H". In this four-terminal device
the current is passed through contacts 1 and 4 and the voltage is
measured between contacts 2 and 3. In the metallic $n$-type regime
(low gate voltage) the voltage signal tends to zero. In the
QSH regime, however, the nonlocal resistance signal
increases to $\approx 6.5$~k$\Omega$, which again fits perfectly
to the result of Laudauer-B\"{u}ttiker considerations: $R_{14,23}
= h/4e^2\approx 6.45$~k$\Omega$. Classically, one would expect
only a minimal signal in this configuration (from Poisson's
equation, assuming diffusive transport, one estimates a signal of
about $40$~$\Omega$), and certainly not one that increases so
strongly when the bulk of the sample is depleted. The signal
measured here is fully nonlocal, and can be taken (as was done
twenty years ago for the QH regime) as definite evidence of the
existence of edge channel transport in the QSH regime.

\section{Three-Dimensional Topological Insulators}
\label{sec:3dti}
The model Hamiltonian for the 2D topological insulator in HgTe QWs also gives a basic template for generalization to 3D, leading to a simple model Hamiltonian for a class of materials: Bi$_2$Se$_3$, Bi$_2$Te$_3$, and Sb$_2$Te$_3$~\cite{zhang2009}. Similar to their 2D counterpart the HgTe QWs, these materials can be described by a simple but realistic model, where SOC drives a band inversion transition at the $\Gamma$ point. In the topologically nontrivial phase, the bulk states are fully gapped, but there is a topologically protected surface state consisting of a single massless Dirac fermion. The 2D massless Dirac fermion is ``helical", in the sense that the electron spin points perpendicularly to the momentum, forming a left-handed helical texture in momentum space.
Similarly to the 1D helical edge states, a single massless Dirac fermion state is ``holographic", in the sense that it cannot occur in a purely 2D system with TR symmetry, but can exist as the boundary of a 3D insulator. TR invariant single-particle perturbations cannot introduce a gap for the surface state. A gap can open for the surface state when a TR breaking perturbation is introduced on the surface. Moreover, the system becomes full insulating, both in the bulk and on the surface. In this case, the topological properties of the fully gapped insulator are characterized by a novel topological magnetoelectric effect.

Soon after the theoretical prediction of the 3D topological insulator in the Bi$_2$Te$_3$, Sb$_2$Te$_3$~\cite{zhang2009} and
Bi$_2$Se$_3$~\cite{zhang2009,xia2009} class of materials, angle-resolved photoemission spectroscopy (ARPES) observed the surface states with a single Dirac cone~\cite{xia2009,chen2009,hsieh2009}. Furthermore, spin-resolved ARPES measurements indeed observed the left-handed helical spin texture of the massless Dirac fermion~\cite{hsieh2009}. These pioneering theoretical and experimental works inspired much of the subsequent developments which we review in this section.

We take advantage of the model simplicity of the Bi$_2$Se$_3$, Bi$_2$Te$_3$, Sb$_2$Te$_3$ class of 3D topological insulators and give a pedagogical introduction based on this particular material system. In the next section, we shall introduce the general theory of the topological insulators. The electronic structure of the Bi$_2$Se$_3$, Bi$_2$Te$_3$, Sb$_2$Te$_3$ class of topological insulators is simple enough to be captured by a simple model Hamiltonian. However, more powerful methods are needed to determine the topological properties of materials with a more complex electronic structure. In this regard, the TBT has played an important role~\cite{fu2007b,moore2007,roy2009}. In particular, a method due to Fu and Kane~\cite{fu2007a} gives a simple algorithm to determine the topological properties of an arbitrarily complex electronic structure with inversion symmetry. This method predicts that Bi$_x$Sb$_{1-x}$ is a topological insulator for a certain range of composition $x$. ARPES measurements~\cite{hsieh2008} have indeed observed topologically nontrivial surface states in this system, giving the first example of a 3D topological insulator. The topological properties of this material have been further investigated both theoretically and experimentally~\cite{teo2008,nishide2010,zhang2009a}. however, the surface states in Bi$_x$Sb$_{1-x}$ are rather complicated, and cannot be described by simple model Hamiltonians. For this reason, we focus on the Bi$_2$Se$_3$, Bi$_2$Te$_3$, Sb$_2$Te$_3$ class of topological insulators in this section.

\subsection{Effective model of the three-dimensional topological insulator}
\label{sec:3dzhang}

\begin{figure}
   \begin{center}
      \includegraphics[width=3.5in]{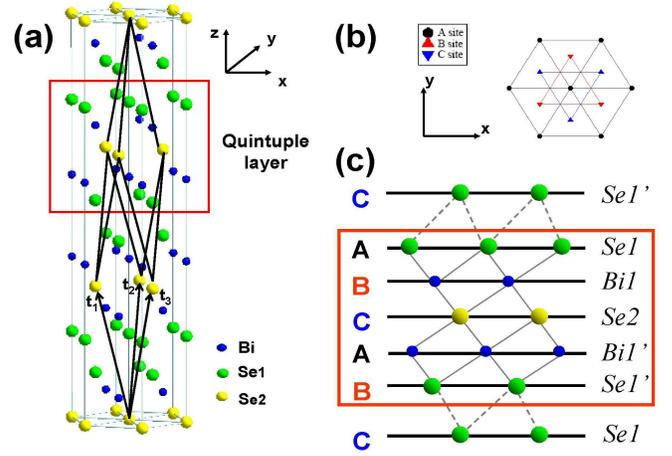}
    \end{center}
    \caption{(a) Crystal structure of Bi$_2$Se$_3$ with
     three primitive lattice vectors denoted by $\mathbf{t}_{1,2,3}$. A quintuple layer with Se$1$-Bi$1$-Se$2$-Bi$1'$-Se$1'$ is indicated by the red box. (b) Top view along the $z$ direction. Triangular
     lattice in one quintuple layer has three inequivalent positions, denoted by A, B and C. (c) Side view of the quintuple layer structure. Along the $z$ direction, Se and Bi atomic layers are stacked in the sequence $\cdots$-C(Se$1'$)-A(Se$1$)-B(Bi$1$)-C(Se$2$)-A(Bi$1'$)-B(Se$1'$)-C(Se$1$)-$\cdots$.
     The Se$1$ (Bi$1$) layer is related to the Se$1'$ (Bi$1'$) layer by inversion,where Se$2$ atoms play the role of inversion center. Adapted from~\onlinecite{zhang2009}.}
    \label{fig:crystal}
\end{figure}

In this review we focus on an effective model for 3D topological insulators~\cite{zhang2009} which, simply by adjusting parameters, is valid for studying the properties of Bi$_2$Se$_3$, Bi$_2$Te$_3$, and Sb$_2$Te$_3$. Bi$_2$Se$_3$, Bi$_2$Te$_3$, and Sb$_2$Te$_3$ share the same rhombohedral crystal
structure with space group $D^5_{3d}$ ($R\bar{3}m$) and five
atoms per unit cell. For example, the crystal structure of Bi$_2$Se$_3$ is shown in Fig.~\ref{fig:crystal}(a), and consists of a layered structure where individual layers form a triangular lattice. The important symmetry axes are a trigonal axis (three-fold rotation symmetry) defined as the $z$ axis, a binary axis (two-fold rotation symmetry) defined as the $x$ axis, and a bisectrix axis (in the reflection plane) defined as the $y$ axis. The material consists of five-atom layers stacked along the $z$ direction, and known as quintuple
layers. Each quintuple layer consists of five atoms per unit cell with two equivalent Se atoms denoted by Se$1$ and Se$1'$ in Fig.~\ref{fig:crystal}(b), two equivalent Bi atoms denoted by Bi$1$ and Bi$1'$ in Fig.~\ref{fig:crystal}(b), and a third Se atom denoted by Se$2$ in Fig.~\ref{fig:crystal}(b). The coupling between two atomic layers within a quintuple layer is strong, while that between quintuple layers is much weaker, and predominantly of the van der Waals type. The primitive lattice vectors $\mathbf{t}_{1,2,3}$ and rhombohedral primitive unit cells are shown in Fig.~\ref{fig:crystal}(a). The Se$2$ site plays the role of an inversion center. Under inversion, Bi$1$ is mapped to
Bi$1'$ and Se$1$ is mapped to Se$1'$.

\begin{figure}
   \begin{center}
    \includegraphics[width=3.5in]{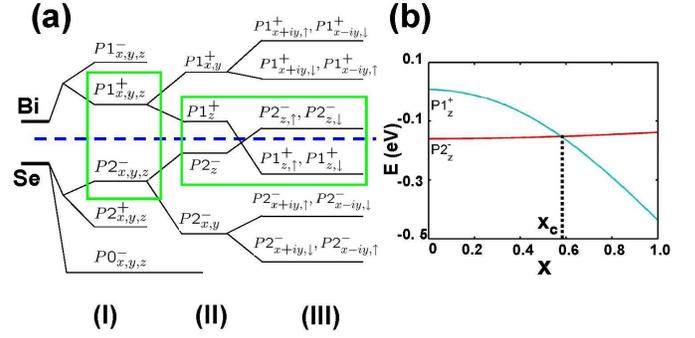}
    \end{center}
    \caption{ (a) Schematic picture of the evolution from the atomic $p_{x,y,z}$ orbitals of Bi and Se into the conduction and valence bands of Bi$_2$Se$_3$ at the $\Gamma$ point. The three different stages (I), (II) and (III) represent the effect of turning on chemical bonding, crystal field splitting, and SOC, respectively (see text). The blue dashed line represents the Fermi energy. (b) The energy levels $|P1^+_z\rangle$ and $|P2^-_z\rangle$ of Bi$_2$Se$_3$ at the $\Gamma$ point versus an artificially rescaled atomic
    SOC $\lambda($Bi$)=x\lambda_0($Bi$)=1.25x$~[eV], $\lambda($Se$)=x\lambda_0($Se$)=0.22x$~[eV] (see text).
    A level crossing occurs between these two states at $x=x_c\simeq 0.6$. Adapted from~\onlinecite{zhang2009}.
    }
    \label{fig:level}
\end{figure}

To get a better understanding of the band structure and orbitals
involved, we start from the atomic energy levels and then consider
the effects of crystal field splitting and SOC on the energy eigenvalues at the $\Gamma$ point in momentum space. This is summarized schematically in three stages (I), (II) and (III) [Fig.~\ref{fig:level}(a)]. Since the
states near the Fermi level are primarily from $p$-orbitals, we
will neglect the $s$-orbitals and start from the atomic
$p$-orbitals of Bi (electronic configuration $6s^26p^3$) and Se ($4s^24p^4$). In stage (I), we consider chemical bonding between Bi and Se atoms within a quintuple layer, which corresponds to the largest energy scale in this problem. First, we can recombine the orbitals in a single unit cell according to their parity. This results in three states (two odd, one even) from each Se $p$-orbital and two states (one odd,
one even) from each Bi $p$-orbital. The formation of chemical
bonds hybridizes the states on the Bi and Se atoms, and pushes
down all the Se states and lifts all the Bi states. In Fig.~\ref{fig:level}(a), these five hybridized states are labeled as
$\left|P1_{x,y,z}^\pm\right\rangle$,
$\left|P2_{x,y,z}^{\pm}\right\rangle$ and
$\left|P0_{x,y,z}^{-}\right\rangle$, where the superscripts $\pm$
stand for the parity of the corresponding states. In stage (II),
we consider the effect of crystal field splitting between
different $p$-orbitals. According to the point group symmetry, the
$p_z$ orbital is split from the $p_x$ and $p_y$ orbitals while the
latter two remain degenerate. After this splitting, the energy
levels closest to the Fermi energy turn out to be the $p_z$ levels
$\left|P1^+_z\right\rangle$ and $\left|P2^-_z\right\rangle$. In
the last stage (III), we take into account the effect of SOC. The
atomic SOC Hamiltonian is given by $H_\mathrm{SO}=\lambda
\mathbf{L}\cdot\mathbf{S}$, with $\mathbf{L},\mathbf{S}$ the orbital and spin angular
momentum, respectively, and $\lambda$ the strength of SOC. The SOC Hamiltonian
mixes spin and orbital angular momenta while preserving the total
angular momentum. This leads to a level repulsion between
$\left|P1_z^+,\uparrow\right\rangle$ and
$\left|P1_{x+iy}^+,\downarrow\right\rangle$, and between similar
combinations. Consequently, the energy of the
$\left|P1_z^+,\uparrow(\downarrow)\right\rangle$ state is pushed
down by the effect of SOC, and the energy of the
$\left|P2_z^-,\uparrow(\downarrow)\right\rangle$ state is pushed
up. If SOC is larger than a critical value $\lambda>\lambda_c$, the order of these two energy levels is reversed. To illustrate this inversion process explicitly, the energy levels $\left|P1_z^+\right\rangle$ and $\left|P2_z^-\right\rangle$ have been calculated~\cite{zhang2009} for a
model Hamiltonian of Bi$_2$Se$_3$ with artificially rescaled
atomic SOC parameters $\lambda($Bi$)=x\lambda_0($Bi$)$,
$\lambda($Se$)=x\lambda_0($Se$)$, as shown in Fig.~\ref{fig:level}(b).
Here $\lambda_0($Bi$)=1.25$~eV and $\lambda_0($Se$)=0.22$~eV are the actual values of the SOC strength for Bi and Se atoms, respectively~\cite{wittel1974}. From Fig.~\ref{fig:level}(b), one can clearly see that a level crossing occurs
between $\left|P1_z^+\right\rangle$ and
$\left|P2_z^-\right\rangle$ when the SOC strength is about 60\% of its actual value. Since these two levels have opposite parity, the
inversion between them drives the system into a topological
insulator phase, similar to the case of HgTe QWs~\cite{bernevig2006c}. Therefore, the mechanism for the occurrence of a 3D
topological insulating phase in this system is closely analogous to the
mechanism for the 2D QSH effect (2D topological insulator) in HgTe~\cite{bernevig2006c}. More precisely, to determine whether or not an inversion-symmetric crystal is a topological insulator, we must have full knowledge of the states at \emph{all} of the eight TR invariant momenta (TRIM)~\cite{fu2007a}. The system is a (strong) topological insulator if and only if the band inversion between states with opposite parity occurs at odd number of TRIM. The parity of the Bloch states at all TRIM have been studied by {\it ab initio} methods for the four materials Bi$_2$Se$_3$, Bi$_2$Te$_3$, Sb$_2$Se$_3$, and Sb$_2$Te$_3$~\cite{zhang2009}. Comparing the Bloch states with and without SOC, one conclude that Sb$_2$Se$_3$ is a trivial insulator, while the other three are topological insulators. For the three topological insulators, the band inversion only occurs at the $\Gamma$ point.

Since the topological nature
is determined by the physics near the $\Gamma$ point, it is possible
to write down a simple effective Hamiltonian to characterize the
low-energy, long-wavelength properties of the system. Starting
from the four low-lying states
$\left|P1^+_z,\uparrow(\downarrow)\right\rangle$ and
$\left|P2^-_z,\uparrow(\downarrow)\right\rangle$ at the $\Gamma$
point, such a Hamiltonian can be constructed by the theory of
invariants~\cite{winklerbook} at a finite wavevector ${\bf k}$. The
important symmetries of the system are TR symmetry $T$,
inversion symmetry $I$, and three-fold rotation symmetry $C_3$
aroung the $z$ axis. In the basis $\left\{\left|P1^+_z,\uparrow\right\rangle,
\left|P2^-_z,\uparrow\right\rangle,
\left|P1^+_z,\downarrow\right\rangle,
\left|P2^-_z,\downarrow\right\rangle\right\}$, the representation
of these symmetry operations is given by $T=i\sigma^y\mathcal{K}\otimes\mathbb{I}_{2\times 2}$, $I=\mathbb{I}_{2\times 2}\otimes\tau_3$ and
$C_3=\exp\left(i\frac{\pi}3\sigma^z\otimes\mathbb{I}_{2\times2}\right)$, where $\mathbb{I}_{n\times n}$ is the $n\times n$ identity matrix, $\mathcal{K}$ is the complex
conjugation operator, and $\sigma^{x,y,z}$ and $\tau^{x,y,z}$ denote
the Pauli matrices in the spin and orbital space, respectively. By
requiring these three symmetries and keeping only terms up to
quadratic order in ${\bf k}$, we obtain the following generic form
of the effective Hamiltonian:
\begin{eqnarray}
    &&H({\bf k})=\epsilon_0({\bf k})\mathbb{I}_{4\times 4}+\nonumber\\
    &&\left(
    \begin{array}{cccc}
        \mathcal{M}({\bf k})&A_1k_z&0&A_2k_-\\
        A_1k_z&-\mathcal{M}({\bf k})&A_2k_-&0\\
        0&A_2k_+&\mathcal{M}({\bf k})&-A_1k_z\\
        A_2k_+&0&-A_1k_z&-\mathcal{M}({\bf k})
    \end{array}
    \right),
    \label{eq:Heff3D}
\end{eqnarray}
with $k_\pm=k_x\pm ik_y$, $\epsilon_0({\bf
k})=C+D_1k_z^2+D_2k_\perp^2$ and $\mathcal{M}({\bf
k})=M-B_1k_z^2-B_2k_\perp^2$. The parameters in the effective model can be determined by fitting the energy spectrum of
the effective Hamiltonian to that of {\it ab initio}
calculations~\cite{zhang2009,zhang2010,liu2010}. The fitting
leads to the parameters displayed in Table~\ref{tab:Bi2Se3Parameters} \cite{liu2010}.

\begin{table}[htb]
  \centering
  \begin{minipage}[t]{0.8\linewidth}
      \caption{ The parameters in the model Hamiltonian (\ref{eq:Heff3D}) obtained from fitting to {\it ab initio} calculation. Adapted from \cite{liu2010}.}
\label{tab:Bi2Se3Parameters}
\begin{tabular}{cccc}
    \hline\hline
    & $Bi_2Se_3$ & $Bi_2Te_3$ & $Sb_2Te_3$ \\\hline
       $A_1 (eV\cdot$\AA) & 2.26 & 0.30 & 0.84 \\\hline
       $A_2 (eV\cdot$\AA) & 3.33 & 2.87 & 3.40 \\\hline
       $C (eV)$ & -0.0083 & -0.18 & 0.001 \\\hline
       $D_1 (eV\cdot$\AA$^2)$ & 5.74 & 6.55 & -12.39 \\\hline
       $D_2 (eV\cdot$\AA$^2)$ & 30.4 & 49.68 & -10.78 \\\hline
       $M (eV)$ & 0.28 & 0.30 & 0.22 \\\hline
       $B_1 (eV\cdot$\AA$^2)$ & 6.86 & 2.79 & 19.64 \\\hline
       $B_2 (eV\cdot$\AA$^2)$ & 44.5 & 57.38 & 48.51 \\
       \hline\hline
  \end{tabular}%
\end{minipage}
\end{table}

Except for the identity term $\epsilon_0({\bf k})$, the Hamiltonian
(\ref{eq:Heff3D}) is similar to the 3D Dirac model with uniaxial
anisotropy along the $z$ direction, but with the crucial difference that the mass term is ${\bf k}$-dependent. From the fact that $M,B_1,B_2>0$ we can see that the order
of the bands $\left|T1^+_z,\uparrow(\downarrow)\right\rangle$ and
$\left|T2^-_z,\uparrow(\downarrow)\right\rangle$ is inverted
around ${\bf k}=0$ compared with large ${\bf k}$, which correctly
characterizes the topologically nontrivial nature of the system.
In addition, the Dirac mass $M$, i.e. the bulk
insulating gap, is $\sim 0.3$~eV, which allows the possibility of
having a room-temperature topological insulator. Such an effective
model can be used for further theoretical study of the
Bi$_2$Se$_3$ system, as long as low-energy properties are concerned.

Corrections to the effective Hamiltonian (\ref{eq:Heff3D}) that are of higher order in ${\bf k}$ can also be considered. To cubic ($k^3$) order, some new terms can break the continuous rotation symmetry around the $z$ axis to a discrete three-fold rotation symmetry $C_3$. Correspondingly, the Fermi surface of the surface state acquires a hexagonal shape~\cite{fu2009c}, which leads to important consequences for experiments on topological insulators such as surface state quasiparticle interference~\cite{tzhang2009,alpichshev2010,lee2009,zhou2009}. A modified version of the effective model (\ref{eq:Heff3D}) taking into account corrections up to $k^3$ has been obtained for the three topological insulators Bi$_2$Se$_3$, Bi$_2$Te$_3$, and Sb$_2$Te$_3$ based on {\it ab initio} calculations~\cite{liu2010}. In this same work~\cite{liu2010}, an eight-band model is also proposed for a more quantitative description of this family of topological insulators.

\subsection{Surface states with a single Dirac cone}
\label{sec:surface}

\begin{figure}
   \begin{center}
\includegraphics[angle=90,width=3.5in]{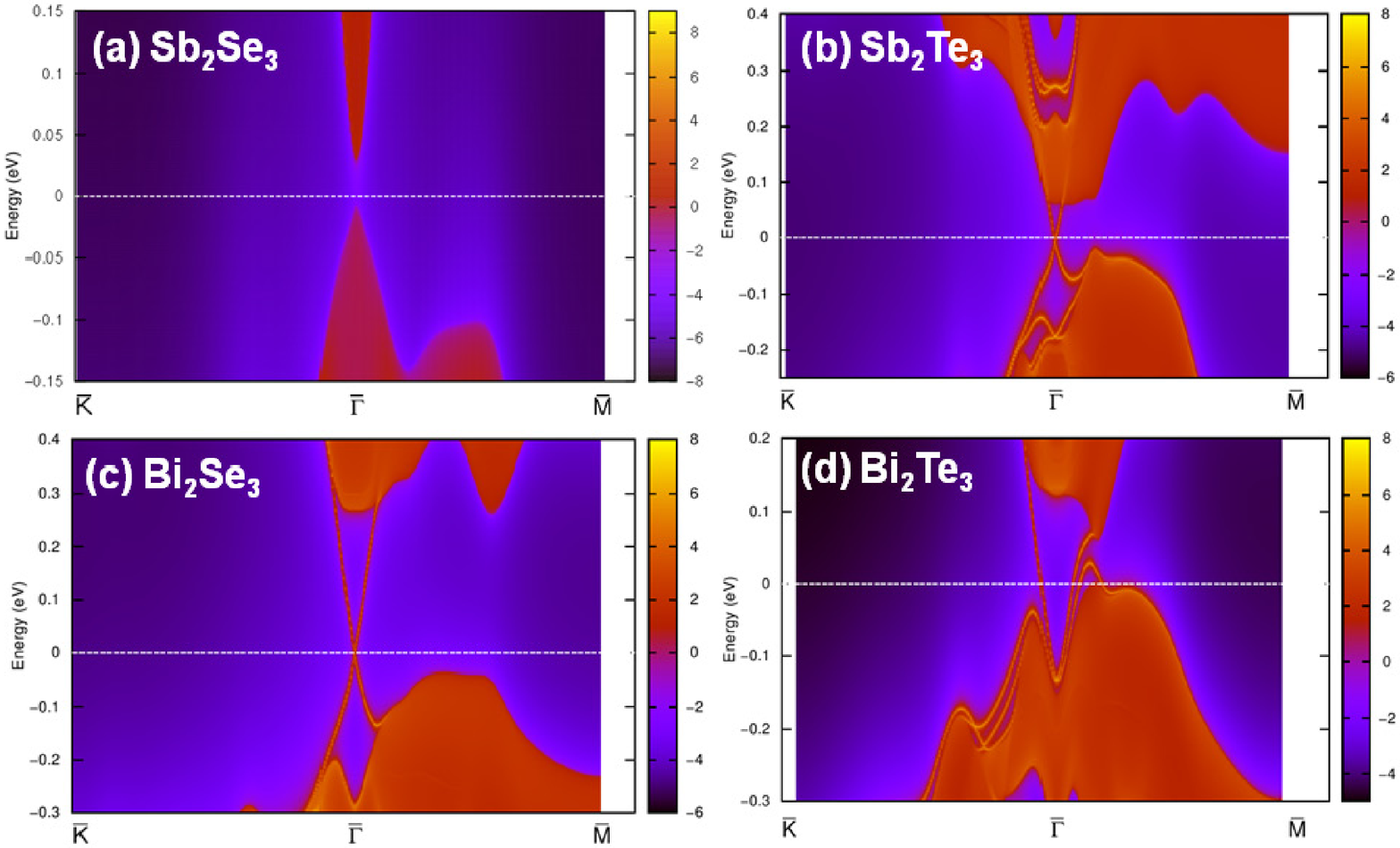}
\includegraphics[angle=0,width=3.5in]{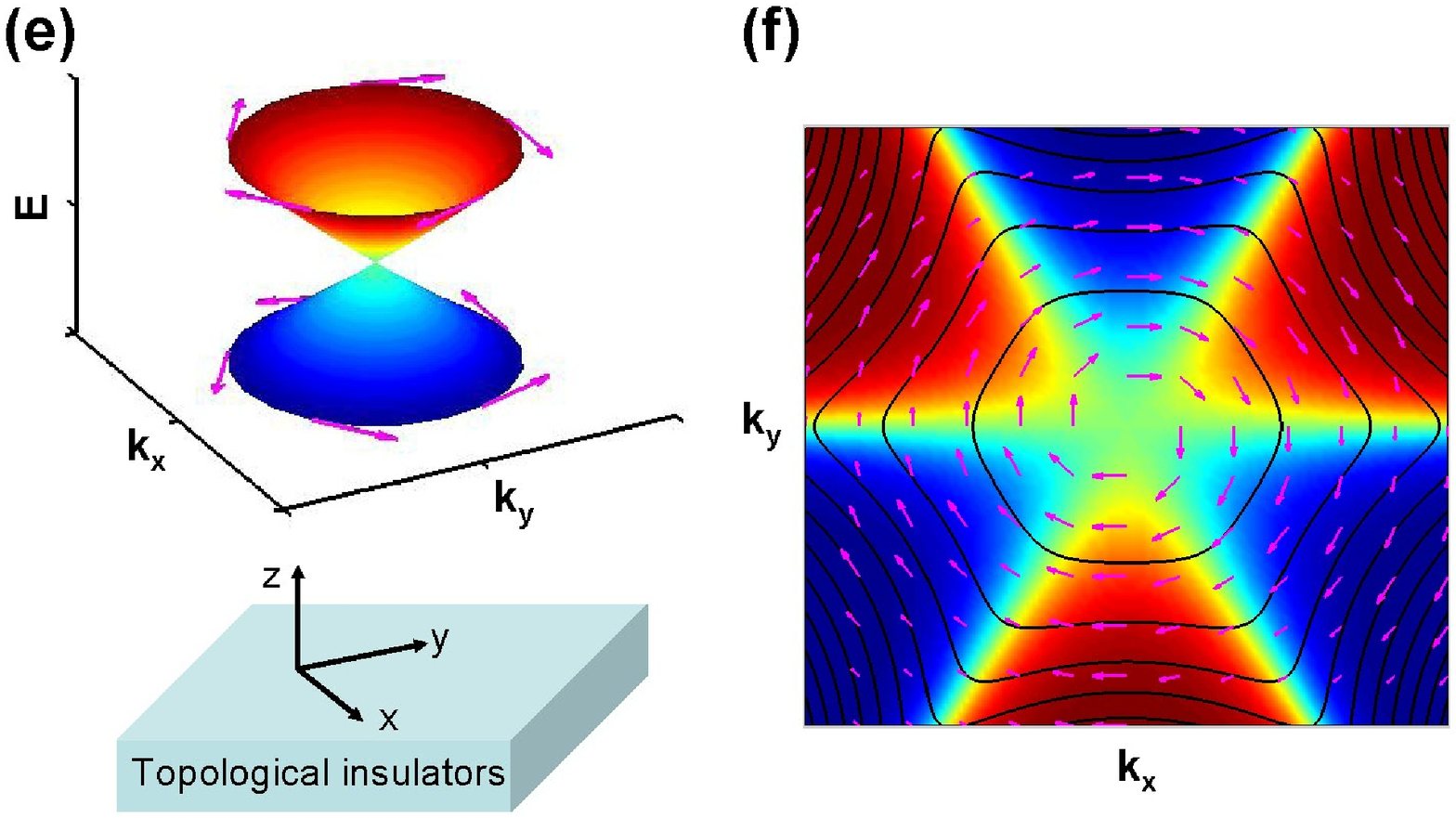}
    \end{center}
    \caption{(a)-(d) Energy and momentum dependence of the local density of states for the Bi$_2$Se$_3$ family of materials on the $[111]$ surface. A warmer color represents a higher local density of states. Red regions indicate
    bulk energy bands and blue regions indicate a bulk energy gap.
    The surface states can be clearly seen around $\Gamma$ point as red lines dispersing inside the bulk gap. (e) Spin polarization of the surface states on the top surface, where the $z$ direction is the surface normal,
    pointing outwards. Adapted from \onlinecite{zhang2009} and \onlinecite{liu2010}.
    }
    \label{fig:ss}
\end{figure}

The existence of topological surface states is one of the most important properties of topological insulators. The surface states can be directly extracted from {\it ab initio} calculations by constructing maximally localized Wannier functions and calculating the local density of states on an open boundary~\cite{zhang2009}. The result for the Bi$_2$Se$_3$ family of materials is shown in Fig.~\ref{fig:ss}(a)-(d), where one can clearly see the single Dirac-cone surface state for the three topologically nontrivial materials. However, to obtain a better understanding of the physical origin of topological surface states, it is helpful to show how the surface states emerge from the effective model (\ref{eq:Heff3D})~\cite{zhang2009,linder2009,liu2010a,lu2010}. The surface states can be obtained in a similar way as the edge states of the BHZ model (Sec.~\ref{sec:edge2d}).

Consider the model Hamiltonian (\ref{eq:Heff3D}) on the half-space $z>0$. In the same way as in the 2D case, we can divide the model Hamiltonian into two parts,
\begin{eqnarray}
    \hat{H}&=&\tilde{H}_0+\tilde{H}_1,\label{Hdecom3d}\\
    \tilde{H}_0&=&\tilde{\epsilon}(k_z)+\left(\begin{array}{cccc}
       \tilde{M}(k_z)&A_1k_z&0&0\\
        A_1k_z&-\tilde{M}(k_z)&0&0\\
        0&0&\tilde{M}(k_z)&-A_1k_z\\
        0&0&-A_1k_z&-\tilde{M}(k_z)
    \end{array}\right),\nonumber\\
    \tilde{H}_1&=&D_2k^2_{\perp}+\left(
    \begin{array}{cccc}
        -B_2k_\perp^2&0&0&A_2k_-\\
        0&B_2k_\perp^2&A_2k_-&0\\
        0&A_2k_+&-B_2k_\perp^2&0\\
        A_2k_+&0&0&B_2k_\perp^2
    \end{array}
    \right),
\end{eqnarray}
with $\tilde{\epsilon}(k_z)=C+D_1k_z^2$ and
$\tilde{M}(k_z)=M-B_1k_z^2$. $\tilde{H}_0$ in
Eq.~(\ref{Hdecom2d}) and Eq.~(\ref{Hdecom3d}) are identical, with
the parameters $A,B,C,D,M$ in Eq.~(\ref{Hdecom2d}) replaced by
$A_1,B_1,C,D_1,M$ in Eq.~(\ref{Hdecom3d}). Therefore, the surface state at $k_x=k_y=0$ is
determined by the same equation as that for the QSH edge states. A surface
state solution exists for $M/B_1>0$. In the same way as in the 2D case,
the surface state has a helicity determined by the sign of
$A_1/B_1$. (Here and below we always consider the case with $B_1B_2>0$, $A_1A_2>0$.)

In analogy to the 2D QSH case, the surface effective model can be
obtained by projecting the bulk Hamiltonian onto the surface states. To the leading order in $k_x,k_y$, the effective surface Hamiltonian $H_{\rm surf}$ has the following matrix form~\cite{zhang2009,liu2010}:
\begin{eqnarray}
H_{\rm
surf}(k_x,k_y)={C}+A_2\left(\sigma^xk_y-\sigma^yk_x\right).
\label{eq:sur_Heff}
\end{eqnarray}
Higher order terms such as $k^3$
terms break the axial symmetry around the $z$ axis down to a three-fold rotation symmetry, which has been studied in the literature~\cite{fu2009c,liu2010}. For $A_2=4.1$~eV$\cdot$\AA, the velocity of the
surface states is given by $v=A_2/\hbar\simeq 6.2\times 10^5$~m/s, which agrees reasonably with {\it ab initio} results
[Fig.~\ref{fig:ss}] $v\simeq 5.0\times 10^5$~m/s.

To understand the physical properties of the surface states, we need to
analyze the form of the spin operators in this system. By using the wave
function from {\it ab initio} calculations and projecting the spin
operators onto the subspace spanned by the four basis states,
we obtain the spin operators for our model Hamiltonian, with matrix elements between surface states given by
$\langle \Psi_\alpha|S_x|\Psi_\beta\rangle=S_{x0}\sigma^{\alpha\beta}_x$, $\langle
\Psi_\alpha|S_y|\Psi_\beta\rangle=S_{y0}\sigma^{\alpha\beta}_y$ and $\langle
\Psi_\alpha|S_z|\Psi_\beta\rangle=S_{z0}\sigma^{\alpha\beta}_z$, with $S_{x(y,z)0}$ some positive
constants. Therefore, we see that the Pauli $\boldsymbol{\sigma}$ matrix in the model Hamiltonian
(\ref{eq:sur_Heff}) is proportional to the physical spin. As discussed above, the spin direction is determined by the sign of the parameter $A_1/B_1$, which depends on material properties such as the atomic SOC. In the Bi$_2$Se$_3$ family of materials, the upper Dirac cone has a left-handed helicity when looking from above the surface [Fig.~\ref{fig:ss}(e),(f)].

From the discussion above, we see that the surface state is described by a 2D massless Dirac Hamiltonian (\ref{eq:sur_Heff}). Another well-known system with a similar property is graphene, a single sheet of graphite~\cite{castroneto2009}. However, there is a key difference between the surface state theory for 3D topological insulators and graphene or any 2D Dirac system, which is the number of Dirac cones. Graphene has four Dirac cones at low energies, due to spin and valley degeneracy. The valley degeneracy occurs because the Dirac cones are not in the vicinity of ${\bf k}=0$ but rather near the two Brillouin zone corners $K$ and $\bar{K}$. This is generic for a purely 2D system: only an even number of Dirac cones can exist in a TR invariant system. In other words, a single 2D Dirac cone without TR symmetry breaking can only exist on the surface of a topological insulator, which is also an alternative way to understand its topological robustness. As long as TR symmetry is preserved, the surface state cannot be gapped out because no purely 2D system can provide a single Dirac cone. Such a surface state is a ``holographic metal" which is 2D but determined by the 3D bulk topological property.

In this section we discussed the surface states of an insulator surrounded by vacuum. This formalism can be straightforwardly generalized to the interface states between two insulators~\cite{fradkin1986,volkov1985}. In these pioneering works, the interface states between PbTe and SnTe were investigated. The interface states consist of four Dirac cones. Therefore, they are topologically trivial and not generally stable under TR invariant perturbations. The surface states of topological insulators are also similar to the domain wall fermions of lattice gauge theory~\cite{kaplan1992}. In fact, domain wall fermions are precisely introduced to avoid the fermion doubling problem on the lattice, which is similar to the concept of a single Dirac cone on the surface of a topological insulator.

The helical spin texture described by the single Dirac cone equation (\ref{eq:sur_Heff}) leads to a general relation between charge current density $\mathbf{j}(\mathbf{x})$ and spin density $\mathbf{S}(\mathbf{x})$ on the surface of the topological insulator~\cite{raghu2010}:
\begin{equation}
\mathbf{j}(\mathbf{x})=v[\psi^\dag(\mathbf{x})\boldsymbol{\sigma}\psi(\mathbf{x})
\times\hat{\mathbf{z}}]
=v\mathbf{S}(\mathbf{x})\times\hat{\mathbf{z}}.
\end{equation}
In particular, the plasmon mode on the surface generally carries spin~\cite{raghu2010,burkov2010}.

\subsection{Crossover from three dimensions to two dimensions}\label{sec:2dto3dTheory}

From the discussion above, one can see that the models describing 2D and 2D topological insulators are quite similar. Both systems are described by lattice Dirac-type Hamiltonians. In particular, when inversion symmetry is present, the topologically nontrivial phase in both models is characterized by a band inversion between two states of opposite parity. Therefore, it is natural to study the relation between these two topological states of matter. One natural question is whether a thin film of 3D topological insulator, viewed as a 2D system, is a trivial insulator or a QSH insulator. Besides theoretical interest, this problem is also relevant to experiments, especially in the Bi$_2$Se$_3$ family of materials. Indeed, these materials are layered and can be easily grown as thin films either by MBE~\cite{yli2009,hdli2010,zhangqpl2009}, catalyst-free vapor-solid growth~\cite{kong2010}, or by mechanical exfoliation~\cite{teweldebrhan2010,shahil2010,hong2010}. Several theoretical works studied thin films of the Bi$_2$Se$_3$ family of topological insulators~\cite{liu2010a,linder2009,lu2010}. Interestingly, thin films of proper thicknesses are predicted to form a QSH insulator~\cite{liu2010a,lu2010}, which may constitute an approach for simpler realizations of the 2D QSH effect.

\begin{figure}
    \begin{center}
        \includegraphics[width=3.3in]{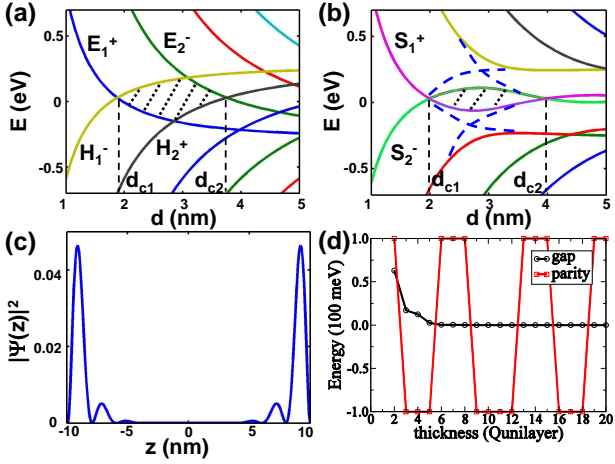}
    \end{center}
    \caption{Energy levels versus quantum well thickness for (a) $A_1=0$~eV$\cdot$\AA, (b) $A_1=1.1$~eV$\cdot$\AA. Other parameters are taken from \onlinecite{zhang2009}. Shaded regions indicate the QSH regime. The blue dashed line in (b) shows how the crossing between $|E_1(H_1)\rangle$ and $|H_2(E_2)\rangle$ evolves into an anti-crossing when $A_1\neq 0$. (c) Probability density in the state $|S^{+}_1\rangle$ (same for $|S^{-}_2\rangle$) for $A_1=1.1$~eV$\cdot$\AA~and $d=20$~nm. (d) Band gap and total parity from {\it ab initio} calculations on Bi$_2$Se$_3$, plotted as a function of the number of quintuple layers. From \onlinecite{liu2010a}.}
    \label{fig:band2}
\end{figure}

Such a crossover from 3D to 2D topological insulators can be studied from two points of view, either from the bulk states of the 3D topological insulator or from the surface states. We first consider the bulk states. A thin film of 3D topological insulator is described by restricting the bulk model (\ref{eq:Heff3D}) to a QW with thickness $d$, outside which there is an infinite barrier describing the vacuum. To establish the connection between the 2D BHZ model (\ref{BHZ}) and the 3D topological insulator model (\ref{eq:Heff3D}), we start from the special case $A_1=0$ and consider a finite $A_1$ later on. For $A_1=0$ and $k_x=k_y=0$, the Hamiltonian (\ref{eq:Heff3D}) becomes diagonal and the Schr\"{o}dinger equation for the infinite QW can be easily solved. The Hamiltonian eigenstates are simply given by $|E_n(H_n)\rangle=\sqrt{\frac{2}{d}}\sin\left(\frac{n\pi z}{d}+\frac{n\pi}{2}\right)|\Lambda\rangle$, with $|\Lambda\rangle=
|P1^+_z,\uparrow(\downarrow)\rangle$ for electron subbands and
$|\Lambda\rangle=|P2^-_z,\uparrow(\downarrow)\rangle$ for hole
subbands. The corresponding energy spectrum is
$E_e(n)=C+M+(D_1-B_1)\left( \frac{n\pi}{d} \right)^2$ and
$E_h(n)=C-M+(D_1+B_1)\left( \frac{n\pi}{d} \right)^2$,
respectively. We assume $M<0$ and $B_1<0$ so that the
system stays in the inverted regime. The energy spectrum is shown in Fig.~\ref{fig:band2}(a). When the width $d$ is small enough, electron subbands $E_n$ have a higher energy than the hole subbands $H_n$ due to quantum confinement effects. Because the bulk bands are inverted at the $\Gamma$ point ($M<0$), the energy of the electron subbands will decrease with increasing $d$ towards their bulk value $M<0$, while the energy of the hole subbands will increase towards $-M>0$. Therefore, there must exist a crossing point between the electron and hole subbands.

When a finite $A_1$ is turned on, the electron and hole bands are hybridized so that some of the crossings between the QW levels are avoided. However, as shown in Fig.~\ref{fig:band2}(b), some level crossings cannot be lifted, which is a consequence of inversion symmetry. When the band index $n$ is increased, the parity of the wave functions alternates for both electron and hole subbands. Moreover, the atomic orbitals forming electron and hole bands are $|P1^+_z,\uparrow(\downarrow)\rangle$ and $|P2^-_z,\uparrow(\downarrow)\rangle$ respectively, which have opposite parity. Consequently, $\left|E_n\right\rangle$ and $\left|H_n\right\rangle$ with the same index $n$ have opposite parity, so that their crossing cannot be avoided by the $A_1$ term. When a finite $k_x,k_y$ is considered, each level becomes a QW subband. The bottom of the lowest conduction band and the top of the highest valence band are indicated by $S_1^+$ and $S_2^-$ in Fig.~\ref{fig:band2}(b). Since these two bands have opposite parity, each level crossing between them is a topological phase transition between trivial and QSH insulator phases~\cite{bernevig2006c,fu2007a}. Since the system must be trivial in the limit $d\rightarrow 0$, we know that the first QSH insulator phase occurs between the first and second level crossing. In the $A_1\rightarrow 0$ limit, the crossing positions are given by the critical well thicknesses $d_{cn}=n\pi\sqrt{\frac{B_1}{|M|}}$. In principle, there is an infinite number of QSH phases between $d_{c,2n-1}$ and $d_{c,2n}$. However, as seen in Fig.~\ref{fig:band2}(b), the gap between $|S^+_1\rangle$ and $|S^-_2\rangle$ decays quickly for large $d$. In the 3D limit $d\rightarrow\infty$, the two states become degenerate and actually form the top and bottom surface states of the bulk crystal [Fig.~\ref{fig:band2}(c)].

This relation between the QW valence and conduction bands and the surface states in the $d\rightarrow\infty$ limit suggests an alternative way to understand the crossover from 3D to 2D, i.e. from the surface states. In the 3D limit the two surfaces are decoupled and are the only low-energy states. The top surface is described by the effective Hamiltonian (\ref{eq:sur_Heff}) while the bottom surface is obtained from the top surface by inversion. Therefore, the complete effective Hamiltonian is given by
\begin{eqnarray}
H_{\rm
surf}(k_x,k_y)
&=&A_2\left(\begin{array}{cccc}
0&ik_-&0&0\\
-ik_+&0&0&0\\
0&0&0&-ik_-\\
0&0&ik_+&0
\end{array}\right).\nonumber
\end{eqnarray}
When a slab of finite thickness is considered, the two surface states start overlapping, such that off-diagonal terms are introduced in the effective Hamiltonian. An effective Hamiltonian consistent with inversion and TR symmetry and incorporating inter-surface tunneling is given by
\begin{eqnarray}
H_{\rm
surf}(k_x,k_y)
&=&A_2\left(\begin{array}{cccc}
0&ik_-&M_\mathrm{2D}&0\\
-ik_+&0&0&M_\mathrm{2D}\\
M_\mathrm{2D}&0&0&-ik_-\\
0&M_\mathrm{2D}&ik_+&0\end{array}\right),\label{slabH}
\end{eqnarray}
where $M_\mathrm{2D}$ is a TR invariant mass term due to inter-surface tunneling, which generally depends on the in-plane momentum. Equation (\ref{slabH}) is unitarily equivalent to the BHZ Hamiltonian (\ref{BHZ}) for HgTe QWs. Whether the Hamiltonian corresponds to a trivial or QSH insulator cannot be determined without studying the behavior of this model at large momenta. Indeed, we are missing a regularization term which would play the role of the quadratic term $Bk^2$ in the BHZ model (Sec.~\ref{sec:bhz}). However, the transitions between trivial and nontrivial phases are accompanied by a sign change in $M_\mathrm{2D}$, independently of the regularization scheme at high momenta. Upon variation of $d$, the sign of the inter-surface coupling $M_\mathrm{2D}$ oscillates because the surface state wave functions oscillate [Fig.~\ref{fig:band2}(b)]. Therefore, we reach the same conclusions as in the bulk approach.

The results above obtained from calculations using an effective model are also confirmed by first-principle calculations. The parity eigenvalues of occupied bands have been calculated as a function of the thickness of the 3D topological insulator film~\cite{liu2010a}, from which the topological nature of the film can be inferred. The result is shown in Fig.~\ref{fig:band2}(d), which confirms the oscillations found in the effective model. The first nontrivial phase appears at a thickness of three quintuple layers, i.e. about $3$ nm for Bi$_2$Se$_3$.

\subsection{Electromagnetic properties}\label{sec:magnetic}

In previous subsections, we have reviewed bulk and surface properties of 3D topological insulators, as well as their relation to 2D topological insulators (QSH insulators), based on a microscopic model. From the effective model of surface states, one can understand their robustness protected by TR symmetry. However, similarly to the quantized Hall response in QH systems, the topological structure in topological insulators should not only lead to robust gapless surface states, but also to unique, quantized electromagnetic response coefficients. The quantized electromagnetic response of 3D topological insulators turns out to be a topological magnetoelectric effect (TME)~\cite{qi2008b,qi2009}, which occurs when TR symmetry is broken on the surface, but not in the bulk. The TME effect is a generic property of 3D topological insulators, which can be obtained theoretically from generic models and from an effective field theory approach~\cite{qi2008b,fu2007a,essin2009}, independently of microscopic details. However, in order to develop a physical intuition for the TME effect, in the section we review this effect and its physical consequences based on the simplest surface effective model, and postpone a discussion in the framework of a general effective theory to Sec.~\ref{sec:general}. We shall also discuss various experimental manifestations of the TME effect.

\subsubsection{Half quantum Hall effect on the surface}
\label{half_QH}
We start by analyzing generic perturbations to the effective surface state Hamiltonian (\ref{eq:sur_Heff}). The only momentum-independent perturbation one can add is $H_1=\sum_{a=x,y,z}m_a\sigma^a$, and the perturbed Hamiltonian has the spectrum $E_{\bf k}=\pm\sqrt{\left(A_2k_y+m_x\right)^2+\left(A_2k_x-m_y\right)^2+m_z^2}$. Thus, the only parameter that can open a gap and destabilize the surface states is $m_z$, and we will only consider this perturbation in the following. The mass term $m_z\sigma^z$ is odd under
TR, as expected from the topological stability of surface states protected by TR symmetry. By comparison, if the surface states consist of even number of Dirac cones, one can check that a TR invariant mass term is indeed possible. For example, if there are two identical Dirac cones, an imaginary coupling between the cones can be introduced, which leads to the gapped TR invariant Hamiltonian
\begin{eqnarray}
H_{\rm surf}'({\bf k})=\left(\begin{array}{cc}A_2\left(\sigma^xk_y-\sigma^yk_x\right)&-im\sigma^z\\im\sigma^z&A_2\left(\sigma^xk_y-\sigma^yk_x\right)\end{array}\right).\nonumber
\end{eqnarray}
From such a difference between an even and an odd number of Dirac cones, one sees that the stability of the surface theory (\ref{eq:sur_Heff}) is protected by a $\mathbb{Z}_2$ topological invariant.

Although the surface state with a single Dirac cone does not remain gapless when a TR breaking mass term $m_z\sigma^z$ is added, an important physical property is induced by such a mass term: a half-integer quantized Hall conductance. As discussed in Sec.~\ref{sec:qah}, the Hall conductance of a generic two-band Hamiltonian $h({\bf k})=d_a({\bf k})\sigma^a$ is determined by Eq.~(\ref{winding2band}), which is the winding number of the unit vector $\hat{\bf d}({\bf k})={\bf d}({\bf k})/|{\bf d}({\bf k})|$ on the Brillouin zone. The perturbed surface state Hamiltonian
\begin{eqnarray}
H_{\rm surf}({\bf k})=A_2\sigma^xk_y-A_2\sigma^yk_x+m_z\sigma^z,\label{TRBsurface}
\end{eqnarray}
corresponds to a vector ${\bf d}({\bf k})=\left(A_2k_y,-A_2k_x,m_z\right)$. At ${\bf k}=0$, the unit vector $\hat{\bf d}({\bf k})=(0,0,m_z/|m_z|)$ points towards the north (south) pole of the unit sphere for $m_z>0$ ($m_z<0$). For $|{\bf k}|\gg |m_z|/A_2$, the unit vector $\hat{\bf d}({\bf k})\simeq A_2(k_y,-k_x,0)/|{\bf k}|$ almost lies in the equatorial plane of the unit sphere. From such a ``meron" configuration one sees that $\hat{\bf d}({\bf k})$ covers half of the unit sphere, which leads to a winding number $\pm 1/2$ and corresponds to a Hall conductance
\begin{eqnarray}
\sigma_H=\frac{m_z}{|m_z|}\frac{e^2}{2h}.\label{halfQH}
\end{eqnarray}
From this formula, it can be seen that the Hall conductance remains finite even in the limit $m_z\rightarrow 0$, and has a jump at $m_z=0$. As a property of the massive Dirac model, such a half Hall conductance has been studied a long time ago in high energy physics. In that context, the effect is termed the ``parity anomaly"~\cite{redlich1984,semenoff1984}, because the massless theory preserves parity (and TR) but an infinitesimal mass term necessarily breaks these symmetries.

The analysis above only applies if the continuum effective model (\ref{eq:sur_Heff}) applies, i.e. if the characteristic momentum $|m_z|/A_2$ is much smaller than the size of the Brillouin zone $2\pi/a$ with $a$ the lattice constant. Since deviations from this Dirac-type effective model at large momenta is not included in the above calculation of the Hall conductance~\cite{fu2007a,lee2009b}, one cannot unambiguously predict the Hall conductance of the surface. In fact, if the effective theory describes a 2D system rather than the surface of a 3D system, additional contributions from large-momentum corrections to the effective model are necessary, since the Hall condutance of any gapped 2D band insulator must be quantized in \emph{integer} units of $e^2/h$~\cite{thouless1982}. For example, the QAH insulator [Eq.~(\ref{QAH})] with mass term $M\rightarrow 0$ is also described by the same effective theory as (\ref{TRBsurface}), but has Hall conductance $0$ or $1$ rather than $\pm 1/2$~\cite{fradkin1986}.

Interestingly, the surface of a 3D topological insulator is different from all 2D insulators, in the sense that such contributions from large momenta vanish due to the requirement of TR symmetry~\cite{qi2008b}. This fact is discussed more rigorously in  Sec.~\ref{sec:general} based on the general effective field theory. Here we present an argument based on the bulk to surface relationship. To understand this, consider the jump in Hall conductance at $m_z=0$. Although deviations from the Dirac effective model at large momenta may lead to corrections to the Hall conductance for a given $m_z$, the change in Hall conductance $\Delta\sigma_H=\sigma_H(m_z\rightarrow 0^+)-\sigma_H(m_z\rightarrow 0^-)=\frac{m_z}{|m_z|}\frac{e^2}h$ is independent of the large-momentum contributions. Indeed, the effect of the mass term $m_z\sigma^z$ on the large-momentum sector of the theory is negligible as long as $m_z\rightarrow 0$. Therefore, any contributions to $\sigma_H$ from large momenta should be continuous functions of $m_z$, and thus cannot affect the value of the discontinuity $\Delta\sigma_H$. On the other hand, since the surface theory with $m_z=0$ is TR invariant, TR transforms the system with mass $m_z$ to that with mass $-m_z$. Consequently, from TR symmetry we have
\begin{eqnarray}
\sigma_H(m_z\rightarrow 0^+)=-\sigma_H(m_z\rightarrow 0^-).\nonumber
\end{eqnarray}
Together with the condition $\Delta\sigma_H=\frac{m_z}{|m_z|}\frac{e^2}h$, we see that the half Hall conductance given by Eq.~(\ref{halfQH}) is robust, and the contribution from large-momentum corrections must vanish. By comparison, in a 2D QAH Hamiltonian discussed in Sec. \ref{sec:qah}, i.e. the upper $2\times 2$ block of Eq.~(\ref{BHZ}), the Hamiltonian with mass $M$ is not the TR conjugate of that with mass $-M$, and the above argument does not apply. Therefore, the half Hall conductance is a unique property of the surface states of 3D topological insulators which is determined by the bulk topology. This property distinguishes the surface states of 3D topological insulators from all pure 2D systems, or topologically trivial surface states.

The analysis above has only considered translationally invariant perturbations to the surface states, but the conclusions remain robust when disorder is considered. A 2D metal without SOC belongs to the orthogonal or unitary symmetry classes of random Hamiltonians, the eigenfunctions of which are always localized when random disorder is introduced. This effect is known as Anderson localization~\cite{abrahams1979}. Anderson localization is a quantum interference effect induced by constructive interference between different backscattering paths. By comparison, a system with TR invariance and SOC belongs to the symplectic class, where the constructive interference becomes destructive. In that case, the system has a metallic phase at weak disorder, which turns into an insulator phase by going through a metal-insulator transition at a certain disorder strength~\cite{hikami1980,evers2008}. Naively, one would expect the surface state with nonmagnetic disorder to be in the symplectic class. However, Nomura, Koshino, and Ryu showed that the surface state is metallic even for an arbitrary impurity strength~\cite{nomura2007}, which is consistent with the topological robustness of the surface state.

On the contrary, with TR symmetry breaking disorder, the system belongs to the unitary class, which exhibits localization for arbitrarily weak disorder strength. While the longitudinal resistivity flows to infinity due to localization, the Hall conductivity flows to the quantized value $\pm e^2/2h$~\cite{nomura2008}. Therefore, the system enters a half QH phase once an infinitesimal TR symmetry breaking perturbation is introduced, independently of the detailed form of the TR breaking perturbation. Physically, TR breaking disorder is induced by magnetic impurities, the spin of which not only contributes a random TR breaking field, but also has its own dynamics. For example, the simplest exchange interaction between impurity spin and surface state can be written as $H_{\rm int}=\sum_iJ_i{\bf S}_i\cdot\psi^\dagger\boldsymbol{\sigma}\psi({\bf R}_i)$ with ${\bf S}_i$ the impurity spin, $\psi^\dagger\boldsymbol{\sigma}\psi({\bf R}_i)$ the spin density of surface electrons at the impurity position ${\bf R}_i$, and $J_i$ the exchange coupling. To understand the physical properties of the topological insulator surface in the presence of magnetic impurities, it is instructive to study the interaction between impurity spins mediated by the surface electrons~\cite{liu2009}. As in a usual Fermi liquid, if the surface state has a finite Fermi wave vector $k_F$, a Ruderman-Kittel-Kasuya-Yosida (RKKY) interaction between the impurity spins is introduced, the sign of which oscillates with wave length $\propto 1/2k_F$\cite{liu2009,ye2010}.
If the Fermi level is close to the Dirac point, i.e. $k_F\rightarrow 0$, the sign of the RKKY interaction does not oscillate but is uniform. The sign of the resulting uniform spin-spin interaction is determined by the coupling to the surface electrons, which turns out to be ferromagnetic. Physically, the interaction is ferromagnetic rather than antiferromagnetic, because a uniform spin polarization can maximize the gap opened on the surface, which is energetically favorable. Due to this ferromagnetic spin-spin interaction, the system can order ferromagnetically when the chemical potential is near the Dirac point~\cite{liu2009}. This mechanism is of great practical importance, because it provides a way to generate a surface TR symmetry breaking field by coating the surface with magnetic
impurities and tuning the chemical potential near the Dirac point~\cite{feng2009,chenwan2010,tran2010,zitko2010,cha2010}.

\subsubsection{Topological magnetoelectric effect}

\begin{figure}
\includegraphics[width=0.44\textwidth]{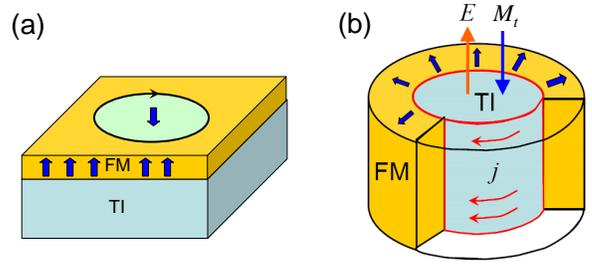}
\caption{(a) Ferromagnetic layer on the surface of topological insulator with a magnetic domain wall, along which a chiral edge state propagates. (b) Relation between surface half QH effect and bulk topological magnetoelectric effect. A magnetization is induced by an electric field due to the surface Hall current. From \onlinecite{qi2008b}.}\label{fig:TMEschematic}
\end{figure}

As discussed above, the surface half QH effect is a unique property of a TR symmetry breaking surface, and is determined by the bulk topology, independently of details of the surface TR symmetry breaking perturbation. A key difference between the surface half QH effect and the usual integer QH effect is that the former cannot be measured by a dc transport experiment. An integer QH system has chiral edge states which contribute to the quantized Hall current while being connected to leads. However, as one can easily convince oneself, it is a simple mathematical fact that the surface of a finite sample of 3D topological insulator is always a closed manifold without an edge. If the whole surface of a topological insulator sample is gapped by magnetic impurities, there are no edge states to carry a dc transport current. If the magnetic impurities form a ferromagnetic phase and there is a domain wall in the magnetic moment, the Hall conductance has a jump at the domain wall due to the formula (\ref{halfQH}). In this case, the jump of Hall conductance is $e^2/h$ across the domain wall, so that a chiral gapless edge state propagates along the domain wall [Fig.~\ref{fig:TMEschematic}(a)]~\cite{qi2008b}. This mechanism provides another route towards the QAH effect without any external magnetic field and the associated LLs. This is very much alike the boundary between two ordinary QH states with Hall conductance $ne^2/h$ and $(n+1)e^2/h$. The wave function for an edge state along a straight domain wall can also be solved for analytically following the same procedure as that used in Sec.~\ref{sec:edge2d}. Interestingly, if one attaches voltage and current leads to the domain wall in the same way as for an ordinary Hall bar, one should observe a Hall conductance of $e^2/h$ rather than $e^2/2h$, since the domain wall chiral state behaves in the same way as the edge state of a $\sigma_H=e^2/h$ QH system. Thus again we see that from dc transport measurements, one cannot observe the half Hall conductance.

Such a difference between integer QH effect and surface half QH effect indicates that the surface half QH effect is actually a new topological phenomenon which, in terms of its observable consequences, is qualitatively different from the usual integer QH effect. Alternatively, the proper detection of this new topological phenomenon actually probes a unique electromagnetic response property of the bulk, the TME~\cite{qi2008,essin2009}. A magnetoelectric effect is defined as a magnetization induced by an electric field, or alternatively, a charge polarization induced by a magnetic field. To understand the relation between surface half QH effect and magnetoelectric effect, consider the configuration shown in Fig.~\ref{fig:TMEschematic}(b), where the side surface of a 3D topological insulator is covered by magnetic impurities with ferromagnetic order, so that the surface is gapped and exhibits a half quantized Hall conductance. When an electric field ${\bf E}$ is applied parallel to the surface, a Hall current ${\bf j}$ is induced [Eq.~(\ref{HallCurrentTME})], which circulates along the surface. This surface current perpendicular to ${\bf E}$ will then induce a magnetic field parallel to ${\bf E}$, so that the system exhibits a magnetoelectric response. The Hall response equation is written as
\begin{eqnarray}
{\bf j}=\frac{m}{|m|}\frac{e^2}{2h}{\bf \hat{n}}\times {\bf E},\label{HallCurrentTME}
\end{eqnarray}
with $\hat{\bf n}$ a unit vector normal to the surface, and the sign of the mass $m/|m|$ is determined by the direction of the surface magnetization. Such a Hall response is equivalent to a magnetization proportional to the electric field:
\begin{eqnarray}
{\bf M}_t=-\frac{m}{|m|}\frac{e^2}{2hc}{\bf E}.\nonumber
\end{eqnarray}
This magnetization is a topological response to the electric field, and is independent of the details of the system. Similarly, a topological contribution to the charge polarization can be induced by a magnetic field. The complete electromagnetic response of the system is described by the following modified constituent equations,
\begin{eqnarray}
{\bf H}={\bf B}-4\pi {\bf M}+2P_3\alpha {\bf E},\nonumber\\
{\bf D}={\bf E}+4\pi{\bf P}-2P_3\alpha {\bf B},\label{constituenteq}
\end{eqnarray}
with $\alpha=e^2/\hbar c$ the fine structure constant, and $P_3\equiv m/2|m|=\pm 1/2$ the quantum of Hall conductance. A detailed explanation of the coefficient $P_3$ and the effective field theory description of the TME effect is discussed in Sec.~\ref{sec:general}. In this section we focus on the physical consequences of the TME effect. We simply note the fact that more generally, for topological insulators $P_3$ can take the value $n+1/2$ with arbitrary integer $n$, since the number of Dirac cones on the surface can be any odd integer.

\subsubsection{Image magnetic monopole effect}

One of the most direct consequences of the TME effect is the image magnetic monopole effect~\cite{qi2009}. Consider bringing an electric charge to the proximity of an ordinary 3D insulator. The electric charge will polarize the dielectric, which can be described by the appearance of an image electric charge inside the insulator. If the same thing is done with a topological insulator, in addition to the image electric charge an image magnetic monopole will also appear inside the insulator.

\begin{figure}
\includegraphics[width=0.25\textwidth]{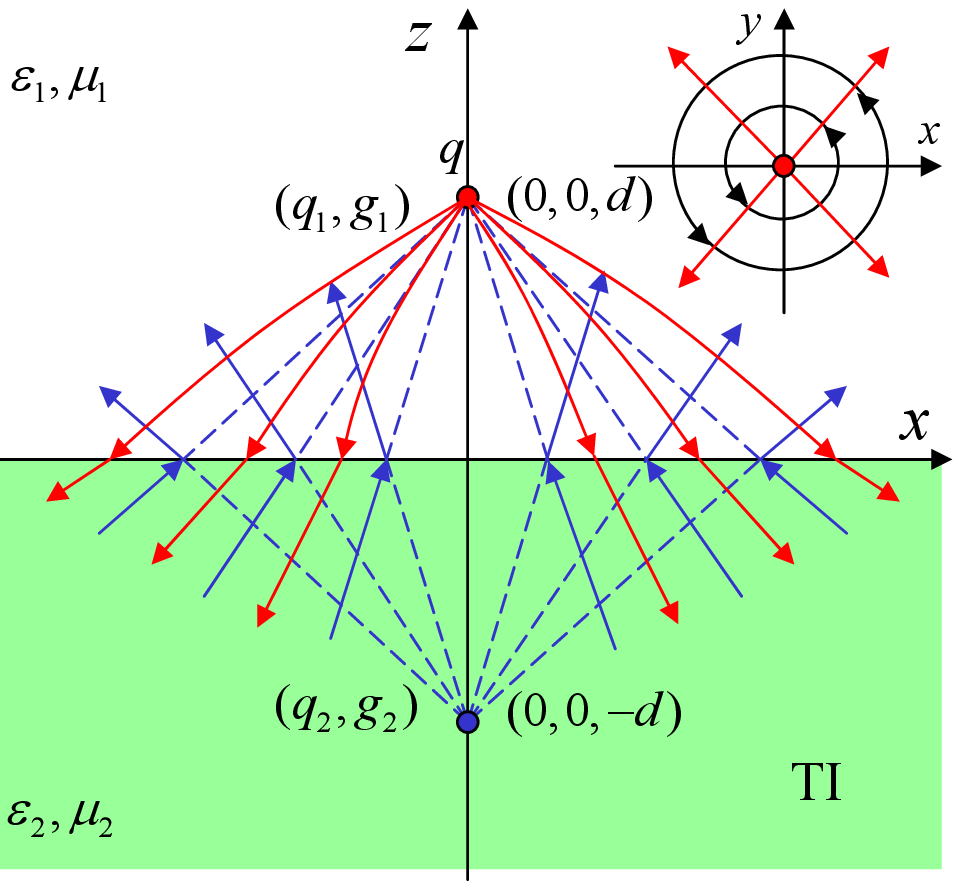}\includegraphics[width=0.25\textwidth]{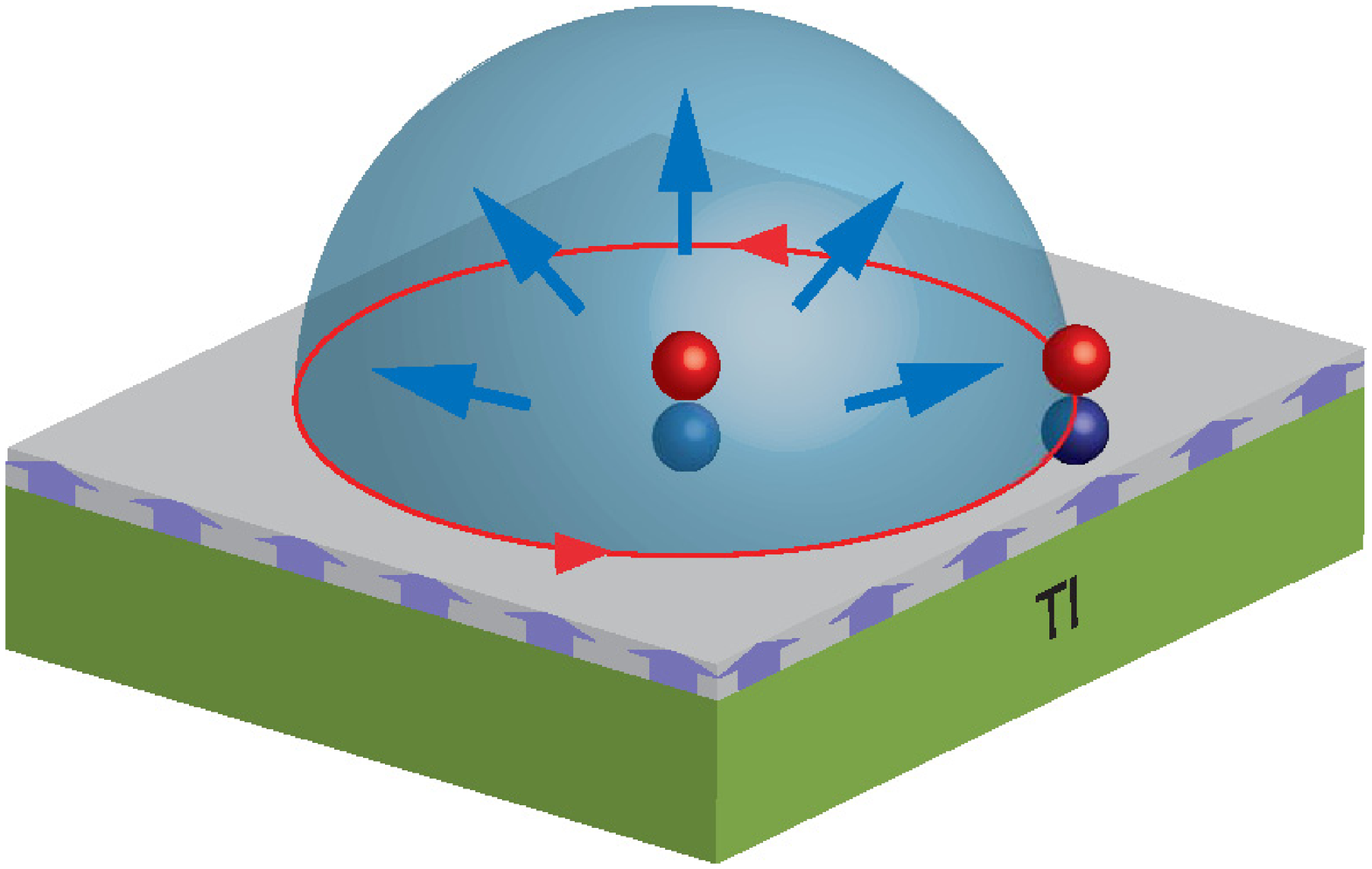}
\caption{(a) Image electric charge and image magnetic monopole due to an external electric point charge. The lower half-space is
occupied by a topological insulator (TI) with dielectric constant
$\epsilon_2$ and magnetic permeability $\mu_2$. The upper half-space
is occupied by a topologically trivial insulator (e.g. vacuum) with
dielectric constant $\epsilon_1$ and magnetic permeability $\mu_1$.
An electric point charge $q$ is located at $(0,0,d)$. Seen from the
lower half-space, the image electric charge $q_1$ and magnetic
monopole $g_1$ are at $(0,0,d)$. Seen from the upper half-space, the
image electric charge $q_2$ and magnetic monopole $g_2$ are at
$(0,0,-d)$. The red (blue) solid lines represent the electric
(magnetic) field lines. The inset is a top-down view showing the
in-plane component of the electric field on the surface (red arrows)
and the circulating surface current (black circles). (b) Illustration of the fractional statistics induced by the image monopole effect. Each electron forms a ``dyon" with its image monopole. When two electrons are exchanged, a Aharonov-Bohm phase factor is obtained, which is deter-
mined by half of the image monopole flux, independently of the
exchange path, leading to the phenomenon of statistical transmutation. From \onlinecite{qi2009}.}\label{fig:monopole1}
  \end{figure}

This image magnetic monopole effect can be studied straightforwardly by
solving Maxwell's equations with the modified constituent equations
(\ref{constituenteq}), in the same way as the image charge problem
in an ordinary insulator. Consider the geometry shown in Fig.~\ref{fig:monopole1}(a). The lower half-space $z<0$ is
occupied by a topological insulator with dielectric constant
$\epsilon_2$ and magnetic permeability $\mu_2$, while the upper
half-space $z>0$ is occupied by a conventional insulator with
dielectric constant $\epsilon_1$ and magnetic permeability
$\mu_1$. An electric point charge $q$ is located at $(0,0,d)$ with
$d>0$. We assume that the surface states are gapped by
some local TR symmetry breaking field $m$, so that the surface half
QH effect and TME exist. The boundary of the topological insulator acts
as a domain wall where $P_3$ jumps from $1/2$ to $0$. In this
semi-infinite geometry, an image point magnetic monopole with flux $g_2$ is located at the mirror position $(0,0,-d)$, together with an image electric point charge $q_2$. Physically, the magnetic field
of such an image monopole configuration is induced by circulating
Hall currents on the surface, which are induced by the electric field of the external charge. A similar effect has been studied in the integer QH
effect~\cite{haldane1983}. Conversely, the electromagnetic field strength inside the topological insulator is described by an image magnetic monopole $g_1$ and electric charge $q_1$ in the upper half-space, at the same point as the external charge. The image magnetic monopole flux and image electric charge $(q_1,g_1)$ and $(q_2,g_2)$ are given by
\begin{eqnarray}
q_1&=&q_2 =
\frac{1}{\epsilon_1}\frac{(\epsilon_1-\epsilon_2)(1/\mu_1+1/\mu_2)-4\alpha^2P_3^2}
{(\epsilon_1+\epsilon_2)(1/\mu_1+1/\mu_2)+4\alpha^2P_3^2}q,\nonumber \\
g_1&=&-g_2 = -\frac{4{\alpha}P_3}
{(\epsilon_1+\epsilon_2)(1/\mu_1+1/\mu_2)+4\alpha^2P_3^2}q.\label{imagemonopole}
\end{eqnarray}
Interestingly, by making use of electric-magnetic duality these expressions can be simplified to more compact forms~\cite{karch2009}.

Moreover, interesting phenomena appear when we consider the dynamics of the external charge. For example, consider a 2D electron gas at a distance $d$ above the surface of the 3D topological insulator. If the motion of the electron is slow enough (with respect to the time scale $\hbar/m$ corresponding to the TR symmetry breaking gap $m$), the image monopole will follow the electron adiabatically, such that the electron forms an electron-monopole composite, i.e. a dyon~\cite{witten1979}. When two electrons wind around each other, each electron perceives the magnetic flux of the image monopole attached to the other electron, which leads to statistical transmutation. The statistical angle is determined by the electron charge and image monopole flux as
\begin{eqnarray}
\theta=\frac{g_1q}{2\hbar c}=\frac{2{\alpha}^2P_3}
{(\epsilon_1+\epsilon_2)(1/\mu_1+1/\mu_2)+4\alpha^2P_3^2}.
\end{eqnarray}
The image monopole can be detected directly by local probes sensitive to small magnetic fields, such as scanning superconducting quantum interference devices (scanning SQUID) and scanning magnetic force microscopy (scanning MFM)~\cite{qi2009}. The current due to the image monopole can also be detected in principle~\cite{zang2010}.

\subsubsection{Topological Kerr and Faraday rotation}

\begin{figure}
          \includegraphics[width=0.24\textwidth]{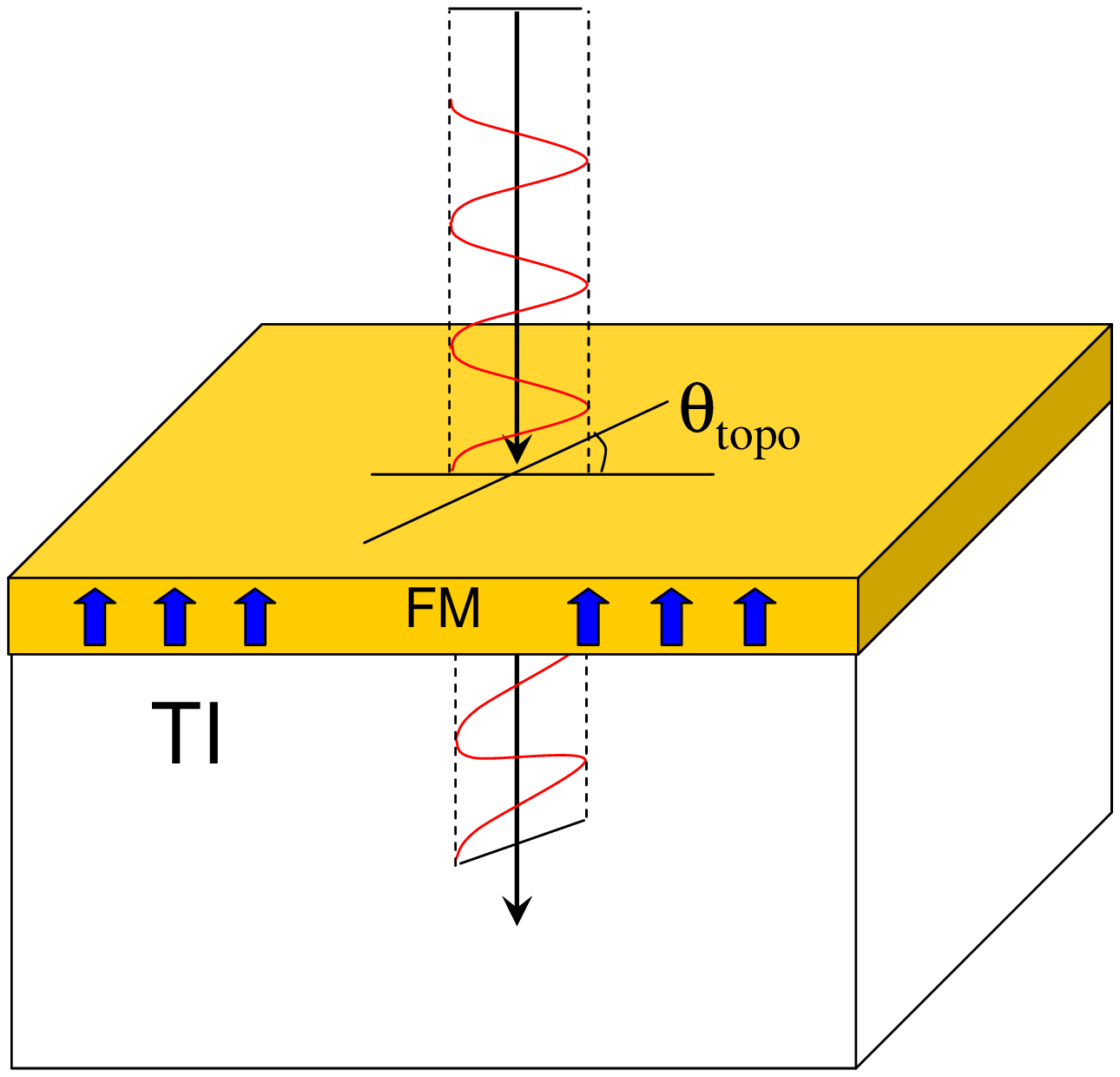}\includegraphics[width=0.24\textwidth]{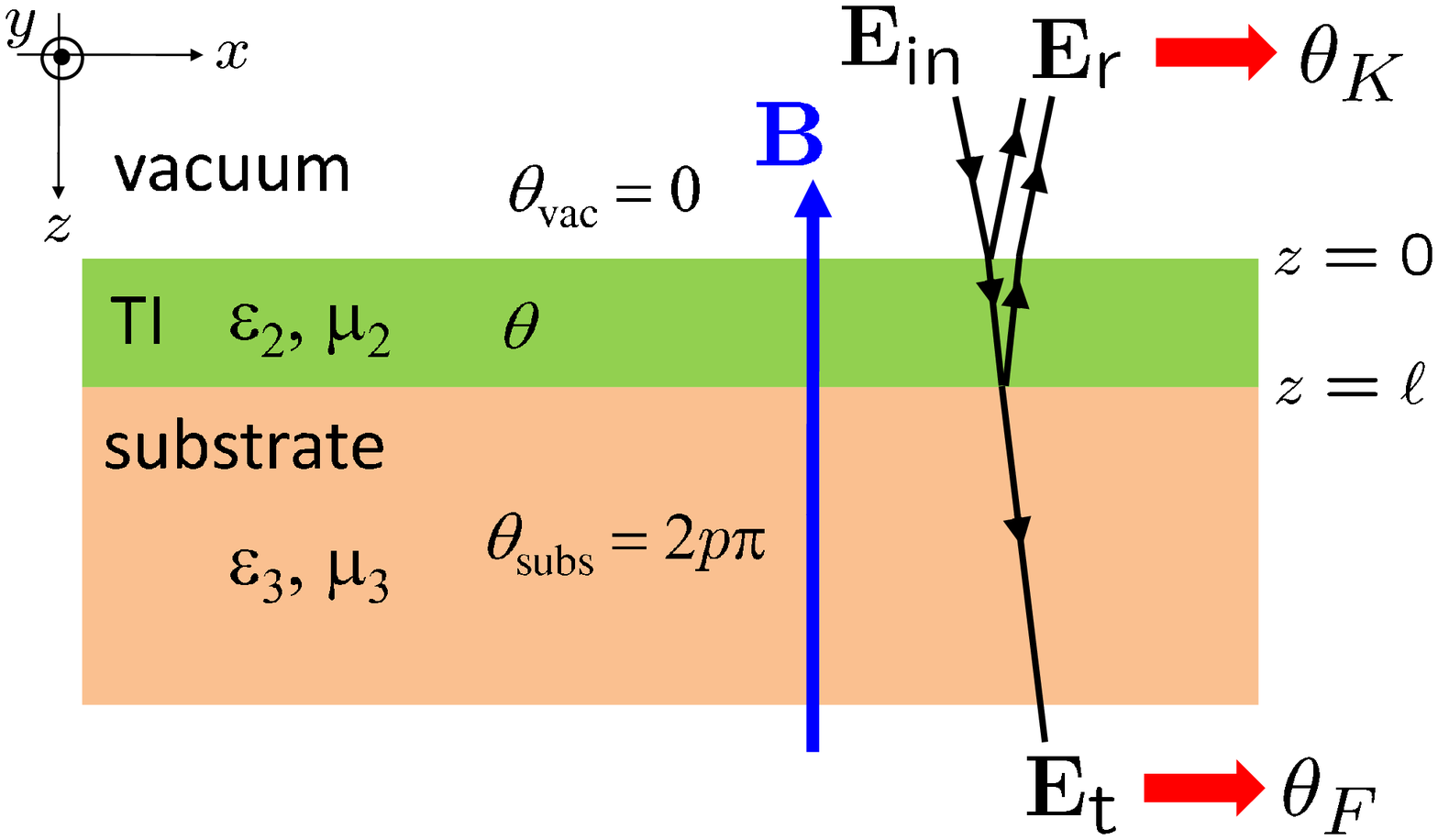}
\caption{(a) Illustration of the Faraday rotation on one surface of topological insulator. (b) A more complicated geometry with Kerr and Faraday rotation on two surfaces of topological insulator. In this geometry, the effect of non-universal properties of the material can be eliminated and the quantized magnetoelectric coefficient $\alpha P_3$ can be directly measured. From \onlinecite{qi2008b} and \onlinecite{maciejko2010}.}\label{fig:Faraday}
  \end{figure}

Another way to detect the TME effect is through the transmission and reflection of polarized light. When linearly polarized light propagates through a medium which breaks TR symmetry, the plane of polarization of the transmitted light may be rotated, which is known as the Faraday effect~\cite{LL_EM}. A similar rotation may occur for light reflected by a TR symmetry breaking surface, which is known as the magneto-optical Kerr effect~\cite{LL_EM}. Since the bulk of the topological insulator is TR invariant, no Faraday rotation will occur in the bulk. However, if TR symmetry is broken on the surface, the TME effect occurs and a unique kind of Kerr and Faraday rotation is induced on the surface. Physically, the plane of polarization of the transmitted and reflected light is rotated because the electric field ${\bf E}_0({\bf r},t)$ of linearly polarized light generates a magnetic field ${\bf B}({\bf r},t)$ in the same direction, due to the TME effect. Similarly as for the image monopole effect, the Faraday and Kerr rotation angles can be calculated by solving Maxwell's equations with the modified constituent equations (\ref{constituenteq}). In the simplest case of a single surface between a trivial insulator and a semi-infinite topological insulator~[Fig.~\ref{fig:Faraday}(a)], the rotation angle for light incident from the trivial insulator is given by~\cite{qi2008b,karch2009,tse2010,maciejko2010}
\begin{eqnarray}
\tan\theta_{K}&=&\frac{4\alpha P_3\sqrt{\epsilon_1/\mu_1}}{\epsilon_2/\mu_2-\epsilon_1/\mu_1+4\alpha^2P_3^2},\\
\tan\theta_F&=&\frac{2\alpha P_3}{\sqrt{\epsilon_1/\mu_1}+\sqrt{\epsilon_2/\mu_2}},
\end{eqnarray}
where $\epsilon_1,\mu_1$ are the dielectric constant and magnetic permeability of the trivial insulator, and $\epsilon_2,\mu_2$ are those of the topological insulator.

Although the surface Faraday and Kerr rotations are induced by the topological property of the bulk, and are determined by the magnetoelectric response with quantized coefficient $\alpha P_3$, the rotation angle is not universal and depends on the material parameters $\epsilon$ and $\mu$. The TME response always coexists with the ordinary electromagnetic response, which makes it difficult to observe the topological quantization phenomenon. However, recently new proposals have been made to avoid the dependence on non-universal material parameters~\cite{tse2010,maciejko2010}. The key idea is to consider a slab of topological insulator of finite thickness with two surfaces, with vacuum on one side and a substrate on the other~[Fig.~\ref{fig:Faraday}(b)]. The combination of Kerr and Faraday angles measured \emph{at reflectivity minima} provides enough information to determine the quantized coefficient $\alpha P_3$~\cite{maciejko2010},
\begin{equation}\label{alphaquant}
\frac{\cot\theta_F+\cot\theta_K}{1+\cot^2\theta_F}=2\alpha P_3,
\end{equation}
provided that both top and bottom surfaces have the same surface Hall conductance $\sigma_H=P_3e^2/h$.
In Eq.~(\ref{alphaquant}), the quantized coefficient $\alpha P_3$ is expressed solely in terms of the measurable Kerr and Faraday angles. This enables a direct experimental measurement of $P_3$ without a separate measurement of the non-universal optical constants $\epsilon,\mu$ of the topological insulator film and the substrate. If the two surfaces have different surface Hall conductances, it is still possible to determine them separately through a measurement of the Kerr and Faraday angles at reflectivity \emph{maxima}~\cite{maciejko2010}.

\subsubsection{Related effects}

The TME has other interesting consequences. The TME effect corresponds to a term $\theta {\bf E\cdot B}$ in the action (see Sec.~\ref{sec:general} for more details), which mediates the transmutation between electric field and magnetic field~\cite{qi2008b}. In the presence of a magnetic monopole, such an electric-magnetic transmutation induces an electric field around the magnetic monopole, so that the monopole carries an electric charge~\cite{witten1979} $q=e\frac{\theta}{2\pi}\frac{g}{\phi_0}$ with $\phi_0=hc/e$ the flux quanta. Such a composite particle carries both magnetic flux and electric charge, and is called a dyon. In principle, the topological insulator provides a physical system which can detect a magnetic monopole through this effect~\cite{rosenberg2010}. For a topological insulator, we have $\theta=\pi$ which corresponds to a half charge $q=e/2$ for a monopole with unit flux. Such a half charge corresponds to a zero energy bound state induced by the monopole. The charge of the monopole is $e/2$ when the bound state is occupied, and $-e/2$ when it is unoccupied. Such a half charge and zero mode is similar to charge fractionalization in 1D systems~\cite{su1979}. If the monopole passes through a hole in the topological insulator, the charge will follow it, which corresponds to a charge pumping effect~\cite{rosenberg2010b}.

All effects discussed up to this point are consequences of the TME in a topological insulator with surface TR symmetry breaking. No effects of electron correlation have been taken into account. When the electron-electron interaction is considered, interesting new effects can occur. For example, if a topological insulator is realized by transition metal compounds with strong electron correlation effects, antiferromagnetic (AFM) long-range order may develop in this material. Since the AFM order breaks TR symmetry and inversion symmetry, the magnetoelectric coefficient $P_3$ defined in Eq.~(\ref{constituenteq}) deviates from its quantized value $n+1/2$. Denoting by ${\bf n}({\bf r},t)$ the AFM N\'{e}el vector, we have $P_3({\bf n})=P_3({\bf n}=0)+\delta P_3({\bf n})$ where $P_3({\bf n}=0)$ is the quantized value of $P_3$ in the absence of AFM order. This change in the magnetoelectric coefficient has interesting consequences when spin-wave excitations are considered. Fluctuations of the N\'{e}el vector $\delta{\bf n}({\bf r},t)$ induce in general fluctuations of $\delta P_3$, leading to a coupling between spin-waves and the electromagnetic field~\cite{li2010}. In high-energy physics, such a particle coupled to the ${\bf E\cdot B}$ term is called an ``axion"~\cite{peccei1977,wilczek2009}. Physically, in a background magnetic field such an ``axionic" spin-wave is coupled to the electric field with a coupling constant tunable by the magnetic field. Consequently, a polariton can be formed by the hybridization of the spin-wave and photon, similar to the polariton formed by optical phonons~\cite{mills1974}. The polariton gap is controlled by the magnetic field, which may realize a tunable optical modulator.

Another interesting effect emerges from electron correlations when a thin film of topological insulator is considered. When the film is thick enough so that there is no direct tunneling between the surface states on the top and bottom surfaces, but not too thick so that the long-range Coulomb interaction between the two surfaces are still important, a inter-surface particle-hole excitation, i.e. an exciton, can be induced~\cite{seradjeh2009}. Denoting the fermion annihilation operator on the two surfaces by $\psi_1,~\psi_2$, the exciton creation operator is $\psi_1^\dagger \psi_2$. In particular, when the two surfaces have opposite Fermi energy with respect to the Dirac point, there is nesting between the two Fermi surfaces, which leads to an instability towards exciton condensation. In the exciton condensate phase, the exciton creation operator acquires a nonzero expectation value $\langle\psi_1^\dagger\psi_2\rangle\neq 0$, which corresponds to an effective inter-surface tunneling. Interestingly, one can consider a vortex in this exciton condensate. Such a vortex corresponds to a complex spatially-dependent inter-surface tunneling amplitude, and is equivalent to a magnetic monopole. According to the Witten effect mentioned above~\cite{rosenberg2010}, such a vortex of the exciton condensate carries charge $\pm e/2$, which provides a way to test the Witten effect in the absence of a real magnetic monopole.

Besides the effects discussed above, there are many other physical effects related to the TME effect, or surface half QH effect. When a magnetic layer is deposited on top of the topological insulator surface, the surface states can be gapped and a half QH effect is induced. In other words, the magnetic moment of the magnetic layer determines the Hall response of the surface states, which can be considered as a coupling between the magnetic moment and the surface electric current~\cite{qi2008b}. Such a coupling leads to the inverse of the half QH effect, which means that a charge current on the surface can flip the magnetic moment of the magnetic layer~\cite{garate2010}. Similar to such a coupling between charge current and magnetic moment, a charge density is coupled to magnetic textures such as domain walls and vortices~\cite{nomura2010}. This effect can be used to drive magnetic textures by electric fields. These effects on a topological insulator surface coupled with magnetic layers are relevant to potential applications of topological insulators in designing new spintronics devices.

\subsection{Experimental results}
\label{sec:3dphys}

\subsubsection{Material growth}

There have been many interesting theoretical proposals for novel
effects in topological insulators, but perhaps the most exciting aspect of the field is the rapid increase in experimental efforts focussed on topological insulators. High-quality materials are being produced in several groups around the world, and of all different types. Bulk materials were first grown for experiments on topological insulators in the Cava group at Princeton University including the Bi$_{1-x}$Sb$_x$ alloy~\cite{hsieh2008} and Bi$_2$Se$_3$, Bi$_2$Te$_3$, Sb$_2$Te$_3$
crystals~\cite{xia2009,hsieh2009b}. Crystalline samples
of Bi$_2$Te$_3$ have also been grown at Stanford University in the Fisher group~\cite{chen2009}. In addition to bulk samples, Bi$_2$Se$_3$ nanoribbons~\cite{peng2010,kong2009,hong2010} have been fabricated in the Cui group at Stanford University, and thin films of Bi$_2$Se$_3$ and Bi$_2$Te$_3$ have been grown by MBE by the Xue group at Tsinghua University~\cite{yzhang2009,yli2009}, as well as other groups~\cite{zhangqpl2009,hdli2010}. Thin films can also be obtained by exfoliation from bulk samples~\cite{hong2010,shahil2010,teweldebrhan2010b}. The stoichiometric compounds Bi$_2$Se$_3$, Bi$_2$Te$_3$, Sb$_2$Te$_3$ are not extremely difficult to grow, which should allow more experimental groups to have access to high-quality topological insulator samples~\cite{butch2010,zhangqpl2009}. due to intrinsic doping from vacancy and anti-site defects,
Bi$_2$Se$_3$ and Bi$_2$Te$_3$~\cite{xia2009,chen2009,hsieh2009b} are shown to contain $n$-type carriers while Sb$_2$Te$_3$~\cite{hsieh2009b} is $p$-type. Consequently, controllable extrinsic doping is required to tune the Fermi energy to the Dirac point of the surface states. For example, Bi$_2$Te$_3$ can be doped with Sn~\cite{chen2009} and
Bi$_2$Se$_3$ can be doped with Sb~\cite{analytis2010} or
Ca~\cite{hor2010,wang2010c}. Furthermore, it is found that doping Bi$_2$Se$_3$ with Cu can induce superconductivity~\cite{hor2009a}, while Fe and Mn dopants may yield ferromagnetism~\cite{hor2010b,xia2008,wray2010,cha2010,chen2010b}.

\subsubsection{Angle-resolved photoemission spectroscopy}\label{sec:ARPES}
ARPES experiments are uniquely positioned to detect the topological surface states. The
first experiments on topological insulators were ARPES experiments
carried out on the Bi$_{1-x}$Sb$_x$ alloy~\cite{hsieh2008}.
The observation of five branches of surface states, together
with the respective spin polarizations determined later
by spin-resolved ARPES~\cite{hsieh2009}, confirms the nontrivial
topological nature of the surface states of Bi$_{1-x}$Sb$_x$.

\begin{figure}
          \includegraphics[width=0.44\textwidth]{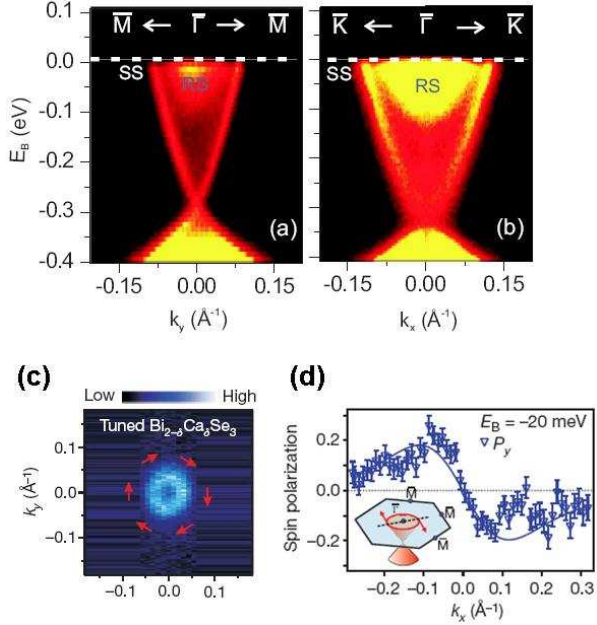}
	  \caption{ ARPES data for the dispersion of the surface states of Bi$_2$Se$_3$,
	  along directions (a) $\bar{\Gamma}-\bar{\rm{M}}$ and (b) $\bar{\Gamma}-\bar{\rm{K}}$
	  in the surface Brillioun zone. Spin-resolved ARPES data
	  is shown along $\bar{\Gamma}-\bar{\rm{M}}$ for a fixed energy in (d),
	  from which the spin polarization in momentum space (c) can be extracted. From \onlinecite{xia2009} and \onlinecite{hsieh2009a}.
	  }\label{fig:BiSeARPES}
  \end{figure}

\begin{figure}
          \includegraphics[width=0.5\textwidth]{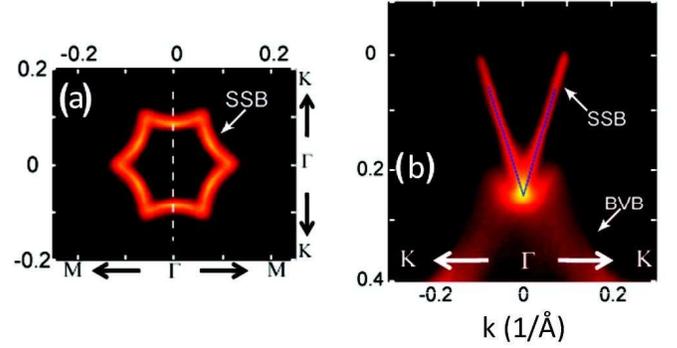}
\caption{ARPES measurement of (a) shape of the Fermi surface
and (b) band dispersion along the $K-\Gamma-K$ direction, for Bi$_2$Te$_3$ nominally doped with 0.67\% Sn. From \onlinecite{chen2009}.}\label{fig:BiTeARPES}
  \end{figure}

ARPES work on Bi$_2$Se$_3$~\cite{xia2009} and Bi$_2$Te$_3$~\cite{hsieh2009a,chen2009} soon followed. Unlike the multiple branches of surface states observed for Bi$_{1-x}$Sb$_x$, these experiments report a remarkably simple surface state spectrum with a single Dirac cone located at the $\Gamma$ point
and a large bulk band gap, in accordance with the theoretical predictions. For Bi$_2$Se$_3$, a single Dirac cone with linear dispersion is
clearly shown at the $\bar{\Gamma}$ point within the band gap
in Fig.~\ref{fig:BiSeARPES}(a) and (b). Figure~\ref{fig:BiSeARPES}(d) shows the $y$ component
of the spin polarization along the $k_x$ ($\bar{\Gamma}-\bar{M}$) direction
measured by spin-resolved ARPES~\cite{hsieh2009a}. The opposite spin polarization in the $y$ direction
for opposite $\mathbf{k}$ indicates the helical nature of the spin polarization
for surface states. As discussed above, Bi$_2$Se$_3$ has a finite density of $n$-type
carriers due to intrinsic doping. Therefore, the above ARPES
data [Fig.~\ref{fig:BiSeARPES}(a),(b)] shows
that the Fermi energy is above the conduction band bottom
and the sample is, in fact, a metal rather than an insulator in the bulk. To obtain a true topological insulating state with the
Fermi energy tuned into the bulk gap, careful control
of external doping is required. Such control was first reported by Chen~\emph{et al.} [Fig.~\ref{fig:BiTeARPES}] for a sample
of Bi$_2$Te$_3$ with 0.67\% Sn doping~\cite{chen2009}. Some recent work on Sb$_2$Te$_3$~\cite{hsieh2009b} supports the theoretical prediction that this material is also a topological insulator~\cite{zhang2009}. This family of materials is moving to the forefront of research on
topological insulators due to the large bulk gap and the
simplicity of the surface state spectrum.

Although the simple model (\ref{Hsurf}) captures most of the surface state physics of these
systems, experiments report a hexagonal surface state Fermi surface [Fig.~\ref{fig:BiTeARPES}], while Eq.~(\ref{Hsurf}) only describes a circular Fermi surface sufficiently close to the Dirac point. However, such a hexagonal warping effect can be easily taken
into account by including an additional term in the surface
Hamiltonian which is cubic in $k$~\cite{fu2009c}. The surface
Hamiltonian for Bi$_2$Te$_3$ can be written
\begin{equation}
H(\mathbf{k})= E_0(\mathbf{k}) + v_\mathbf{k}(k_x \sigma^y - k_y \sigma^x)  +  \frac{ \lambda}{2} (k_+^3 + k_-^3) \sigma^z,
\end{equation}
where $E_0(\mathbf{k}) = \mathbf{k}^2/(2m^*)$ breaks the particle-hole symmetry, the Dirac velocity $v_\mathbf{k}=v(1+\alpha \mathbf{k}^2)$ acquires a quadratic dependence on $\mathbf{k}$, and $\lambda$ parameterizes the amount of hexagonal warping~\cite{fu2009c}.

\begin{figure}
          \includegraphics[width=0.48\textwidth]{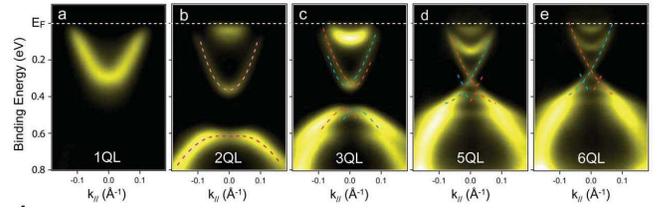}
\caption{ARPES data for Bi$_2$Se$_3$ thin films of thickness (a) 1QL (b) 2QL (c) 3QL (d) 5QL (e) 6QL, measured at room temperature (QL stands for quintuple layer). From \onlinecite{yzhang2009}.}\label{fig:MBEARPES}
  \end{figure}

In addition to its usefulness for studying bulk crystalline samples, ARPES has also been used to characterize the thin films of Bi$_2$Se$_3$ and Bi$_2$Te$_3$~\cite{yzhang2009,yli2009,sakamoto2010}. The thin films were grown to initiate a study of the crossover~\cite{liu2010a} from a 3D topological insulator to a 2D QSH state (Sec.~\ref{sec:2dto3dTheory}). In Fig.~\ref{fig:MBEARPES}, ARPES spectra are shown for several thicknesses of a Bi $_2$Se$_3$ thin film, which show the evolution of the surface states.

\subsubsection{Scanning tunneling microscopy}

\begin{figure}[!t]
\begin{center}
\includegraphics[width=0.8\columnwidth]{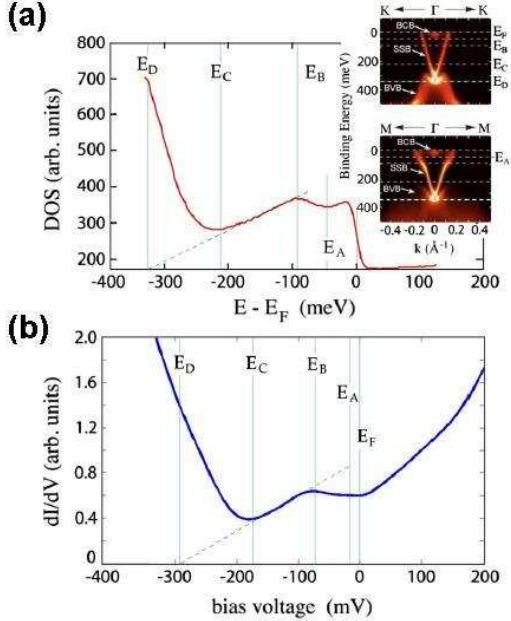}
\end{center}
\caption{Good agreement is found between (a) the integrated density of states from ARPES and (b) a typical scanning tunneling spectroscopy
spectrum. $E_F$ is the Fermi level, $E_A$ the bottom of the bulk conduction band, $E_B$ the point where the surface states become warped, $E_C$ the top of the bulk valence band, and $E_D$ the Dirac point. From \onlinecite{alpichshev2010}.} \label{fig:STMARPES}
\end{figure}

In addition to the ARPES characterization of 3D topological insulators,
scanning tunneling microscopy (STM) and scanning tunneling spectroscopy (STS)
provide another kind of surface-sensitive technique to probe the topological surface states.
A set of materials have been investigated in STM/STS experiments:
Bi$_{1-x}$Sb$_x$~\cite{roushan2009}, Bi$_2$Te$_3$~\cite{alpichshev2010,tzhang2009}, and Sb~\cite{gomes2009}. (Although Sb is topologically nontrivial, it is
a semi-metal instead of an insulator.) The comparison between STM/STS and ARPES was
first performed for Bi$_2$Te$_3$~\cite{alpichshev2010}, where it was found that the integrated density of states obtained from ARPES
[Fig.~\ref{fig:STMARPES}(a)] agrees well with the differential conductance $dI/dV$ obtained from STS measurements [Fig.~\ref{fig:STMARPES}(b)]. From such a comparison, different characteristic energies ($E_F$, $E_A$, $E_B$, $E_C$ and $E_D$ in Fig.~\ref{fig:STMARPES}) can be easily and unambiguously identified.

\begin{figure}[!t]
\begin{center}
\includegraphics[width=0.96\columnwidth]{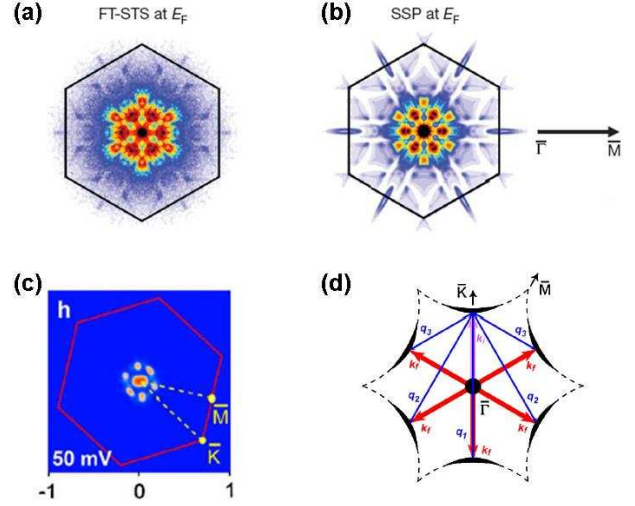}
\end{center}
\caption{\label{fig:STM}
(a) Measured interference pattern in momentum space for impurities on the surface of Bi$_x$Sb$_{1-x}$. (b) Pattern calculated from ARPES
data on Bi$_x$Sb$_{1-x}$, which agrees well with the interference pattern in (a). (c) Similar interference pattern and (d) possible
scattering wave vectors for Bi$_2$Te$_3$. From \onlinecite{roushan2009} and \onlinecite{tzhang2009}.}
\end{figure}

Besides the linear Dirac dispersion which has already been well established by ARPES experiments,
STM/STS can provide further information about the topological nature of the surface states,
such as the interference patterns of impurities or edges~\cite{roushan2009,alpichshev2010b,gomes2009,tzhang2009}. When there are impurities on the surface of a topological insulator,
the surface states will be scattered and form an interference
pattern around the impurities. Fourier transforming the interference pattern into momentum space,
one can quantitatively extract the scattering intensity
for a fixed energy and scattering wave vector. With such information one
can determine what types of scattering events are suppressed. Figure~\ref{fig:STM}(a) and (c) shows the interference pattern
in momentum space for Bi$_x$Sb$_{1-x}$~\cite{roushan2009} and Bi$_2$Te$_3$~\cite{tzhang2009}, respectively. In order to analyze the interference pattern~\cite{lee2009}, we take Bi$_2$Te$_3$ as an example [Fig.~\ref{fig:STM}(c),(d)]. The surface Fermi surface of Bi$_2$Te$_3$
is shown in Fig.~\ref{fig:STM}(d), for which the possible scattering events are dominated by the wave vectors $\mathbf{q}_1$ along the $\bar{K}$ direction,
$\mathbf{q}_2$ along the $\bar{M}$ direction and $\mathbf{q}_3$ between the $\bar{K}$ and $\bar{M}$ directions. However, from Fig.~\ref{fig:STM}(c) we see that there is a peak along the $\bar{\Gamma}-\bar{M}$ direction, while scattering along the $\bar{\Gamma}-\bar{K}$ direction is suppressed. This observation coincides with the theoretical prediction that backscattering between $\mathbf{k}$ and $-\mathbf{k}$ is forbidden due to TR symmetry, which supports the topological nature of the surface states. Other related theoretical analysis are also consistent~\cite{guo2010c,zhou2009,biswas2010}. A similar analysis can be applied to the surface of Bi$_x$Sb$_{1-x}$, and the obtained pattern [Fig.~\ref{fig:STM}(b)] also agrees well with the experimental data [Fig.~\ref{fig:STM}(a)]~\cite{roushan2009}.
More recently, STM experiments have further demonstrated that the topological surface states can penetrate barriers while maintaining their extended nature~\cite{seo2010}.

\begin{figure}[!t]
\begin{center}
\includegraphics[width=0.96\columnwidth]{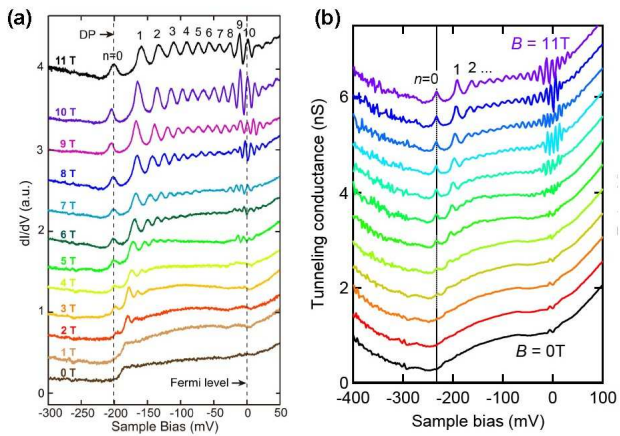}
\end{center}
\caption{ Tunneling spectra for the surface of Bi$_2$Se$_3$ in a magnetic field, showing a series of peaks attributed to the occurrence of surface Landau levels. From \onlinecite{cheng2010} and \onlinecite{hanaguri2010}.\label{fig:SurfaceLL}}
\end{figure}

Another important result of STM/STS measurements is the observation of surface state LLs in a magnetic field~\cite{cheng2010,hanaguri2010}. As shown in Fig.~\ref{fig:SurfaceLL}(a) and (b), discrete LLs appear as a series of peaks in the differential conductance spectrum ($dI/dV$), which supports the 2D nature of the surface states. Further analysis on the dependence of the LLs on the magnetic field $B$ shows that the energy of the LLs is proportional to $\sqrt{nB}$ where $n$ is the Landau level index, instead of the usual linear-in-$B$ dependence. This unusual dependence provides additional evidence for the existence of surface states consisting of massless Dirac fermions. Furthermore, the narrow peaks in the spectrum also indicate the good quality of the sample surface.

\subsubsection{Transport}

In addition to the above surface-sensitive techniques, a large effort has been devoted to transport measurements including dc transport~\cite{analytis2010a,eto2010,butch2010,tang2010,steinberg2010,chen2010a} and measurements in the microwave~\cite{analytis2010} and infrared regimes~\cite{butch2010,sushkov2010,laforge2010}, which are necessary steps towards the direct measurement of topological effects such as the TME, and for future device applications. However, transport experiments on topological insulators turn out to be much more difficult than surface-sensitive measurements such as ARPES and STM. The main difficulty arises from the existence of a finite residual bulk carrier density. Materials such as Bi$_{1-x}$Sb$_x$ or Bi$_2$Se$_3$ are predicted to be topological insulators if they are perfectly crystalline. However, real materials always have impurities and defects such as anti-sites and vacancies. Therefore, as-grown materials are not truly insulating but have a finite bulk carrier density. As discussed in Sec.~\ref{sec:ARPES}, such a residual bulk carrier density is also observed in ARPES for Bi$_2$Se$_3$ and Bi$_2$Te$_3$~\cite{hsieh2009a,chen2009}. From the ARPES results it seems that the residual carrier density can be compensated for by chemical doping~\cite{chen2009}. Nevertheless, in transport experiments the compensation of bulk carriers appears to be much more difficult. Even samples which appear as bulk insulators in ARPES experiments still exhibit some finite bulk carrier density in transport measurements~\cite{analytis2010a}, which suggests the existence of an offset between bulk and surface Fermi levels. Another difficulty in transport measurements is that a cleaved surface rapidly becomes heavily
$n$-doped when exposed to air. This leads to further discrepancies between the surface condition observed in transport and surface-sensitive measurements.

\begin{figure}[!t]
\begin{center}
\includegraphics[width=0.6\columnwidth]{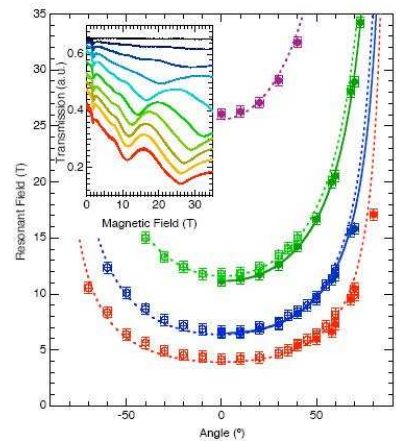}
\end{center}
\caption{Angular dependence of the cyclotron resonance field for a $71$~GHz microwave. The inset shows examples of transmission data
offset for clarity (top to bottom: $90^{\circ}$ to $0^{\circ}$ in steps of $10^{\circ}$). From \onlinecite{ayala2010}.}\label{fig:cyclotron}
\end{figure}

In spite of the complexity described above, the signature of 2D surface states in transport experiments has been recently reported~\cite{analytis2010a,ayala2010}. For example, Fig.~\ref{fig:cyclotron} shows results obtained by microwave spectroscopy~\cite{ayala2010} on Bi$_2$Se$_3$, where it is found that the cyclotron resonance frequency only scales with perpendicular magnetic field $B_\perp$, suggesting the 2D nature of the resonance. Similarly, the dependence of Shubnikov-de Haas oscillations on the angle of the magnetic field can help to distinguish the 2D surface states from the 3D bulk states, both for Bi$_2$Se$_3$~\cite{ayala2010} and Bi$_x$Sb$_{1-x}$~\cite{taskin2009}. Signatures of the topological surface states have also been searched for in the temperature dependence of the resistance~\cite{checkelsky2009,checkelsky2010,analytis2010}, the magnetoresistance~\cite{tang2010}, and weak antilocalization effects~\cite{chen2010a,checkelsky2010}. However, besides the experiments mentioned above with positive evidence for the existence of topological surface states, some experiments show that the transport data can be entirely explained by bulk carriers~\cite{eto2010,butch2010}. The resolution of this controversy requires further improvements in experiments and sample quality.

\begin{figure}[!t]
\begin{center}
\includegraphics[width=0.8\columnwidth]{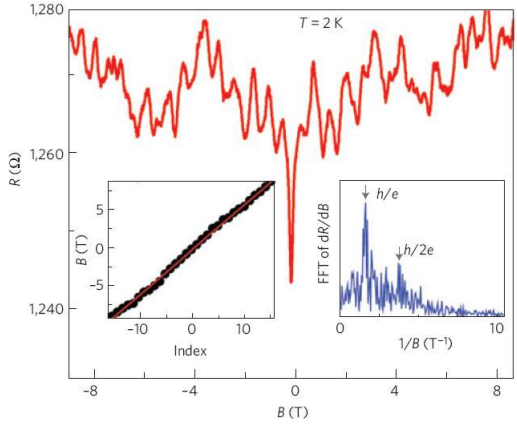}
\end{center}
\caption{ Magnetoresistance for fields up to $\pm9$~T. Left inset: magnetic fields at which well-developed resistance minima are observed. Right inset: fast Fourier transform of the resistance derivative $dR/dB$, where peaks correspond to $hc/e$ and $hc/2e$ oscillations are labeled. From \onlinecite{peng2010}. \label{fig:ABoscillation}
}
\end{figure}

\begin{figure}[!t]
\begin{center}
\includegraphics[width=0.95\columnwidth]{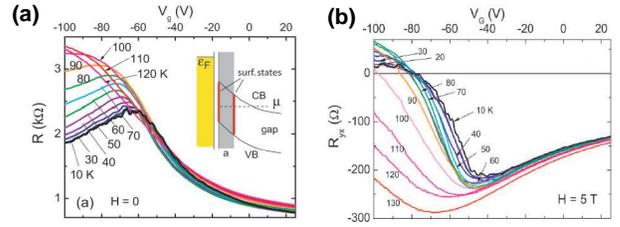}
\end{center}
\caption{(a) Gate voltage dependence of the resistance in zero magnetic field for different temperatures. (b) Gate voltage dependence of the Hall resistance $R_{yx}$ in a magnetic field of $5$~T for different temperatures. From \onlinecite{checkelsky2010}. \label{fig:gatecontrol}
}
\end{figure}

To reach the intrinsic topological insulator state without bulk carriers, various efforts have been made to reduce the bulk carrier density. One approach consists in compensating the bulk carriers by chemical doping, e.g. doping Bi$_2$Se$_3$ with Sb~\cite{analytis2010}, Ca~\cite{hor2009}, or doping Bi$_2$Te$_3$ with Sn~\cite{chen2009}. Although chemical doping is an efficient way to reduce the bulk carrier density, the mobility will be usually reduced due to foreign dopants. However, we note that the substitution of the isovalent Bi with Sb can reduce
the carrier density but still keep high mobilities~\cite{analytis2010}. Also, it is difficult to achieve accurate tuning of the carrier density by chemical doping, because each different chemical doping level needs to be reached by growing a new sample. The second method consists in suppressing the contribution of bulk carriers to transport by reducing the sample size down to the nanoscale, such as quasi-1D nanoribbons~\cite{peng2010,kong2009,steinberg2010}, or quasi-2D thin film~\cite{yli2009,checkelsky2010,chen2010a}. In Fig.~\ref{fig:ABoscillation}, the magnetoresistance of a nanoribbon exhibits a primary $hc/e$ oscillation, which corresponds to Aharonov-Bohm oscillations of the surface state around the surface of the nanoribbon~\cite{peng2010}. This oscillation also indicates that the bulk carrier density has been reduced greatly so that the contribution of the surface states can be observed. The Aharonov-Bohm oscillation has also been investigated theoretically~\cite{zhang2010b,bardarson2010}. An important advantage of a sample of mesoscopic size is the possibility of tuning the carrier density by an external gate voltage. Gate control of the carrier density is rather important because it can tune the bulk carrier density continuously while preserving the quality of the sample. Gate control of the carrier density has been indeed observed in nanoribbons of Bi$_2$Se$_3$~\cite{steinberg2010}, mechanically exfoliated thin films~\cite{checkelsky2010}, or epitaxially grown thin films~\cite{chen2010a}. In particular, tuning of the carrier polarity from $n$-type to $p$-type has been reported~\cite{checkelsky2010}, where the change in polarity corresponds to a sign change of the Hall resistance $R_{yx}$ in a magnetic field [Fig.~\ref{fig:gatecontrol}(b)].

\subsection{Other topological insulator materials}
\label{materials}

The topological materials HgTe, Bi$_2$Se$_3$, Bi$_2$Te$_3$ and Sb$_2$Te$_3$ not only provide us with a prototype material for 2D and 3D topological insulators, but also give us a rule of thumb to search for new topological insulator materials. The nontrivial topological property of topological insulators originates from the inverted band structure induced by SOC. Therefore, it is more likely to find topological insulators in materials which consist of covalent compounds with narrow band gaps and heavy atoms with strong SOC. Following such a guiding principle, a large number of topological insulator materials have been proposed recently, which can be roughly classified into several different groups.

The first group is similar to the tetradymite semiconductors, where the atomic $p$-orbitals of Bi or Sb play an essential role. Thallium-based III-V-VI$_2$
ternary chalcogenides, including TlBiQ$_2$ and TlSbQ$_2$ with Q = Te, Se and S, belong to this class~\cite{yan2010a,lin2010b}. These materials have the same rhombohedral crystal structure (space group $D^5_{3d}$) as the tetradymite semiconductors, but are genuinely 3D, in contrast to the layered tetradymite compounds. These materials have recently been experimentally observed to be topological insulators~\cite{chen2010,sato2010}.

A typical material of the second group is distorted bulk HgTe. In contrast to conventional zincblende semiconductors, HgTe has an inverted bulk band structure with the $\Gamma_8$ band being higher in energy than the $\Gamma_6$ band. However, HgTe by itself is a semi-metal with the Fermi energy at the touching point between the light-hole and heavy-hole $\Gamma_8$ bands. Consequently, in order to get a topological insulator, the crystal structure of HgTe should be distorted along the $[111]$ direction to open a gap between the heavy-hole and light-hole bands~\cite{dai2008}. A similar band structure also exists in ternary Heusler compounds~\cite{chadov2010,lin2010a}, and around fifty of them are found to exhibit band inversion. These materials become 3D topological insulators upon distortion, or they can be grown in quantum well form similar to HgTe/CdTe to realize the 2D or the QSH insulators. Due to the diversity of Heusler materials, multifunctional topological insulators can be realized with additional properties ranging from superconductivity to magnetism and heavy-fermion behavior.

Besides the above two large groups of materials, there are also some other theoretical proposals of new topological insulator materials with electron correlation effects. An example is the case of Ir-based materials. The QSH effect has been proposed in Na$_2$IrO$_3$~\cite{shitade2009}, and topological Mott insulator phases have been proposed in Ir-based pyrochlore oxides Ln$_2$Ir$_2$O$_7$ with Ln = Nd, Pr~\cite{pesin2010,wan2010,guo2009,yang2010}. Furthermore, a topological structure has also been considered in Kondo insulators, with a possible realization in SmB$_6$ and CeNiSn~\cite{dzero2010}.

\section{General Theory of Topological Insulators}
\label{sec:general}
The TFT~\cite{qi2008b} and the TBT~\cite{kane2005b,fu2007b,moore2007,roy2009,fu2007a} are two different general theories of the topological insulators. The TBT is valid for the non-interacting system without disorder. The TBT has given simple and important criteria to evaluate which band insulators are topologically non-trivial. The TFT is generally valid for interacting systems including disorder, and the it identifies the physical response associated with the topological order. Remarkably, the TFT reduces exactly to the TBT in the non-interacting limit. In this section, we review both general theories, and also discuss their connections.

\subsection{Topological field theory}

We are generally interested in the long-wavelength and low-energy
properties of a condensed matter system. In this case, the details
of the microscopic Hamiltonian are not important, and we would like
to capture essential physical properties in terms of a low-energy
effective field theory. For conventional broken-symmetry states, the
low-energy effective field theory is fully determined by the order
parameter, symmetry and dimensionality~\cite{anderson1997}.
Topological states of quantum matter are similarly described by a
low-energy effective field theory. In this case, the effective field
theory generally involve topological terms which describe the
universal topological properties of the state. The coefficient of
the topological term can be generally identified as the topological
order parameter of the system. A successful example is the TFT of
the QH effect~\cite{zhang1992}, which captures the
universal topological properties such as the quantization of the
Hall conductance, the fractional charge and statistics of the
quasiparticles and the ground state degeneracy on a topologically
nontrivial spatial manifold. In this section, we shall describe the TFT of the TR invariant topological insulators.

\subsubsection{Chern-Simons insulator in $2+1$ dimensions}

\begin{figure}
\includegraphics[width=6.5cm, height=3.5cm]{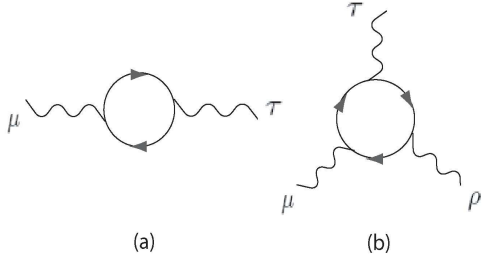}
\caption{Fermion loop diagrams leading to the Chern-Simons term. (a) The
$(2+1)$D Chern-Simons term is calculated from a loop diagram with two external
photon lines. (b) The $(4+1)$D Chern-Simons term is calculated from a loop
diagram with three external photon lines. \label{feynman}}
\end{figure}

We start from the previously mentioned QH
system in $(2+1)$D, the TFT for which is given as~\cite{zhang1992}
\begin{eqnarray}
S_{\rm eff}&=&\frac{C_1}{4\pi}\int d^2x\int dt
\,A_\mu\epsilon^{\mu\nu\tau}\partial_\nu A_\tau,\label{2dCS}
\end{eqnarray}
where the coefficient $C_1$ is generally given by~\cite{wang2010b}
\bea C_1 = \frac{\pi}{3} \int \frac{d^3k}{(2\pi)^3} {\rm
Tr}[\epsilon^{\mu \nu \rho} G\partial_\mu G^{-1}G\partial_\nu G^{-1}
G\partial_\rho G^{-1}], \label{C1} \eea and $G(k)\equiv G({\bf k},
\omega)$ is the imaginary-time single-particle Green's function of a
fully interacting insulator, and $\mu,\nu,\rho = 0,1,2\equiv
t,x,y$. For a general interacting system, assuming that $G$ is
nonsingular, we have a map from the three dimensional momentum space to
the space of nonsingular Green's functions, belonging to the group
${\rm GL}(n,\mathbb{C})$, whose third homotopy group is labeled by an
integer~\cite{wang2010b}: \bea \pi_3({\rm GL}(n,\mathbb{C}))\cong\mathbb{Z}.
\label{homotopy} \eea The winding number for this homotopy class is
exactly measured by $C_1$ defined in Eq.~(\ref{C1}). Here
$n\geq 3$ is the number of bands.
In the noninteracting limit,
$C_1$ in Eq.(\ref{2dCS}) can be calculated explicitly from a single
Feynman diagram [Fig.~\ref{feynman}(a)]~\cite{niemi1983,golterman1993,volovik2002,qi2008b}, and one obtains $C_1$ as given in Eq. (\ref{C1}), but with $G$
replaced by the noninteracting Green's function $G_0$. Carrying out
the frequency integral explicitly, one obtains the TKNN invariant
expressed as an integral of the Berry curvature~\cite{thouless1982}, \bea
C_1=\frac1{2\pi}\int dk_x\int dk_y f_{xy}({\bf k})\in \mathbb{Z},
\label{TKNN} \eea with \bea f_{xy}({\bf k})&=&\frac{\partial
a_y({\bf k})}{\partial
k_x}-\frac{\partial a_x({\bf k})}{\partial k_y},\nonumber\\
a_i({\bf k})&=&-i\sum_{\alpha\in ~{\rm occ}}\left\langle \alpha{\bf
k}\right|\frac{\partial}{\partial k_i}\left|\alpha{\bf
k}\right\rangle,~i=x,y.\nonumber \eea Under TR, we have
$A_0 \rightarrow A_0,\, {\bf A} \rightarrow - {\bf A} $, from which
we see that the $(2+1)$D CS field theory in Eq. (\ref{2dCS}) breaks TR symmetry. All the low-energy response of the QH system can be derived from this TFT. For instance, from the effective Lagrangian in
Eq.~(\ref{2dCS}), taking a functional derivative with respective to $A_\mu$, we obtain the current \bea j_\mu = \frac{C_1}{2\pi}
\epsilon^{\mu\nu\tau}\partial_\nu A_\tau. \eea The spatial component
of this current is given by
\bea j_i= \frac{C_1}{2\pi}\epsilon^{ij}E_j, \eea
while the temporal component is given by
\bea j_0 =
\frac{C_1}{2\pi}\epsilon^{ij}\partial_i A_j = \frac{C_1}{2\pi} B.
\eea This is exactly the QH response with Hall conductance
$\sigma_H = C_1/(2\pi)$, implying that an electric field induces a
transverse current, and a magnetic field induces charge accumulation.
The Maxwell term contains more derivatives than the CS term and is therefore less relevant at low energies in the renormalization
group sense. Therefore, all the topological response of the QH state is exactly contained in the low-energy TFT of Eq.~(\ref{2dCS}).

\subsubsection{Chern-Simons insulator in $4+1$ dimensions}

The TFT of the QH effect does not only capture the universal low-energy
physics, it also points out a way to generalize the TR symmetry breaking
QH state to TR invariant topological states. The CS field
theory can be generalized to all odd dimensional spacetimes~\cite{nakahara1990}. This
observation lead Zhang and Hu to discover a generalization of the QH
insulator state~\cite{zhang2001} which is TR invariant, and
defined in $(4+1)$D. It is the fundamental TR invariant
insulator state from which all the lower-dimensional cases are derived, and is described by the TFT~\cite{bernevig2002} \bea S_{\rm
eff}=\frac{C_2}{24\pi^2}\int
d^4xdt\,\epsilon^{\mu\nu\rho\sigma\tau}A_\mu\partial_\nu
A_\rho\partial_\sigma A_\tau.  \label{4dCS} \eea Under TR, we have $A_0 \rightarrow A_0$, ${\bf A} \rightarrow
-{\bf A},$  and this term is explicitly TR invariant. Generally, the coefficient $C_2$ is expressed in terms of the Green's function of an interacting system as~\cite{wang2010b}
\bea C_2 &\equiv& \frac{\pi^2}{15} \int \frac{d^{5}k}{(2\pi)^{5}}
\textrm{Tr}[\epsilon^{\mu \nu \rho \sigma \tau}
G\partial_{\mu}G^{-1} G\partial_{\nu}G^{-1} G\partial_{\rho}G^{-1}
\nonumber \\ &\,& \times G\partial_{\sigma}G^{-1}
G\partial_{\tau}G^{-1}],   \label{C2} \eea which labels the
homotopy group~\cite{wang2010b} \bea \pi_5({\rm GL}(n,\mathbb{C}))\cong
\mathbb{Z}, \eea similarly to the case of the $(2+1)$D CS term. For a noninteracting system, $C_2$ can be calculated from a single
Feynman diagram [Fig.~\ref{feynman}(b)] and one obtains $C_2$ as
given in Eq.~(\ref{C2}), with $G$ replaced by the noninteracting
Green's function $G_0$. Explicit integration over the frequency gives
the second Chern number~\cite{qi2008b},
\begin{eqnarray}
C_2&=&\frac1{32\pi^2}\int d^4k\epsilon^{ijk\ell}{\rm
tr}[f_{ij}f_{k\ell}],\label{2ndtknn}
\end{eqnarray}
with \begin{eqnarray} f^{\alpha\beta}_{ij}&=&\partial_i
a^{\alpha\beta}_j-\partial_j
a^{\alpha\beta}_i+i\left[a_i,a_j\right]^{\alpha\beta},\nonumber\\
a_i^{\alpha\beta}({\bf k})&=&-i\left\langle \alpha,{\bf
k}\right|\frac{\partial}{\partial k_i }\left|\beta,{\bf
k}\right\rangle,\nonumber
\end{eqnarray}
where $i,j,k,\ell=1,2,3,4\equiv x,y,z,w)$.

Unlike the $(2+1)$D case, the CS term is less relevant than the
non-topological Maxwell term in $(4+1)$D, but is still of primary
importance when understanding topological phenomena such as the chiral
anomaly in a $(3+1)$D system, which can be regarded as the boundary of a
$(4+1)$D system~\cite{qi2008b}. Similar to the $(2+1)$D QH case, the physical
response of $(4+1)$D CS insulators is given by \bea j^\mu =
\frac{C_2}{2\pi^2}\epsilon^{\mu\nu\rho\sigma\tau}\partial_\nu A_\rho
\partial_\sigma A_\tau, \label{4dresponse} \eea
which is the nonlinear response to the external field $A_\mu$. To understand this response better, we consider a special field
configuration~\cite{qi2008b}:
\begin{eqnarray}
A_x=0,~A_y=B_{z} x,~A_z=-E_zt,~A_w=A_t=0,\label{EBchoice}
\end{eqnarray}
where $x,y,z,w$ are spatial coordinates and $t$ is time. The
only nonvanishing components of the field strength are
$F_{xy}=B_{z}$ and $F_{zt}=-E_z$. According to Eq.~(\ref{4dresponse}), this field configuration induces the current
\begin{eqnarray}
j_w=\frac{C_2}{4\pi^2}B_{z}E_z.\nonumber
\end{eqnarray}
If we integrate the equation above over the $x,y$ dimensions, with
periodic boundary conditions and assuming that $E_z$ does not depend
on $x,y$, we obtain
\begin{eqnarray}
\int dxdy\,j_w=\frac{C_2}{4\pi^2}\left(\int dxdy B_{z}\right)E_z\equiv
\frac{C_2N_{xy}}{2\pi}E_z,
\end{eqnarray}
where $N_{xy}=\int dxdyB_{z}/2\pi$ is the number of flux quanta
through the $xy$ plane, which is always quantized to be an integer.
This is exactly the 4D generalization of the QH effect mentioned earlier~\cite{zhang2001}. Therefore, from this example we can
understand the physical response associated with a nonvanishing
second Chern number. In a $(4+1)$D insulator with second Chern
number $C_2$, a quantized Hall conductance $C_2N_{xy}/2\pi$ in the
$zw$ plane is induced by a magnetic field with flux $2\pi N_{xy}$ in
the perpendicular ($xy$) plane.

We have discussed the CS insulators in $(2+1)$D and $(4+1)$D. Actually,
these discussions can be straightforwardly generalized to higher
dimensions. In doing so, it is worth noting that there is a even-odd
alternation of the homotopy groups of ${\rm GL}(n,\mathbb{C})$:
we have $\pi_{2k+1}({\rm GL}(n,\mathbb{C}))\cong\mathbb{Z}$, while $\pi_{2k}({\rm
GL}(n,\mathbb{C}))=0$. This is the mathematical
mechanism underlying the fact that CS insulators of integer
class exist in even spatial dimensions, but do not exist in odd
spatial dimensions. Reduced to noninteracting insulators, this
becomes the alternation of Chern numbers. As a number characteristic of complex fiber bundles, Chern numbers exist only in even spatial
dimensions. This is an example of the relationship between homotopy
theory and homology theory. We shall see another example of this relationship in the following section: both the Wess-Zumino-Witten (WZW) terms and the CS terms are well-defined only modulo an integer.

\subsubsection{Dimensional reduction to the three-dimensional $\mathbb{Z}_2$ topological insulator}

\begin{figure}
\includegraphics[width=6.5cm, height=3.0cm]{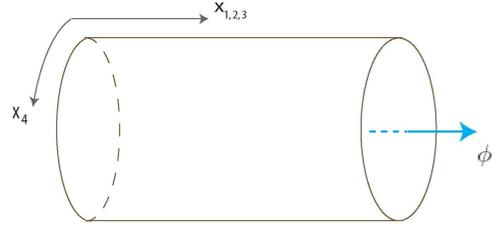}
\caption{Dimensional reduction from $(4+1)$D to $(3+1)$D. The $x_4$
direction is compactified into a small circle, with a finite flux
$\phi$ threading through the circle due to the gauge field $A_4$. To
preserve TR symmetry, the total flux can be either $0$ or
$\pi$, resulting in a $\mathbb{Z}_2$ classification of 3D topological insulators. }
\label{dimreduction}
\end{figure}

The 4D generalization of the QH effect gives the fundamental TR invariant topological insulator from which all lower-dimensional
topological insulators can be derived systematically by a
procedure called dimensional reduction~\cite{qi2008b}. Starting from
the $(4+1)$D CS field theory in Eq.~(\ref{4dCS}), we consider field
configurations where $A_\mu (x)=A_\mu (x_0,x_1,x_2,x_3)$ is
independent of the ``extra dimension" $x_4\equiv w$, for
$\mu=(x_0,x_1,x_2,x_3)$, and $A_4\equiv A_w$ depending on all
coordinates $(x_0,x_1,x_2,x_3,x_4)$. We consider the geometry where
the ``extra dimension" $x_4$ forms a small circle. In
this case, the $x_4$ integral in Eq.~(\ref{4dCS}) can be carried out
explicitly. After restoring the unit of electron charge $e$ and flux $hc/e$ following the
convention in electrodynamics, we obtain an effective TFT in $(3+1)$D: \bea
S_\theta=\frac{\alpha}{32\pi^2} \int d^3xdt\,\theta (x,t)
\epsilon^{\mu\nu\rho\tau}F_{\mu\nu}F_{\rho\tau}(x,t),\label{axion}
\eea where $\alpha=e^2/\hbar c\simeq1/137$ is the fine structure constant and \bea \theta(x,t) \equiv C_2 \phi(x,t) = C_2\oint
dx_4\,A_4 (x,t,x_4), \label{theta} \eea which can be interpreted as
the flux due to the gauge field $A_4 (x,t,x_4)$ through the compact extra dimension [Fig.~\ref{dimreduction}].  The field $\theta(x,t)$ is called the axion field in the
field theory literature~\cite{wilczek1987}. In order to preserve the spatial and temporal
translation symmetry, $\theta(x,t)$ can be chosen as a constant
parameter rather than a field. Furthermore, we already explained
that the original $(4+1)$D CS TFT is TR invariant. Therefore, it
is natural to ask how can TR symmetry be preserved in the
dimensional reduction. If we choose $C_2=1$, then $\theta=\phi$ is
the magnetic flux threading the compactified circle, and the physics
should be invariant under a shift of $\theta$ by $2\pi$. TR transforms $\theta$ to $-\theta$. Therefore, there are two and only
two values of $\theta$ which are consistent with TR
symmetry, namely $\theta=0$ and $\theta=\pi$. In the latter case,
TR transforms $\theta=\pi$ to $\theta=-\pi$, which is
equivalent to $\theta=\pi$ mod $2\pi$. We therefore conclude that there are two different classes of TR invariant topological insulators in 3D, the topologically trivial class with $\theta=0$, and the topologically nontrivial class with $\theta=\pi$.

As just seen, it is most natural to view the 3D topological insulator as a dimensionally reduced version of the 4D topological insulator. However, for most physical systems in 3D, we are generally given an interacting Hamiltonian, and would like to define a topological order parameter that can be evaluated directly for any 3D model Hamiltonian. Since the $\theta$ angle can only take the two values $0$ and $\pi$ in the presence of TR symmetry, it can be naturally defined as the topological order
parameter itself. For a generally interacting system, it is given
by~\cite{wang2010b}: \bea P_{3}\equiv \frac{\theta}{2\pi} &=&
\frac{\pi}{6} \int_{0}^{1}du \int \frac{d^{4}k}{(2\pi)^{4}}
\textrm{Tr}\,\epsilon^{\mu \nu \rho \sigma} [G\partial_{\mu}G^{-1}
G\partial_{\nu}G^{-1} \nonumber
\\  &\,& \times G\partial_{\rho}G^{-1} G\partial_{\sigma}G^{-1}
G\partial_{u}G^{-1}], \label{P3} \eea where the momentum ${\bf
k}=(k_1,k_2,k_3)$ is integrated over the 3D Brillouin zone and the
frequency $k_0$ is integrated over $(-\infty,+\infty)$. $G(k,u=0)\equiv G(k_0,{\bf k},u=0)\equiv G(k_0,{\bf k})$ is the
imaginary-time single-particle Green's function of the fully
interacting many-body system, and $G(k,u)$ for $u\neq 0$ is a smooth
extension of $G(k,u=0)$, with a fixed reference value $G(k, u=1)$
corresponding to the Green's function of a topologically trivial
insulating state. $G(k, u=1)$ can be chosen as a diagonal matrix
with $G_{\alpha \alpha} = (i k_{0} - \Delta)^{-1} $ for empty bands
$\alpha$ and $G_{\beta \beta} = (i k_{0} + \Delta)^{-1} $ for filled
bands $\beta$, where $\Delta > 0$ is independent of ${\bf k}$. Even
though $P_{3}$ is a physical quantity in 3D, a WZW~\cite{witten1983} type of extension parameter $u$ is introduced in its definition, which plays the role of $k_4$ in the formula (\ref{C2}) defining the 4D TI. The definition in momentum-space is analogous to the real-space WZW term. Similar to the WZW term, $P_3$ is only well-defined modulo an integer, and can only take the quantized values of $0$ or $1/2$ modulo an integer for an TR invariant insulator~\cite{wang2010b}.

Essential for the definition of $P_3$ in Eq.~(\ref{P3}) is the TR invariance identity~\cite{wang2010b}: \bea G(k_{0},-{\bf k})=T
G(k_{0},{\bf k})^{T} T^{-1}, \label{GT} \eea which is crucial for
the quantization of $P_3$. Therefore, we see that unlike the integer-class CS insulators, the $\mathbb{Z}_2$ insulators are \emph{symmetry-protected} topological insulators. This can be clearly seen from the viewpoint of the topological order parameter. In fact, the quantization of $P_3$ is protected by
TR symmetry. In other words, if TR symmetry in
Eq.~(\ref{GT}) is broken, then $P_3$ can be tuned continuously, and
can be adiabatically connected from $1/2$ to $0$. This is fundamentally different from the CS insulators, for which the coefficient of the CS term given by Eq.~(\ref{C2}) is always quantized to be an integer, regardless of the presence or absence of symmetries. This integer, if
nonzero, cannot be smoothly connected to zero provided that the
energy gap remains open. There exists a more
exhaustive classification scheme for topological insulators in various
dimensions~\cite{qi2008b,kitaev2009,ryu2010} which takes into account the constraints imposed by various symmetries.

For a noninteracting system, the full Green's function $G$ in the
expression for $P_3$ [Eq.~(\ref{P3})] is replaced by the noninteracting Green's function $G_0$. Furthermore, the frequency integral can be carried out explicitly. After some manipulations, one finds a simple and
beautiful formula \bea P_{3}=\frac{1}{16\pi^{2}} \int d^{3}k\,\epsilon^{ijk} \textrm{Tr}\{[f_{ij}({\bf k})-\frac{2}{3}i
a_{i}({\bf k})a_{j}({\bf k})]a_{k}({\bf k})\}, \label{P3CS} \eea
which expresses $P_3$ as the integral of the CS form over the 3D momentum space. For explicit models of topological insulators, such as the model by Zhang {\it
et al.}~\cite{zhang2009} discussed in Sec.~\ref{sec:3dzhang}, one can evaluate this formula explicitly to obtain
\begin{eqnarray}
P_3=\theta/2\pi=1/2
\end{eqnarray}
in the topologically nontrivial state~\cite{qi2008b}. Essin, Moore and Vanderbilt also calculated $P_3$ for a variety of interesting
models~\cite{essin2009}. In a generic system without time-reversal or inversion symmetry, there may be non-topological contributions to the effective action (\ref{axion}) which modifies the formula of $P_3$ given in Eq. (\ref{P3CS}). However, such corrections vanishes when time-reversal symmetry or inversion symmetry is present, so that the quantized value $P_3=1/2$ (mod $1$) in topological insulator remains robust~\cite{essin2010,malashevich2010}.

Similar to the case of $(4+1)$D CS insulators, there is also an important difference between the $\theta$ term for $(3+1)$D topological insulators and the $(2+1)$D CS term for QH systems, which we shall briefly discuss~\cite{maciejko2010}. In $(2+1)$D,
the topological CS term dominates over the non-topological Maxwell
term at low energies in the renormalization group flow, as a
simple result of dimensional analysis. However, in $(3+1)$D, the $\theta$
term has the same scaling dimension as the Maxwell term, and is therefore equally important at low energies. The full set of modified Maxwell's equations including the topological term (\ref{theta}) is given by
\begin{eqnarray}
\frac1{4\pi}\partial_\nu F^{\mu\nu}+\partial_\nu
\mathcal{P}^{\mu\nu}+\frac{\alpha}{4\pi}\epsilon^{\mu\nu\sigma\tau}\partial_\nu
(P_3F_{\sigma\tau})=\frac1c j^\mu,
\end{eqnarray}
which can be written in component form as
\begin{eqnarray}
& & \nabla\cdot {\bf D}=4\pi\rho+2\alpha (\nabla P_3\cdot{\bf
B}),\nonumber\\
& & \nabla\times{\bf H}-\frac1c\frac{\partial{\bf D}}{\partial t}
=\frac{4\pi}c{\bf j}-2\alpha\left( (\nabla P_3\times {\bf
E})+\frac1c\left(\partial_tP_3\right){\bf B}\right),\nonumber\\
& & \nabla\times {\bf E} + \frac1c\frac{\partial {\bf B}}{\partial
t}= 0,\nonumber\\
& & \nabla\cdot{\bf B}=0,\label{topoMaxwell}
\end{eqnarray}
where ${\bf D}={\bf E}+4\pi{\bf P}$ and ${\bf H}={\bf B}-4\pi {\bf
M}$ only include the non-topological contributions. Alternatively, one can use the ordinary Maxwell's equations with modified constituent equations (\ref{constituenteq}). These set of modified Maxwell equations are called the axion electrodynamics in field theory~\cite{wilczek1987}.

Even though the conventional Maxwell term and the topological term
are both present, there exist experimental designs which can in principle extract the purely topological contributions~\cite{maciejko2010}.
Furthermore, the topological response is completely captured by the
TFT, which we shall discuss. Starting from the TFT [Eq.~\ref{axion}],
we take a functional derivative with respect to $A_\mu$, and obtain the current as
\begin{eqnarray}
j^\mu=\frac{1}{2\pi}\epsilon^{\mu\nu\sigma\tau}\partial_\nu
P_3\partial_\sigma A_\tau,\label{motioneqn3d}
\end{eqnarray}
which is the general topological response of $(3+1)$D insulators. It
is worth noting that we do not assume TR invariance here,
otherwise $P_3$ should be quantized to be integer or half-integer.
In fact, here we assume a inhomogeneous $P_3(x,t)$. It is
interesting to notice that this electromagnetic response looks very similar to the 4D response in Eq.~\ref{4dresponse}, with the only difference that $A_4$ is replaced by $P_3$ in Eq.~(\ref{motioneqn3d}). This is a manifestation of dimensional reduction at the level of the electromagnetic response. A more systematic treatment on this topic in phase space can be also be performed~\cite{qi2008b}. The physical consequences Eq.~(\ref{motioneqn3d}) can be understood by studying the following two cases.

\noindent\emph{1. Half-QH effect on the surface of a 3D topological insulator.} Consider a system in which $P_3=P_3(z)$ only depends on $z$. For example, this can be realized by the lattice Dirac model~\cite{qi2008b} with $\theta=\theta(z)$~\cite{fradkin1986,wilczek1987}. In this
case, Eq.~(\ref{motioneqn3d}) becomes
\begin{eqnarray}
j^\mu=\frac{\partial_zP_3}{2\pi}\epsilon^{\mu\nu\rho}\partial_\nu
A_\rho,~\mu,\nu,\rho=t,x,y,~\nonumber
\end{eqnarray}
which describes a QH effect in the $xy$ plane with Hall
conductivity $\sigma_{xy}=\partial_zP_3/2\pi$. A uniform electric
field $E_x$ along the $x$ direction induces a current density along the
$y$ direction $j_y=(\partial_zP_3/2\pi)E_x$, the integration of
which along the $z$ direction gives the Hall current \bea J_y^{\rm
2D}=\int_{z_1}^{z_2} dz j_y=\frac{1}{2\pi}\left(\int_{z_1}^{z_2}
dP_3\right)E_x, \nonumber \eea which corresponds to a 2D QH
conductance
\begin{eqnarray}
\sigma^{\rm 2D}_{xy}=\int_{z_1}^{z_2}dP_3/2\pi.\label{sigmaHP3}
\end{eqnarray}
For a interface between a topologically nontrivial insulator with
$P_3=1/2$ and a topologically trivial insulator with $P_3=0$, which
can be taken as the vacuum, the Hall conductance is $\sigma_H =
\Delta P_3 = \pm 1/2$. Aside from an integer ambiguity, the QH
conductance is exactly quantized, independent of the details of the interface. As discussed in section \ref{half_QH}, the half quantum Hall effect on the surface is a reflection of the bulk topology with $P_3=1/2$, and can not be determined purely from the low energy surface models.

\noindent\emph{2. Topological magnetoelectric effect induced by a
temporal gradient of $P_3$.} Having considered a time-independent $P_3$, we now consider the case
when $P_3=P_3(t)$ is spatially uniform, but time-dependent. Equation~(\ref{motioneqn3d}) now becomes
\begin{eqnarray}
j^i=-\frac{\partial_tP_3}{2\pi}\epsilon^{ijk}\partial_jA_k,~i,j,k=x,y,z,
\nonumber
\end{eqnarray}
which can be simply written as
\begin{eqnarray}
\mathbf{j}=-\frac{\partial_tP_3}{2\pi}\mathbf{B}.\label{MEeffect}
\end{eqnarray}
Because the charge polarization $\mathbf{P}$ satisfies $\mathbf{j}=\partial_t\mathbf{P}$, we can integrate Eq.~\ref{MEeffect} in a static, uniform magnetic field $B$ to get
$\partial_t\mathbf{P}=-\partial_t\left(P_3\mathbf{B}/2\pi\right)$, so that
\begin{eqnarray}
\mathbf{P}=-\frac{\mathbf{B}}{2\pi}\left(P_3+{\rm
const.}\right).\label{MEeffect2}
\end{eqnarray}
This equation describes the charge polarization induced by a
magnetic field, which is a magnetoelectric effect. The prominent
feature here is that it is exactly quantized to a half-integer for a
TR invariant topological insulator, which is called the topological magnetoelectric effect~\cite{qi2008b} (TME).

Another related effect originating from the TFT is the Witten
effect~\cite{witten1979, qi2008b}. For this discussion, we assume
that there are magnetic monopoles. For a uniform $P_3$, Eq.~(\ref{MEeffect}) leads to \bea \nabla\cdot
\mathbf{j}=-\frac{\partial_tP_3}{2\pi}\nabla\cdot\mathbf{B}.\eea Even if
magnetic monopole does not exist as elementary particles, for a
lattice system, the monopole density $\rho_m=\nabla\cdot
\vec{B}/2\pi$ can still be nonvanishing, and we obtain
\begin{eqnarray}
\partial_t\rho_e=\left(\partial_tP_3\right)\rho_m.\label{monopolecharge}
\end{eqnarray}
Therefore, when $P_3$ is adiabatically changed from zero to
$\Theta/2\pi$, the magnetic monopole will acquire an electric charge of
\begin{eqnarray}
Q_e=\frac{\Theta}{2\pi} Q_m,
\end{eqnarray}
where $Q_m$ is the magnetic charge. Such a relation was first derived by Witten in the context of the
topological term in quantum chromodynamics~\cite{witten1979}, and later discussed in the context of topological insulators~\cite{qi2008b,rosenberg2010}. This effect could also appear under a different guise in topological exciton condensation~\cite{seradjeh2009}, where a $e/2$ charge is induced by a vortex in the exciton condensate, which serves as the ``magnetic monopole''.

\subsubsection{Further dimensional reduction to the two-dimensional $\mathbb{Z}_2$ topological insulator}

Now we turn our attention to $(2+1)$D TR invariant $\mathbb{Z}_2$ topological insulators. Similar to the WZW-type topological order parameter $P_3$ in $(3+1)$D, there is also a topological order parameter defined for $(2+1)$D TR invariant insulators. The main difference between $(2+1)$D and $(3+1)$D is that in $(2+1)$D we need two WZW extension parameters $u$ and $v$, in contrast to a single parameter $u$ in $(3+1)$D, which is a manifestation
of the fact that both descend from the fundamental $(4+1)$D topological insulator~\cite{zhang2001}. For a general interacting insulator, the $(2+1)$D topological order parameter is expressed as~\cite{wang2010b}
\bea P_2 &=& \frac{1}{120}\epsilon^{\mu\nu\rho\sigma\tau
}\int_{-1}^{1}du\int_{-1}^{1}dv\int\frac{d^3k}{\left(2\pi\right)^3}{\rm
Tr}[G\partial_{\mu}G^{-1} \nonumber
\\ &\,& \times G\partial_{\nu}G^{-1}G\partial_{\rho}G^{-1}
G\partial_{\sigma}G^{-1}G\partial_{\tau}G^{-1} \nonumber
\\ &=& 0 \,\,{\rm or}\, \, 1/2 \,\,\,({\rm mod}\,\mathbb{Z}), \label{P2}
\eea where $\epsilon^{\mu\nu\rho\sigma\tau } $  is the totally
antisymmetric tensor in five dimensions, taking value $1$ when the variables are an even permutation of $(k_0,k_1,k_2,u,v)$. The cases $P_2=0$ and $P_2=1/2$ modulo an integer correspond to topologically trivial and nontrivial TR invariant insulators in $(2+1)$D, respectively. This topological order parameter is valid for interacting QSH systems in $(2+1)$D, including states in the Mott
regime~\cite{raghu2008}. $P_2$ can be physically measured by the
fractional charge at the edge of the QSH state~\cite{qi2008}.

\subsubsection{General phase diagram of topological Mott insulator and topological Anderson insulator}

So far, we have introduced topological order parameters for TR invariant topological insulators in 4D, 3D and 2D. These
topological order parameters are defined in terms of the full single-particle Green's function. A caution in order is that these topological order
parameters are not applicable to fractional states with ground state
degeneracy~\cite{bernevig2006a,levin2009,maciejko2010b,swingle2010}, for which a TFT approach is still possible, but simple topological
order parameters are harder to find. In 3D, fractional topological insulators are characterized by a topological order parameter $P_3$ that is a rational multiple of $1/2$~\cite{maciejko2010b,swingle2010}. Such states are consistent with TR symmetry if fractionally charged excitations and ground state degeneracies on spatial manifolds of nontrivial topology are present. When TR symmetry is broken on the surface, a fractional TME gives rise to half of a fractional QH effect on the surface~\cite{maciejko2010b,swingle2010}.

\begin{figure}
\includegraphics[width=5.5cm, height=3.5cm]{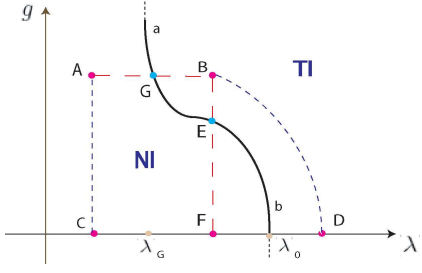}
\caption{Phase diagram in the $(\lambda, g)$ plane. The dark curve
$ab$ is the phase boundary separating normal insulators (NI) and
topological insulators (TI). All phases are gapped, except on $ab$. The true parameter space is in fact infinite dimensional, but this 2D diagram illustrates the main features. From \onlinecite{wang2010b}.} \label{connection}
\end{figure}

Next we shall discuss the physical consequences implied by the
topological order parameter such as $P_2$ and $P_3$. The discussion
we shall present is very general and its applicability does not
depend on the spatial dimensions. Furthermore, since the topological
order parameters are expressed in terms of the full Green's function
of an interacting system, they can be useful to general interacting
systems. Suppose we have a family of Hamiltonians labeled by several
parameters. To be specific, we consider a typical phase
diagram~\cite{wang2010b} [Fig.~(\ref{connection})] for an
interacting Hamiltonian $H=H_0(\lambda)+H_1(g)$, where $H_0$ is the
noninteracting part including terms such as $t_{ij}c_i^\dag c_j$,
and $H_1$ is the electron-electron interaction part including
terms such as the Hubbard interaction $g n_{i\downarrow}
n_{i\uparrow}$. These two parts are
determined by single-particle parameters
$\lambda=(\lambda_1,\lambda_2,\cdots)$ and coupling constants
$g=(g_1,g_2,\cdots)$. When $(\lambda, g)$ are smoothly
tuned, the ground state also evolves smoothly, as long as the energy gap
remains open and the topological order parameters such as $P_2$ and
$P_3$ remain unchanged. Only when the gap closes and the full Green's
function $G$ becomes singular, these topological order parameters
have a jump, as indicated by the curve $ab$ in Fig.~(\ref{connection}).
The most important point in Fig.~\ref{connection} can be illustrated
by considering the vertical line $BF$. Starting from a noninteracting
state $F$, one adiabatically tunes on interactions, and eventually there is a phase transition at $E$ to the topological insulator (TI) state. The interacting state $B$, which is an interaction-induced topological insulator state, has a different topological order parameter from the corresponding noninteracting normal insulator (NI) state $F$. The difference in the topological order parameter thus provides a criterion for distinguishing topological classes of insulators in the presence of general interactions.

There have already been several theoretical proposals of strongly interacting topological insulators, i.e. topological Mott insulators~\cite{raghu2008}. 2D topologically nontrivial insulating states have been obtained from the combination of a trivial noninteracting band structure and interaction terms~\cite{raghu2008,guo2009,weeks2010}. Such topological insulators can be regarded as topologically nontrivial states arising from dynamically generated SOC~\cite{wu2004,wu2007}. The effect of interactions on the the QSH state has also been recently studied~\cite{rachel2010}. In 3D, strong topological insulators with topological excitations have been obtained~\cite{zhang2009b,pesin2010}. Topological insulators have been suggested to exist in transition metal oxides~\cite{shitade2009}, where the correlation effect is strong. It was also proposed that one could achieve the topological insulator state in Kondo insulators~\cite{dzero2010}. All these topological Mott insulator states can be understood in the framework of the topological order
parameter expressed in terms of the full Green's
function~\cite{wang2010b}. Interaction-induced topological insulator states such as the topological Mott insulators proposed in Ref. \cite{raghu2008} correspond to regions represented by the point $B$ in Fig.~\ref{connection}, which has a trivial noninteracting unperturbed Hamiltonian $H_0(B)$ but acquires a nontrivial topological order parameter due to the interaction part $H_1(B)$ of the Hamiltonian. The previously discussed topological order parameters are useful for determining the phase diagrams of interacting systems.

For disordered systems, the topological order parameters are
still applicable, provided that we use the disorder-averaged
Green's functions. In this case, Fig.~\ref{connection} can be
regarded as a simple phase diagram of disordered systems, with
$g$ interpreted as the disorder strength. The representative
point $B$ is a disorder-induced topological insulator state. The disorder-induced TI state has been studied recently~\cite{li2009,groth2009,obuse2008,guo2010,shindou2009,jiang2009,ostrovsky2009,loring2010,imura2009,guo2010b,olshanetsky2010}. Therefore, the topological order parameters previously discussed have the ability to describe both interacting and disordered systems.

\subsection{Topological band theory}\label{sec:z2}

We shall now give a brief introduction to TBT. Even though this theory is only valid for non-interacting systems, it has become an important tool in the discovery of new topological materials. Unfortunately, evaluating the $\mathbb{Z}_2$ invariants for a generic band structure is in general a difficult problem. Several approaches have been explored in the literature including spin Chern
numbers~\cite{sheng2006,fukui2007,prodan2009}, topological invariants constructed from Bloch wave functions~\cite{kane2005b,moore2007,roy2009,fu2006},
and discrete indices calculated from single-particle states at TRIM in
the Brillouin zone~\cite{fu2007a}. We will focus on the last method
for its simplicity~\cite{fu2007a}.

\begin{figure}
\begin{center}
\includegraphics[width=0.96\columnwidth]{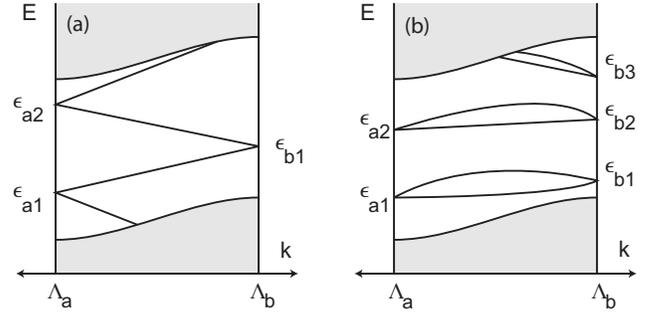}
\end{center}
\caption{Schematic representation of the surface energy levels of a
crystal in either 2D or 3D, as a function of surface
crystal momentum on a path connecting TRIM $\Lambda_a$ and
$\Lambda_b$. The shaded region shows the bulk continuum states, and
the lines show discrete surface (or edge) bands localized near one
of the surfaces.  The Kramers degenerate surface states at
$\Lambda_a$ and $\Lambda_b$ can be connected to each other in two
possible ways, shown in (a) and (b), which reflect the change in
TR polarization $\Delta P_\theta$ of the cylinder between
those points. Case (a) occurs in topological insulators, and
guarantees the surface bands cross any Fermi energy inside the bulk
gap. From \onlinecite{fu2007b}.}\label{fig:kaneInversion2}
\end{figure}

This basic quantity in this approach is the matrix element of the TR operator $T$ between states with TR conjugate momenta $\mathbf{k}$ and $-\mathbf{k}$~\cite{fu2006}, \bea
B_{\alpha \beta}(\mathbf{k}) = \langle -\mathbf{k},\alpha | T |\mathbf{k},\beta
\rangle. \eea Since $B_{\alpha\beta}$ is defined as a matrix element between Bloch states at TR conjugate momenta, it is expected that this quantity contains some information about the
band topology of TR invariant topological insulators. At the TRIM $\Gamma_i$, $B({\bf k}=\Gamma_i)$ is antisymmetric, so that the following quantity can be defined\cite{fu2006}:
\bea \delta_{i} = \frac{
\sqrt{\textrm{det}[B(\Gamma_{i})]} }{ \textrm{Pf}[B(\Gamma_{i})] }.
\eea
in which ${\rm Pf}$ stands for the Pfaffian of an antisymmetric matrix. Since ${\rm Pf}[B(\Gamma_i)]^2=\textrm{det}[B(\Gamma_{i})]$, we have $\delta_i=\pm 1$. It should be noticed that the wavefunctions $\left|{\bf k},\alpha\right\rangle$ must be chosen continuously in BZ to avoid ambiguity in the definition of $\delta_i$.
In 1D, there are only two TRIM, and a
``TR polarization''~\cite{fu2006} can be defined as the
product of $\delta_i$, \bea \pi \equiv (-1)^{P_\theta} = \delta_1
\delta_2, \eea which is a $\mathbb{Z}_2$ analog to the charge
polarization~\cite{thouless1983,zak1989,kingsmith1993,resta1994}.
A further analogy between
the charge polarization and the TR polarization suggests
the form of the $\mathbb{Z}_2$ invariant for TR invariant topological insulators. If an angular parameter $\theta$ is tuned from $0$ to $2\pi$, the change in the charge polarization $P$ after such a cycle is expressed as the first Chern number $C_1$ in the $(k,\theta)$ space. In fact, the
same $C_1$ would gives the TKNN invariant~\cite{thouless1983} if
$\theta$ were regarded as a momentum. By analogy with $P$, the $\mathbb{Z}_2$
invariant for $(2+1)$D topological insulators can be defined as \bea (-1)^{\nu_\mathrm{2D}} =
(-1)^{P_\theta(k_2=0)- P_\theta(k_2=\pi)},  \label{dPtheta} \eea
where $P_\theta(k_2) = \delta_1 \delta_2$, $\delta_i$ is defined at the TRIM $k_1 = 0 \, {\rm or} \, \pi $, and $k_2$ is regarded as a
parameter. Expanding Eq.~(\ref{dPtheta}) gives \bea
(-1)^{\nu_\mathrm{2D}}=\prod_{i=1}^{4} \delta_i,\label{Z2in2D} \eea where
$i=1,2,3,4$ labels the four TRIM in the 2D Brillouin zone.
$(-1)^{\nu_\mathrm{2D}} = +1 $ implies a trivial insulator while
$(-1)^{\nu_\mathrm{2D}} = -1 $ implies a topological insulator. Furthermore, as a TR polarization, $\nu_\mathrm{2D}$ also determines the way in which Kramers pairs of surface states are connected [Fig.~(\ref{fig:kaneInversion2})], which suggests that bulk topology and edge physics are intimately related. This is another example of the ``holographic principle'' for topological phenomena in condensed matter physics.

\begin{figure}
\begin{center}
\includegraphics[width=0.96\columnwidth]{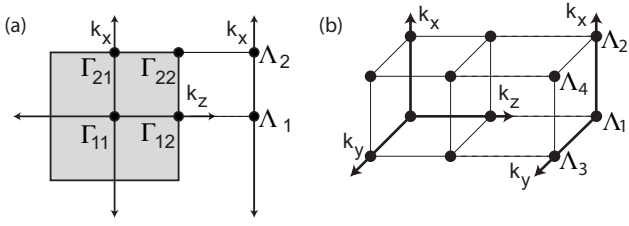}
\end{center}
 \caption{(a) The 2D bulk Brillouin zone projected onto the 1D edge Brillouin zone. The two edge TRIM $\Lambda_1$ and $\Lambda_2$ are projections of pairs of the four bulk TRIM $\Gamma_{i=(a\mu)}$.
(b) Projection of the TRIM of the 3D Brillouin zone onto a 2D surface Brillouin zone. From \onlinecite{fu2007a}.}
 \label{fig:kaneInversion1}
 \end{figure}

We now discuss 3D topological insulators. It is interesting to note that in TBT the natural route is ``dimensional increase'', in contrast to the ``dimensional reduction'' procedure of TFT. From this dimensional
increase, the 3D (strong) topological invariant is naturally
defined as~\cite{fu2007b,fu2007a}
\begin{equation}
(-1)^{\nu_\mathrm{3D}}=\prod_{i=1}^{8} \delta_i.\label{Z2in3D}
\end{equation}
In addition to the strong invariant, it has been shown that the
product of any four $\delta_i$'s for which the $\Gamma_i$ lie in the
same plane is also gauge invariant, and defines topological
invariants characterizing the band structure~\cite{fu2007a}. This
fact leads to the definition of three additional
invariants in 3D known as weak topological
invariants~\cite{fu2007b,fu2007a,moore2007,roy2009a}. These $\mathbb{Z}_2$ invariants can be arranged as a 3D vector with elements $\nu_k$ given by
\begin{equation}
(-1)^{\nu_k} = \prod_{n_k=1;n_{j\ne k}=0,1} \delta_{i=(n_1n_2n_3)},
\label{weakTI}
\end{equation}
where $(\nu_1\nu_2\nu_3)$ depend on the choice of reciprocal lattice
vectors and are only strictly well defined when a well-defined
lattice is present. It is useful to view these invariants as
components of a mod 2 reciprocal lattice vector,
\begin{equation}
{\bf G}_\nu =
\nu_1 {\bf b}_1 + \nu_2 {\bf b}_2 + \nu_3 {\bf b}_3.
\label{weakGvector}
\end{equation}
When $\nu_0=0$, states are classified according to ${\bf G}_\nu$, and
are called \emph{weak} topological insulators~\cite{fu2007b} when
the weak indices $\nu_k$ are odd.

Heuristically these states can be interpreted as stacked QSH states.
As an example, consider planes of QSH stacked in the $z$ direction.
When the coupling between the layers is zero, the band dispersion
will be independent of $k_z$. It follows that the four $\delta_i$'s
associated with the plane $k_z=\pi/a$ will have product $-1$ and will be the same as the four associated with the plane
$k_z=0$.  The topological invariants will then be given by $\nu_0=0$
and ${\bf G}_\nu = (2\pi/a)\hat{\mathbf{z}}$. This structure will remain when hopping between the layers is introduced. More generally, when QSH
states are stacked in the ${\bf G}$, direction the invariant will be
${\bf G}_\nu = {\bf G}$ mod 2. This implies that QSH states stacked
along different directions ${\bf G}_1$ and ${\bf G}_2$ are
equivalent if ${\bf G}_1={\bf G}_2$ mod 2~\cite{fu2007a}. As for the
surface states, when the coupling between the layers is zero, it is
clear that the gap in the 2D system implies there will be no surface
states on the top and bottom surfaces; only the side surfaces will
have gapless states. We can also think about the stability of the
surface states for the weak insulators. In fact, weak topological insulators are unstable with respect to disorder. We can heuristically see that they are less stable than the strong insulators in the following way. If we stack an odd number of QSH layers, there would at least be one delocalized surface branch. However, the surface states for an even
number of layers can be completely localized by disorder or
perturbations. Despite this instability, it has been shown~\cite{ran2009} that the weak topological invariants guarantee the existence
of gapless modes on certain crystal defects. For a dislocation with
Burgers vector ${\bf b}$ it was shown that there will be gapless
modes on the dislocation if ${\bf G}_\nu\cdot {\bf b}=(2n+1)\pi$ for
integer $n.$

\begin{figure}
\begin{center}
\includegraphics[width=0.96\columnwidth]{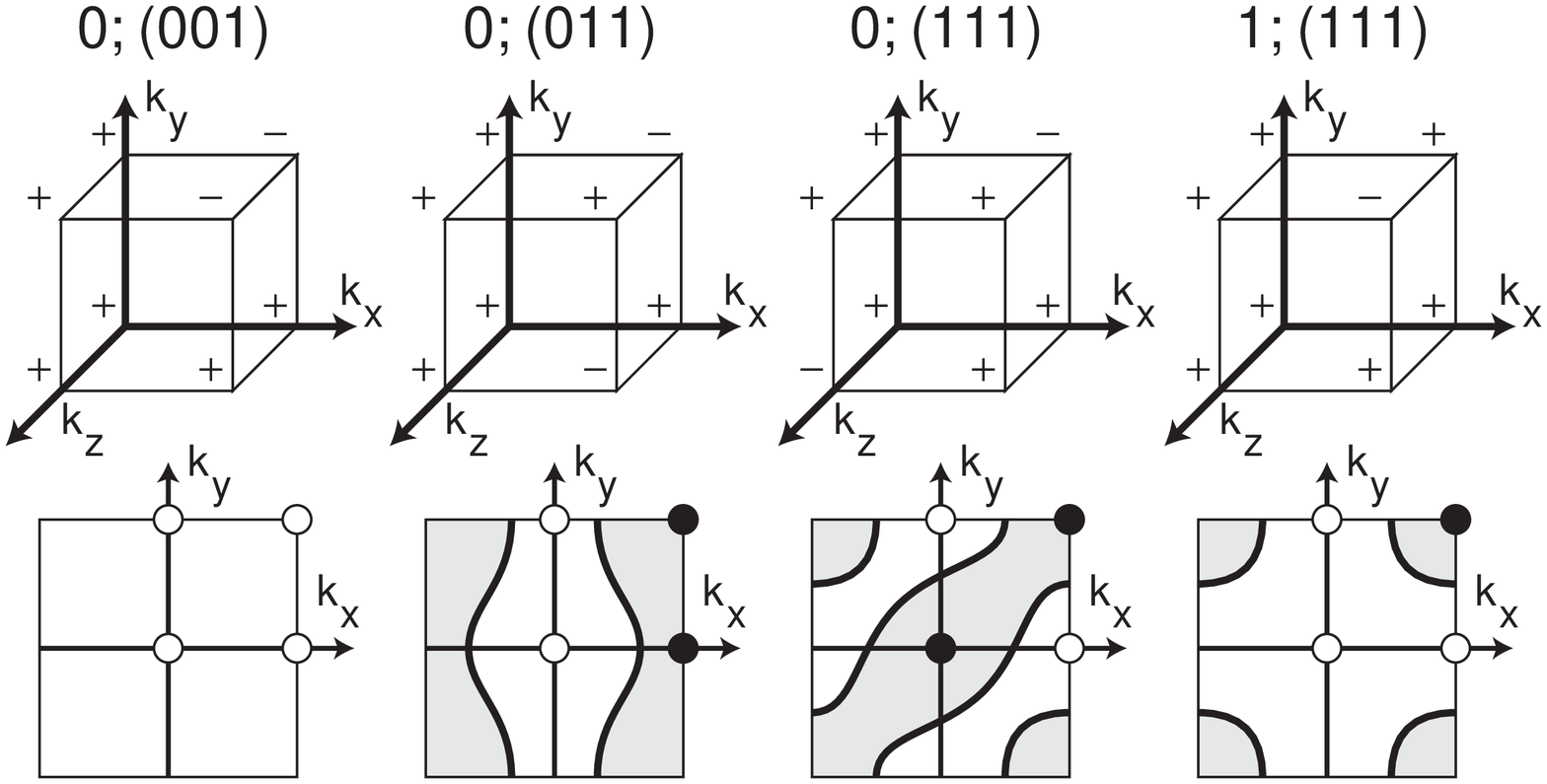}
\end{center}
 \caption{Diagrams depicting four different phases indexed by
 $\nu_0; (\nu_1\nu_2\nu_3)$.  The top panel depicts the signs of
 $\delta_i$ at the points $\Gamma_i$ on the vertices of a cube.  The
 bottom panel characterizes the band structure of a $001$ surface for
 each phase. The solid and open circles depict the TR
 polarization $\pi_a$ at the surface momenta $\Lambda_a$,  which
 are projections of pairs of $\Gamma_i$ which differ only in their
 $z$ component.  The  thick lines indicate possible Fermi arcs which enclose specific $\Lambda_a$. From \onlinecite{fu2007b}.
}
\label{fig:kaneInversion3}
\end{figure}

Similar to the 2D topological insulator, there are connections between the bulk invariants of 3D topological insulator and the corresponding 2D surface state spectrum. As a sample, Fig.~(\ref{fig:kaneInversion3}) shows four different
topological classes for 3D band structures labeled with the
corresponding $(\nu_0; \nu_1\nu_2\nu_3)$. The eight $\Gamma_i$ are
represented as the vertices of a cube in momentum space, with the
corresponding $\delta_i$ shown as $\pm$ signs.  The lower panel
shows a characteristic surface Brillouin zone for a $001$ surface
with the four $\Lambda_a$ labeled by either filled or solid circles,
depending on the value of $\pi_a = \delta_{i=(a1)}\delta_{i=(a2)}$.
Generically it is expected that the surface band structure will
resemble Fig.~\ref{fig:kaneInversion2}(b) on paths connecting two
filled circles or two empty circles, and will resemble Fig.~\ref{fig:kaneInversion2}(a) on paths connecting a filled circle to an empty circle~\cite{fu2007a}. This consideration determines the 2D
surface states qualitatively.

If an insulator has inversion symmetry, there is a simple
algorithm to calculate the $\mathbb{Z}_2$ invariant~\cite{fu2007a}: indeed, the replacement in Eq.~(\ref{Z2in2D}) and Eq.~(\ref{Z2in3D}) of $\delta_i$ by
\begin{equation}
\delta_i = \prod_{m=1}^N \xi_{2m}(\Gamma_i), \label{deltaParity}
\end{equation}
gives the correct $\mathbb{Z}_2$ invariants. Here $\xi_{2m}(\Gamma_i)=\pm 1$ is the parity eigenvalue of the $2m$th occupied energy band at
$\Gamma_i$ [Fig.~\ref{fig:kaneInversion1}], which shares the
same eigenvalue $\xi_{2m} = \xi_{2m-1}$ with its degenerate Kramers partner~\cite{fu2007a}. The product is only over half of the occupied
bands. Since the definition of the $\delta_i$ relies on parity
eigenvalues, the $\delta_i$ are only well-defined in this case when inversion symmetry is present~\cite{fu2007a}. However, for insulators without inversion symmetry, this algorithm is very useful. In fact, if we can deform a given insulator to an inversion-symmetric insulator
and keep the energy gap open along the way, the resultant $\mathbb{Z}_2$
invariants are the same as the initial ones due to topological
invariance, but can be calculated from parity~\cite{fu2007a}.

\subsection{Reduction from topological field theory to topological band theory}

We now briefly discuss the relation between the TFT and TBT. On the one
hand, the TFT approach is very powerful to reveal various aspects of the low-energy physics, and it also provide a deep understanding of the universality among different systems. Furthermore, in contrast to TBT, TFT is valid for interacting systems. On the other hand, from a practical viewpoint, we also need fast algorithms to calculate topological
invariants, which is the goal of TBT. An intuitive understanding of
the TBT of $\mathbb{Z}_2$ topological insulators is as follows. For integer-class CS topological insulators, the topological invariant $C_n$ is expressed as the integral of Green's
functions (or Berry curvature, in the noninteracting limit).
Therefore, the knowledge of Bloch states over the whole Brillouin
zone is needed to calculate $C_n$. For TR invariant $\mathbb{Z}_2$
topological insulators, the TR symmetry constraint enables us to determine the topological class of a given insulator with less information: we do not need the information over the entire Brillouin zone. For insulators with inversion symmetry, the parity at several high-symmetry points completely determines the topological class~\cite{fu2007a}, which explains the success of TBT. As
naturally expected, the TBT approach is related to the TFT approach.
In fact, it has been recently proved~\cite{wang2010a} that the TFT
description can be exactly reduced to the TBT in the noninteracting
limit. We now outline this proof~\cite{wang2010a}. Starting from the expression for $P_3$ in Eq.~(\ref{P3}) and Eq.~(\ref{P3CS}), one can show that
\begin{eqnarray}
2P_{3}(\textrm{mod}\, 2) &=& -\frac{1}{24\pi^{2}} \int d^{3}k\,
\epsilon ^{ijk} \textrm{Tr}[(B\partial_{i}B^{\dagger})(
B\partial_{j}B^{\dagger}) \nonumber \\ && \times (
B\partial_{k}B^{\dagger} )] \, (\textrm{mod}\, 2). \label{deg}
\end{eqnarray}
By some topological argument, this expression for $P_3$ is shown to give the degree $\deg f$ of certain map~\cite{dubrovin1985} from the Brillouin zone three-torus $T^3$ to the $SU(2)$ group manifold. There are two seemingly different expressions for $\deg f$, one of which is of integral form as given by Eq.~(\ref{deg}), while the other is of discrete form and given simply by the number of points mapped to a arbitrarily chosen image in $SU(2)$. Due to TR symmetry, if we choose the image
point as one of the two antisymmetric matrices in $SU(2)$ ({\it e.g.} $i\sigma_y$), we have
an interesting ``pair annihilation'' of those points other than the eight
TRIM~\cite{wang2010a}. The final result is exactly the $\mathbb{Z}_2$ invariant from TBT. The explicit relation between TFT and TBT is~\cite{wang2010a}
\begin{eqnarray}
(-1)^{2P_{3}} = (-1)^{\nu_\mathrm{3D}}.
\end{eqnarray}
\section{Topological Superconductors and Superfluids}
\label{sec:TSC}

Soon after their discovery, the study of TR invariant topological insulators was generalized to TR invariant topological superconductors and superfluids~\cite{qi2009b,roy2008,schnyder2008,kitaev2009}. There is
a direct analogy between superconductors and insulators because the
Bogoliubov-de Gennes (BdG) Hamiltonian for the quasiparticles of a
superconductor is analogous to the Hamiltonian of a band insulator,
with the superconducting gap corresponding to the band gap of the
insulator.

$^3$He-B is an example of such a topological superfluid state. This
TR invariant state has a full pairing gap in the bulk, and gapless surface states consisting of a single Majorana
cone~\cite{qi2009b,roy2008,schnyder2008,chung2009}. In fact, the BdG
Hamiltonian for $^3$He-B is identical to the model Hamiltonian of a
3D topological insulator~\cite{zhang2009}, and investigated extensively in Sec.~\ref{sec:3dzhang}. In 2D, the classification of topological
superconductors is very similar to that of topological insulators.
TR breaking superconductors are classified by an
integer~\cite{volovik1988,read2000}, similar to quantum Hall
insulators~\cite{thouless1982}, while TR invariant superconductors are
classified~\cite{qi2009b,roy2008,schnyder2008,kitaev2009} by a $\mathbb{Z}_2$
invariant in 1D and 2D, but by an integer ($\mathbb{Z}$) invariant in 3D\cite{kitaev2009,schnyder2008}.

Besides the TR invariant topological superconductors, the TR breaking topological
superconductors have also attracted a lot of interest recently, because of
their relation with non-Abelian statistics and their potential application
to topological quantum computation. The TR breaking topological
superconductors are described by an integer ${\cal N}$. The vortex
of a topological superconductor with odd topological quantum number ${\cal
N}$ carries an odd number of Majorana zero modes~\cite{volovik1999},
giving rise to non-Abelian statistics~\cite{read2000,ivanov2001}
which could provide a platform for topological quantum
computing~\cite{nayak2008}. The simplest model for an ${\cal N}=1$
chiral topological superconductor is realized in the $p_{x}+ip_{y}$ pairing state of
spinless fermions~\cite{read2000}. A spinful version of the chiral
superconductor has been predicted to exist in
Sr$_2$RuO$_{4}$~\cite{mackenzie2003}, but the experimental
situation is far from definitive. Recently, several new proposals to
realize Majorana fermion states in conventional superconductors
have been investigated by making use of strong SOC~\cite{fu2008,sau2010,qi2010}.

\subsection{Effective models of time-reversal invariant superconductors}
\label{sec:tscmodels}

The simplest way to understand TR invariant topological superconductors is
through their analogy with topological insulators. The 2D chiral
superconducting state is the superconductor analog of the QH state.
A QH state with Chern number $N$ has $N$ chiral edge states, while a
chiral superconductor with topological quantum number ${\cal N}$ has
${\cal N}$ chiral Majorana edge states. Since the positive and
negative energy states of the BdG Hamiltonian of a superconductor describe the same physical degrees of freedom, each chiral Majorana
edge state has half the degrees of freedom of the chiral edge state
of a QH system. Therefore, the chiral superconductor is the ``minimal"
topological state in 2D. The analogy between a chiral superconductor and a QH state is illustrated in the upper panels of Fig.~\ref{fig:SCTIanalogy}. Following the same analogy, one can consider
the superconducting analog of QSH state --- a ``helical" superconductor
in which fermions with up spins are paired in the $p_x+ip_y$
state, and fermions with down spins are paired in the $p_x-ip_y$
state. Such a TR invariant state has a full gap in the bulk, and
counter-propagating helical Majorana states at the edge. In
contrast, the edge states of the TR invariant topological insulator are
helical \emph{Dirac} fermions with twice the degrees of freedom. As
is the case for the QSH state, a mass term for the edge states is
forbidden by TR symmetry. Therefore, such a superconducting
phase is topologically protected in the presence of TR
symmetry, and can be described by a $\mathbb{Z}_2$ topological quantum
number~\cite{qi2009b,roy2008,schnyder2008,kitaev2009}. The four types
of 2D topological states of matter discussed here are summarized
in Fig.~\ref{fig:SCTIanalogy}.

As a starting point, we first consider the Hamiltonian of the simplest
nontrivial TR breaking superconductor, the $p+ip$ superconductor~\cite{read2000} for spinless fermions:
\begin{eqnarray}
H=\frac12\sum_p
\left(c^\dagger_{\bf p},c_{- \bf p}\right)\left(\begin{array}{cc}\epsilon_{\bf
p}&\Delta p_+\\
\Delta^* p_-&-\epsilon_{\bf p}\end{array}\right)
\left(\begin{array}{c}c_{\bf p}\\c^\dagger_{-\bf p}\end{array}\right),\label{Hpip}
\end{eqnarray}
with $\epsilon_{\bf p}={\bf p}^2/2m-\mu$ and $p_\pm=p_x\pm ip_y$.
In the weak pairing phase with $\mu>0$, the $p_x+ip_y$
\emph{chiral} superconductor is known to have chiral Majorana edge
states propagating on each boundary, described by the Hamiltonian
\begin{eqnarray}
H_{\rm edge}=\sum_{k_y\geq 0}v_F k_y\psi_{-k_y}\psi_{k_y},
\end{eqnarray}
where $\psi_{-k_y}=\psi_{k_y}^\dagger$ is the quasiparticle creation
operator~\cite{read2000} and the boundary is taken parallel to the
$y$ direction. The strong pairing phase $\mu<0$ is trivial, and the
two phases are separated by a topological phase transition at
$\mu=0$.

In the BHZ model for the QSH state in HgTe~\cite{bernevig2006c}, if we ignore the coupling terms between spin up and spin down electrons, the system is a direct product of two independent QH systems in which spin up and spin down electrons have opposite Hall conductance. In the same way, the simplest model for the topologically nontrivial TR invariant
superconductor in 2D is given by the following Hamiltonian:
\begin{eqnarray}
H=\frac12\sum_p
\tilde{\Psi}^\dagger\left(\begin{array}{cccc}\epsilon_{\bf
p}&\Delta p_+& 0 & 0\\
\Delta^* p_-&-\epsilon_{\bf p}& 0 & 0\\
0 & 0 &\epsilon_{\bf p}&-\Delta^* p_-\\
0 &0 &-\Delta p_+&-\epsilon_{\bf
p}\end{array}\right)\tilde{\Psi},\label{Hdoublepip}
\end{eqnarray}
with $\tilde{\Psi}({\bf p})\equiv \left(c_{\uparrow {\bf
p}},c^\dagger_{\uparrow {-\bf p}}, c_{\downarrow {\bf
p}},c^\dagger_{\downarrow {-\bf p}}\right)^T$. From Eq.~(\ref{Hdoublepip}) we see that
spin up (down) electrons form $p_x+ip_y$ ($p_x-ip_y$) Cooper pairs,
respectively. Comparing this model Hamiltonian (\ref{Hdoublepip}) for
the topological superconductor with the BHZ model of the HgTe
topological insulator [Eq.~\ref{BHZ}], we first see that the
term proportional to the identity matrix in the BHZ model is absent
here, reflecting the generic particle-hole symmetry of the BdG
Hamiltonian for superconductors. On the other hand, the terms
proportional to the Pauli matrices $\sigma^a$ are {\it identical} in
both cases. Therefore, a topological superconductor can be viewed as
a topological insulator with particle-hole symmetry. The topological
superconductor Hamiltonian also has half as many degrees of freedom
as the topological insulator. The model Hamiltonian
(\ref{Hdoublepip}) is expressed in terms of the Nambu spinor
$\tilde{\Psi}({\bf p})$ which artificially doubles the degrees of
freedom as compared to the topological insulator Hamiltonian. Bearing
these differences in mind, in analogy with the QSH system, we know
that the edge states of the TR invariant system described by the Hamiltonian (\ref{Hdoublepip}) consist of spin up and spin down quasiparticles with opposite chiralities:

\begin{figure}
\begin{center}
\includegraphics[width=0.45\textwidth] {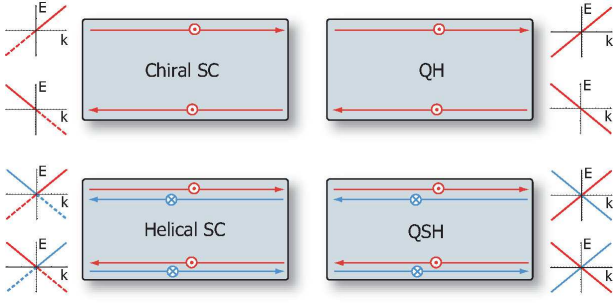}
\end{center}
\caption{(Top row) Schematic comparison of 2D chiral
superconductor and QH state. In both systems, TR symmetry is
broken and the edge states carry a definite chirality. (Bottom
row) Schematic comparison of 2D TR invariant topological superconductor
and QSH insulator. Both systems preserve TR symmetry and have
a helical pair of edge states, where opposite spin states
counter-propagate. The dashed lines show that the edge states of
the superconductors are Majorana fermions so that the $E<0$ part
of the quasiparticle spectrum is redundant. In terms of the edge
state degrees of freedom, we have symbolically ${\rm QSH}={\rm (QH)}^2={\rm
(Helical~SC)}^2={\rm (Chiral~SC)}^4$. From \onlinecite{qi2009b}.
 } \label{fig:SCTIanalogy}
\end{figure}

\begin{eqnarray}
H_{\rm edge}=\sum_{k_y\geq
0}v_Fk_y\left(\psi_{-k_y\uparrow}\psi_{k_y\uparrow}-\psi_{-k_y\downarrow}\psi_{k_y\downarrow}\right).\label{Hdoubleedge}
\end{eqnarray}
The quasiparticle operators
$\psi_{k_y\uparrow},~\psi_{k_y\downarrow}$ can be expressed in terms
of the eigenstates $u_{k_y}(x),v_{k_y}(x)$ of the BdG Hamiltonian as
\begin{align}
\psi_{k_y\uparrow}&=\int
d^2x\left(u_{k_y}(x)c_{\uparrow}(x)+v_{k_y}(x)c_{\uparrow}^\dagger(x)\right),\nonumber\\
\psi_{k_y\downarrow}&=\int
d^2x\left(u_{-k_y}^*(x)c_{\downarrow}(x)+v_{-k_y}^*(x)c_{\downarrow}^\dagger(x)\right),\nonumber
\end{align}
from which the TR transformation of the quasiparticle
operators can be determined to be
$T\psi_{k_y\uparrow}T^{-1}=\psi_{-k_y\downarrow},~T\psi_{k_y\downarrow}T^{-1}=-\psi_{-k_y\uparrow}$.
In other words, $(\psi_{k_y\uparrow},~\psi_{-k_y\downarrow})$
transforms under TR as a Kramers doublet, which forbids a gap in the edge
state spectrum when TR is preserved by preventing the mixing of spin up and spin down modes. To see this explicitly, notice that the
only $k_y$-independent term that can be added to the edge
Hamiltonian (\ref{Hdoubleedge}) is
$im\sum_{k_y}\psi_{-k_y\uparrow}\psi_{k_y\downarrow}$, with $m$ real. However, such a term is odd under TR, which implies
that any backscattering between quasiparticles is forbidden by
TR symmetry. The discussion above is exactly parallel to the $\mathbb{Z}_2$
topological characterization of QSH system. In
fact, the Hamiltonian (\ref{Hdoublepip}) has exactly the same form
as the four-band effective Hamiltonian of the QSH effect in HgTe quantum wells~\cite{bernevig2006c}. The edge states of the QSH insulator consist of an odd number of Kramers pairs, which remain gapless under any small
TR invariant perturbation~\cite{wu2006,xu2006}. Such a ``helical
liquid" with an odd number of Kramers pairs at the Fermi energy
cannot be realized in any bulk 1D system, and can only appear {\it
holographically} as the edge theory of a 2D QSH
insulator~\cite{wu2006}. Similarly, the edge state theory Eq.~(\ref{Hdoubleedge}) can be called a ``helical Majorana liquid", and
can only exist on the boundary of a $\mathbb{Z}_2$ topological
superconductor. Once such a topological phase is established, it is
robust under any TR invariant perturbations such as Rashba-type SOC and $s$-wave pairing, even if spin rotation symmetry is broken.
The edge helical Majorana liquid can be detected by electric transport through a quantum point contact between two topological superconductors~\cite{asano2010}.

The 2D Hamiltonian (\ref{Hdoublepip}) describes a spin-triplet
pairing, the spin polarization of which is correlated with the
orbital angular momentum of the pair. Such a correlation can be
naturally generalized to 3D where spin polarization and orbital
angular momentum are both vectors. The Hamiltonian of such a 3D
superconductor is given by
\begin{eqnarray}
H=\frac12\sum_p \Psi^\dagger \left(\begin{array}{cc}\epsilon_{\bf
p}\mathbb{I}_{2\times 2}&i\sigma^2\sigma^\alpha\Delta^{\alpha
j}p_j\\\mathrm{h.c.}&-\epsilon_{\bf
p}\mathbb{I}_{2\times 2}\end{array}\right)\Psi,\label{HHe3B}
\end{eqnarray}
where we use a different basis $\Psi({\bf
p})\equiv \left(c_{\uparrow {\bf p}}, c_{\downarrow {\bf p}},
c^\dagger_{\uparrow {-\bf p}}, c^\dagger_{\downarrow {-\bf
p}}\right)^T$. $\Delta^{\alpha j}$ is a $3\times 3$ matrix with
$\alpha=1,2,3$ and $j=x,y,z$. Interestingly, an example of such a
Hamiltonian is given by the well-known $^3$He-B phase, for which
the order parameter $\Delta^{\alpha j}$ is determined by an
orthogonal matrix $\Delta^{\alpha j}=\Delta u^{\alpha j}$, $u\in SO(3)$~\cite{vollhardt1990}. Here and below we ignore the
dipole-dipole interaction term~\cite{leggett1975}, since it does not
affect any essential topological properties. Performing a spin
rotation, $\Delta^{\alpha j}$ can be diagonalized to $\Delta^{\alpha
j}=\Delta \delta^{\alpha j}$, in which case the Hamiltonian
(\ref{HHe3B}) can be expressed as:
\begin{eqnarray}
H=\frac12\int d^2x\,\Psi^\dagger
\left(\begin{array}{cccc}\epsilon_{\bf p}& 0 &\Delta p_+&-\Delta
p_z\\0 & \epsilon_{\bf p}&-\Delta p_z&-\Delta
p_-\\\Delta^*p_-&-\Delta^*p_z&-\epsilon_{\bf
p}& 0 \\-\Delta^*p_z&-\Delta^*p_+& 0 &-\epsilon_{\bf
p}\end{array}\right)\Psi.\label{HHe3B2}
\end{eqnarray}
Compared with the model Hamiltonian Eq.~(\ref{eq:Heff3D}) for the simplest 3D topological insulators~\cite{zhang2009}, we see that the Hamiltonian (\ref{HHe3B2}) has the same
form as that for Bi$_2$Se$_3$ (up to a basis transformation), but with complex fermions
replaced by Majorana fermions. The kinetic energy term ${\bf
p}^2/2m-\mu$ corresponds to the momentum dependent mass term $M({\bf
p})=M-B_1p_z^2-B_2p_\parallel^2$ of the topological insulator. The weak pairing phase $\mu>0$ corresponds
to the nontrivial topological insulator phase, and the strong pairing phase $\mu<0$ corresponds to the trivial insulator. From this analogy, we see
that the superconductor Hamiltonian in the weak pairing phase describes
a topological \emph{superconductor} with gapless surface states
protected by TR symmetry. Different from the topological insulator, the surface
states of the topological superconductor are Majorana fermions
described by
\begin{eqnarray}
H_{\rm surf}=\frac12\sum_{\bf k}v_F\psi_{-\bf
k}^T\left(k_x\sigma_y-k_y\sigma_x\right)\psi_{\bf k},
\end{eqnarray}
with the Majorana condition $\psi_{-\bf k}=\sigma_x\psi_{\bf k}^{\dagger T}$.
We see that this Hamiltonian for the surface Majorana fermions of a
topological superconductor takes the same form as the
surface Dirac Hamiltonian of a topological insulator in this special basis. However,
because of the generic particle-hole symmetry of the BdG Hamiltonian
for superconductors, the possible particle-hole symmetry breaking
terms for surface Dirac fermions such as a finite chemical potential is absent for surface Majorana
fermions. Because of the particle-hole symmetry and TR symmetry,
the spin lies strictly in the plane perpendicular to the surface normal, and the
integer winding number of the spin around the momentum is now a well-defined quantity. This integer winding number gives a $\mathbb{Z}$ classification of the 3D topological superconductor \cite{schnyder2008,kitaev2009}. The surface
state remains gapless under any small TR invariant perturbation, since the only available mass term $m\sum_{\bf k}\psi_{-\bf
k}^T\sigma^y\psi_{\bf k}$ is TR odd. The Majorana surface
state is spin-polarized, and can thus be detected by its special
contribution to the spin relaxation of an electron on the surface of
$^3$He-B, similar to the measurement of electron spin
correlation in a solid state system by nuclear magnetic resonance~\cite{chung2009}.

\begin{figure}
\begin{center}
\includegraphics[width=0.4\textwidth]{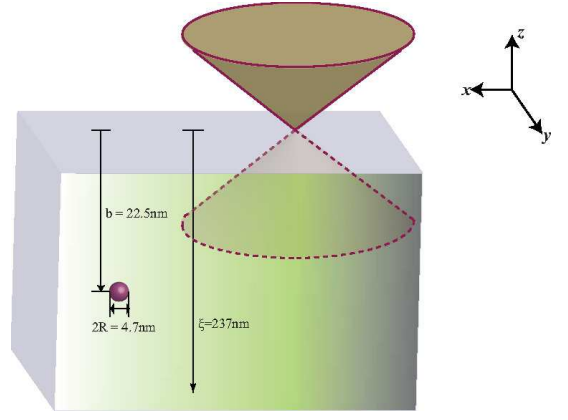}
\end{center}
\caption{Setting for detecting the Majorana surface states of the $^3$He-B phase, which
consist of a single Majorana cone. When electrons are injected into
$^3$He-B, they exist as ``bubbles''. If the injected electrons are
spin-polarized, the spin will relax by interaction with the
surface Majorana modes, and this relaxation is strongly anisotropic. From \onlinecite{chung2009}.
} \label{FIG:bubble}
\end{figure}

\subsection{Topological invariants}
\label{sec:invariants}

From the discussion above, we see that the model Hamiltonian for the topological superconductor is the same as
that for the topological insulator, but with the
additional particle-hole symmetry. The simultaneous presence of both
TR and particle-hole symmetry gives a different
classification for the 2D and 3D topological superconductors, in
that the 3D TR invariant topological superconductors are classified by
integer ($\mathbb{Z}$) classes, and the 2D TR invariant topological superconductors are
classified by the $\mathbb{Z}_2$ classes. To define an integer-valued
topological invariant~\cite{schnyder2008}, we start from a generic
mean-field BdG Hamiltonian for a 3D TR invariant superconductor, which can be
written in momentum space as
\begin{eqnarray}
H=\sum_{\bf k}\left[\psi_{\bf k}^\dagger
h_{\bf k}\psi_{\bf k}+\frac12\left(\psi_{\bf k}^\dagger
\Delta_{\bf k}\psi_{-\bf k}^{\dagger
T}+\mathrm{h.c.}\right)\right],\nonumber
\end{eqnarray}
In a different basis we have $H=\sum_{\bf k}\Psi_{\bf k}^\dagger
H_{\bf k}\Psi_{\bf k}$ with
\begin{eqnarray}
\Psi_{\bf k}&=&\frac1{\sqrt{2}}\left(\begin{array}{c}\psi_{\bf k}
-i{\mathcal{T}}\psi^\dagger_{-{\bf k}}\\\psi_{\bf k}+i{\mathcal{T}}\psi^\dagger_{-{\bf k}}
\end{array}\right),\nonumber\\
H_{\bf k}&=&\frac12\left(\begin{array}{cc}
0 & h_{\bf k}+i{\mathcal{T}}\Delta_{\bf k}^\dagger\\h_{\bf k}-i{\mathcal{T}}
\Delta_{\bf k}^\dagger& 0\end{array}\right).\label{HBdG}
\end{eqnarray}
In general, $\psi_{\bf k}$ is a vector with $N$ components, and
$h_{\bf k}$ and $\Delta_{\bf k}$ are $N\times N$ matrices. The matrix
${\mathcal{T}}$ is the TR matrix satisfying ${\mathcal{T}}^\dagger h_{\bf k}
{\mathcal{T}}=h_{-{\bf k}}^T$, ${\mathcal{T}}^2=-\mathbb{I}$ and ${\mathcal{T}}^\dagger
{\mathcal{T}}=\mathbb{I}$, with $\mathbb{I}$ the identity matrix. We have chosen a special basis in which the BdG
Hamiltonian $H_{\bf k}$ has a special off-diagonal form. It should be
noted that such a choice is only possible when the system has both
TR symmetry and particle-hole symmetry. These two
symmetries also require ${\mathcal{T}}\Delta_{\bf k}^\dagger$ to be Hermitian,
which makes the matrix $h_{\bf k}+i{\mathcal{T}}\Delta_{\bf k}^\dagger$ generically
non-Hermitian. The matrix $h_{\bf k}+i{\mathcal{T}}\Delta_{\bf k}^\dagger$ can be
decomposed by a singular value decomposition as
$h_{\bf k}+i{\mathcal{T}}\Delta_{\bf k}^\dagger=U_{\bf k}^\dagger D_{\bf k} V_{\bf k}$ with
$U_{\bf k}, V_{\bf k}$ unitary matrices and $D_{\bf k}$ a diagonal matrix with
nonnegative elements. One can see that the
diagonal elements of $D_{\bf k}$ are actually the positive eigenvalues
of $H_{\bf k}$. For a fully gapped superconductor, $D_{\bf k}$ is positive
definite, and we can adiabatically deform it to the identity
matrix $\mathbb{I}$ without closing the superconducting gap.
During this deformation, the matrix $h_{\bf k}+i{\mathcal{T}}\Delta^\dagger_{\bf k}$
is deformed to a unitary matrix $Q_{\bf k}=U_{\bf k}^\dagger V_{\bf k}\in U(N)$. The integer-valued
topological invariant characterizing topological superconductors
is defined as the winding number of $Q_{\bf k}$~\cite{schnyder2008}:
\begin{eqnarray}
N_W=\frac1{24\pi^2}\int d^3k\,\epsilon^{ijk}{\rm
Tr}\left[{Q^\dagger_{\bf k}\partial_iQ_{\bf k}Q^\dagger_{\bf
k}\partial_jQ_{\bf k}Q^\dagger_{\bf k}\partial_kQ_{\bf
k}}\right].\label{Windingnumber}
\end{eqnarray}

\begin{figure}
   \begin{center}
    \includegraphics[width=0.48\textwidth]{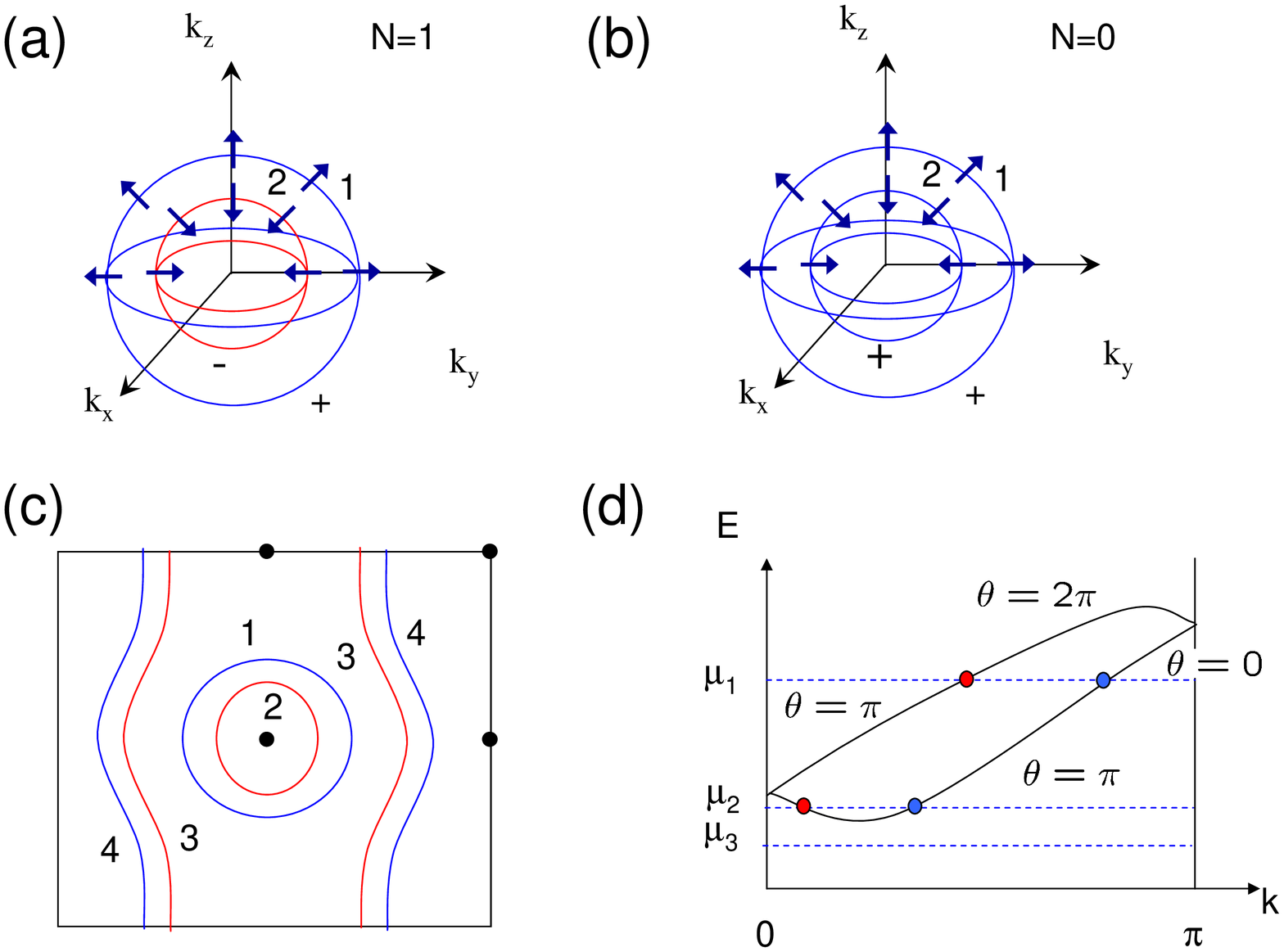}
    \end{center}
    \caption{(a,b) Superconducting pairing on two Fermi surfaces of a 3D superconductor. (c) An example of 2D TR invariant topological superconductor. (d) 1D TR invariant topological superconductor. Adapted from \onlinecite{qi2010b}.
    }
    \label{fig:FSinv}
\end{figure}

We note that the topological invariant (\ref{Windingnumber}) is
expressed as an integral over the entire Brillouin zone, similar to its counterpart for topological insulators. However, there is a key
difference. Whereas the insulating gap is well defined over the
entire Brillouin zone, the superconducting pairing gap in the BdG
equation is only well defined close to the Fermi surface. Indeed, superconductivity arises from a Fermi surface instability, at least in the BCS limit. Therefore, one would like to define
topological invariants for a topological superconductor strictly
in terms of Fermi surface quantities. The desired topological
invariant can be obtained by reducing the winding number in
Eq.~(\ref{Windingnumber}) to a integral over the Fermi surface~\cite{qi2010b}:
\begin{eqnarray}
N_W=\frac12\sum_{s}{\rm sgn}(\delta_s)C_{1s},\label{WindingFS}
\end{eqnarray}
where $s$ is summed over all disconnected Fermi surfaces and ${\rm
sgn}(\delta_s)$ denotes the sign of the pairing amplitude on the $s$th Fermi
surface. $C_{1s}$ is the first Chern number of the $s$th Fermi
surface (denoted by ${\rm FS}_s$):
\begin{eqnarray}
C_{1s}=\frac1{2\pi}\int_{{\rm
FS}_s}d\Omega^{ij}\left(\partial_ia_{sj}({\bf
k})-\partial_ja_{si}({\bf k})\right),
\end{eqnarray}
with $a_{si}=-i\left\langle s{\bf k}\right|\partial/\partial
k_i\left|s{\bf k}\right\rangle$ the adiabatic connection defined for
the band $\left|s{\bf k}\right\rangle$ which crosses the Fermi
surface, and $d\Omega^{ij}$ the surface element $2$-form of the
Fermi surface.

As an example, we consider a two-band model with noninteracting
Hamiltonian $h_{\bf k} = {\bf k}^2 /2m - \mu+\alpha{\bf
k}\cdot\sigma$, for which there are two Fermi surfaces with opposite
Chern number $C_{\pm}=\pm 1$ [Fig.~\ref{fig:FSinv}(a),(b)]. If we
choose $\Delta_{\bf k} = i\Delta_0 \sigma^y$, which has the same
$\delta_s$ for both Fermi surfaces, we obtain
$N_W=0$ [Fig.~\ref{fig:FSinv}(b)]. If we instead choose $\Delta_{\bf k}
= i\Delta_0 \sigma^y \sigma\cdot{\bf k}$, we obtain
$N_W=1$ [Fig.~\ref{fig:FSinv}(a)]. In the latter case, if we take the
$\alpha \rightarrow 0$ limit, we arrive at the result $N_W = 1$ for the
$^3$He-B phase, which indicates that $^3$He-B is topologically
nontrivial~\cite{volovik2003}.

For 2D TR invariant superconductors, a procedure of dimensional reduction leads to the following simple Fermi surface topological invariant:
\begin{eqnarray}
N_\mathrm{2D}=\prod_{s}\left({\rm sgn}(\delta_s)\right)^{m_s}.\label{Z22d}
\end{eqnarray}
The criterion (\ref{Z22d}) is quite simple: a 2D TR invariant
superconductor is topologically nontrivial (trivial) if there is an odd (even) number of Fermi surfaces, each of which encloses one TR invariant point in the Brillouin zone and has negative pairing. As an example, see
Fig.~\ref{fig:FSinv}(c), where Fermi surfaces 2 and 3 have
negative pairing. Fermi surfaces 3 and 4 enclose an even number of TR invariant momenta, which do not affect the $\mathbb{Z}_2$ topological invariant. There is only one Fermi surface, surface 2, which encloses on odd number of TR invariant momenta
and has negative pairing. As a result, the $\mathbb{Z}_2$ topological invariant is $(-1)^1 = -1$.

For 1D TR invariant superconductors, a further dimensional reduction can be carried out to give
\begin{eqnarray}
N_\mathrm{1D}=\prod_s\left({\rm sgn}(\delta_s)\right)\label{Z21d}
\end{eqnarray}
where $s$ is summed over all the Fermi points between $0$ and $\pi$.
In geometrical terms, a 1D TR invariant superconductor is nontrivial
(trivial) if there is an odd number of Fermi points between $0$ and
$\pi$ with negative pairing. We illustrate this formula in
Fig.~\ref{fig:FSinv}(d), where the sign of pairing on the red (blue)
Fermi point is $-1$ ($+1$), so that the number of Fermi points with
negative pairing is $1$ if the chemical potential $\mu=\mu_1$ or $\mu=\mu_2$, and $0$ if $\mu=\mu_3$. The superconducting states
with $\mu=\mu_1$ and $\mu=\mu_2$  can be adiabatically deformed to
each other without closing the gap. However, the superconductor with
$\mu = \mu_3$ can only be obtained from that with $\mu_2$ through a
topological phase transition, where the pairing order parameter
changes sign on one of the Fermi points. It is easy to see from this
example that there are two classes of 1D TR invariant superconductors.

\subsection{Majorana zero modes in topological superconductors}

\subsubsection{Majorana zero modes in $p+ip$ superconductors}

Besides the new TR invariant topological superconductors,
the TR breaking topological superconductors have
attracted a lot of interest because of their relevance to non-Abelian
statistics and topological quantum computation. In a $p+ip$ superconductor described by Eq.~(\ref{Hpip}), it can be shown that
the core of a superconducting vortex contains a localized quasiparticle with exactly
zero energy~\cite{volovik1999}. The corresponding quasiparticle operator
$\gamma$ is a Majorana fermion obeying $\left[\gamma,H\right]=0$ and $\gamma^\dagger=\gamma$. When two vortices wind around each other,
the two Majorana fermions $\gamma_1$, $\gamma_2$ in their cores
transform nontrivially. Because the phase of the charge-$2e$ order
parameter winds by $2\pi$ around each vortex, an electron acquires a Berry phase of $\pi$ when winding once around a vortex. Since the Majorana fermion operator is a superposition of electron
creation and annihilation operators, it also acquires a $\pi$ phase
shift, i.e. a minus sign when winding around another vortex.
Consequently, when two vortices are exchanged, the Majorana operators
$\gamma_1$, $\gamma_2$ must transform as
$\gamma_1\rightarrow\gamma_2$, $\gamma_2\rightarrow -\gamma_1$. The
additional minus sign may be associated with $\gamma_1$ or
$\gamma_2$, but not both, so that after a full winding we have
$\gamma_{1(2)}\rightarrow -\gamma_{1(2)}$. Since two Majorana
fermions $\gamma_1$ and $\gamma_2$ define one complex fermion
operator $\gamma_1+i\gamma_2$, the two vortices actually share two
internal states labeled by $i\gamma_1\gamma_2=\pm 1$. When there are
$2N$ vortices in the system, the core states span a $2^N$-dimensional Hilbert space. The braiding of vortices leads to non-Abelian unitary transformations in this Hilbert space, implying that the
vortices in this system obey non-Abelian
statistics~\cite{read2000,ivanov2001}. Because the internal states of
the vortices are not localized on each vortex but shared in a nonlocal fashion between the vortices, the coupling of the internal state to the
environment is exponentially small. As a result, the superposition of
different internal states is immune to decoherence, which is ideal
for the purpose of quantum computation. Quantum computation with
topologically protected q-bits is generally known as topological
quantum computation, and is currently an active field of
research~\cite{nayak2008}.

Several experimental candidates for $p$-wave superconductivity have
been proposed, among which Sr$_2$RuO$_4$, which is considered as the most promising candidate for 2D chiral superconductivity~\cite{mackenzie2003}. However, many properties of this system remain unclear, such as whether this superconducting phase is gapped and whether there are gapless edge states.

Fortunately, there is an alternate route towards topological
superconductivity without $p$-wave pairing. In 1981, Jackiw and
Rossi~\cite{jackiw1981} showed that adding a Majorana mass term to a
single flavor of massless Dirac fermions in $(2+1)$D would lead to a
Majorana zero mode in the vortex core. Such a Majorana mass term can
be naturally interpreted as the pairing field due to the proximity
coupling to a conventional $s$-wave superconductor. There are now
three different proposals to realize this route towards topological
superconductivity: the superconducting proximity effect on the 2D surface state of the 3D topological insulator~\cite{fu2008}, on the 2D TR breaking topological insulator~\cite{qi2010}, and on semiconductors with strong Rashba SOC~\cite{sau2010}. We shall review these three proposals in the following.

\subsubsection{Majorana fermions in surface states of the topological insulator}
\label{sec:proximity}

Fu and Kane~\cite{fu2008} proposed a way to realize the
Majorana zero mode in a superconducting vortex core by making use of
the surface states of 3D topological insulators. Consider a topological insulator such as Bi$_2$Se$_3$, which has a single Dirac cone on the surface with Hamiltonian from Eq.~\ref{eq:sur_Heff}:
\begin{equation}
H = \sum_\mathbf{p} \psi^\dagger \left[v
(\boldsymbol{\sigma}\times\mathbf{p})\cdot\hat{\mathbf{z}}-\mu
\right]\psi,
\label{Hsurf}
\end{equation}
where
$\psi=\left(\psi_\uparrow,~\psi_\downarrow\right)^T$
and we have taken into account a finite chemical potential $\mu$. Consider now the superconducting proximity effect of a conventional
$s$-wave superconductor on the 2D surface states, which leads to the pairing term
$H_\Delta=\Delta\psi_\uparrow^\dagger\psi_\downarrow^\dagger+\mathrm{h.c.}$
The BdG Hamiltonian is given by $H_{\rm BdG}=\frac12\sum_\mathbf{p}\Psi^\dag H_{\bf p}\Psi$, where $\Psi^\dag\equiv\left(\begin{array}{cc}\psi^\dagger & \psi\end{array}\right)$ and
\begin{eqnarray*}
H_{\bf p}\equiv\left(\begin{array}{cc}v
(\boldsymbol{\sigma}\times\mathbf{p})\cdot\hat{\mathbf{z}}
 -\mu&i\sigma^y\Delta
\\-i\sigma^y\Delta^*&-v
(\boldsymbol{\sigma}\times\mathbf{p})\cdot\hat{\mathbf{z}}
+\mu\end{array}\right).
\end{eqnarray*}
The vortex core of such a superconductor has been shown to have a single
Majorana zero mode, similar to a $p+ip$
superconductor~\cite{jackiw1981,fu2008}. To understand this
phenomenon, one can consider the case of finite $\mu$, and introduce
a TR breaking mass term $m\sigma^z$ in the surface state
Hamiltonian (\ref{Hsurf}). As discussed in Sec.~\ref{sec:surface}, this opens a gap of magnitude $|m|$ on the surface. Considering the case
$\mu>m>0,~\mu-m\ll m$, the Fermi level in the normal state lies near the
bottom of the parabolic dispersion, and we can consider a
``nonrelativistic approximation" to the massive Dirac Hamiltonian,
\begin{eqnarray}
H &=&\sum_\mathbf{p} \psi^\dagger \left[v
(\boldsymbol{\sigma}\times\mathbf{p})\cdot\hat{\mathbf{z}} +m\sigma^z-\mu\right]\psi\nonumber\\
&\simeq &\int d^2x\,\psi_+^\dagger\left(\frac{{\bf p}^2}{2m}+m-\mu\right)\psi_+,
\end{eqnarray}
where $\psi_+$ is the positive energy branch of the surface
states. In momentum space, $\psi_{+\bf p}=u_{\bf
p}\psi_\uparrow+v_{\bf p}\psi_\downarrow$ with $u_{\bf
p}=\sqrt{\frac12+\frac{m}{2\sqrt{{\bf p}^2+m^2}}}$ and $v_{\bf
p}=\frac{{ p_+}}{|{\bf p}|}\sqrt{\frac12-\frac{m}{2\sqrt{{\bf
p}^2+m^2}}}$. Considering the projection of the pairing term
$H_\Delta$ onto the $\psi_+$ band, we obtain
\begin{eqnarray}
H_\Delta&\simeq &\sum_{\bf p}\psi_{+,{\bf p}}^\dagger
\psi_{+,-{\bf p}}^\dagger \Delta u_{\bf p}v_{\bf p}+\mathrm{h.c.}\nonumber\\
&\simeq& \sum_{\bf p}\frac{\Delta p_+}{2m}\psi_{+,{\bf p}}^\dagger \psi_{+,-{\bf p}}^\dagger+\mathrm{h.c.}
\end{eqnarray}
We see that in this limit, the surface Hamiltonian is the same
as that of a spinless $p+ip$ superconductor [Eq.~(\ref{Hpip})]. When
the mass $m$ is turned on from zero to a finite value, it can be shown that as long as $m<\mu$, the superconducting gap near the Fermi
surface remains finite, so that the Majorana zero mode we obtained
in the limit $0<\mu-m\ll m$ must remain at zero energy for the
original $m=0$ system. Once we have shown the existence of a Majorana
zero mode at finite $\mu$, taking the $\mu\rightarrow 0$ limit for a
finite $\Delta$ also leaves the superconducting gap open, so that
the Majorana zero mode is still present at $\mu=0$.

From the analogy with the $p+ip$ superconductor shown above, we also see that the non-Abelian statistics of vortices with
Majorana zero modes apply to this new system as well. A key difference
between this system and a chiral $p+ip$ superconductor is that the
latter necessarily breaks TR symmetry while the former
can be TR invariant. Only a conventional $s$-wave superconductor is
required to generate the Majorana zero modes in this proposal and in
the other proposals discussed in the following subsection. This is an important advantage compared to previous proposals requiring an unconventional $p+ip$ pairing mechanism.

There is also a lower dimensional analog of this nontrivial surface state superconductivity. When the edge states of 2D QSH insulator are in proximity with an $s$-wave superconductor and a ferromagnetic insulator, one Majorana fermion appears at each domain wall between ferromagnetic region and superconducting region (\cite{fu2009d}).
The Majorana fermion in this system can only move along the 1D QSH edge, so that non-Abelian statistics is not well-defined.
Because an electron cannot be backscattered on the QSH edge, the scattering of the edge electron by a superconducting region induced by proximity effect is always perfect Andreev reflection \cite{adroguer2010,guigou2010,sato2010}.

\subsubsection{Majorana fermions in semiconductors with Rashba spin-orbit coupling}

From the above analysis, we see that conventional $s$-pairing in
the surface Hamiltonian (\ref{Hsurf}) induces topologically nontrivial
superconductivity with Majorana fermions. There is a 2D
system which is described by a Hamiltonian very similar to Eq.~(\ref{Hsurf}), i.e. a 2D electron gas with Rashba SOC. The Hamiltonian is $H=\int
d^2x~\psi^\dagger\left(\frac{{\bf p}^2}{2m}+\alpha (\boldsymbol{\sigma}\times\mathbf{p})\cdot\hat{\mathbf{z}}-\mu\right)\psi$, which differs from the surface
state Hamiltonian only by the spin-independent term ${\bf p}^2/2m$.
Consequently, when conventional $s$-wave pairing is introduced,
each of the two spin-split Fermi surfaces forms a nontrivial
superconductor. However, the Majorana fermions from these two Fermi
surfaces annihilate each other so that the $s$-wave superconductor in the Rashba system is trivial. It was
pointed out recently~\cite{sau2010} that a nontrivial superconducting phase can be obtained by introducing a TR breaking term $M\sigma^z$ into the Hamiltonian, which splits the degeneracy near $\mathbf{k}=0$. If the chemical potential is tuned to $|\mu|<|M|$, the inner Fermi surface disappears. Therefore, superconductivity is only induced by pairing on the outer Fermi surface, and becomes topologically nontrivial. Physically, one cannot induce a TR breaking mass term by applying a magnetic field in the perpendicular direction, because the magnetic field may destroy superconductivity. Two ways to realize a TR breaking
mass term have been proposed: by
applying an in-plane magnetic field and making use of the
Dresselhaus SOC~\cite{alicea2010}, or by exchange coupling to a
ferromagnetic insulating layer~\cite{sau2010}. The latter
proposal requires a heterostructure consisting of a superconductor, a 2D electron gas with Rashba SOC, and a magnetic insulator.

This mechanism can also be generalized to the 1D semiconductor wires with Rashba SOC coupling in proximity with a superconductor~\cite{oreg2010,wimmer2010}. Despite the 1D nature of the wires, non-Abelian statistics is still possible by making use of wire networks~\cite{alicea2010b}.

\subsubsection{Majorana fermions in quantum Hall and quantum anomalous Hall insulators}

\begin{figure}[t!]
\begin{center}
\includegraphics[width=0.45\textwidth] {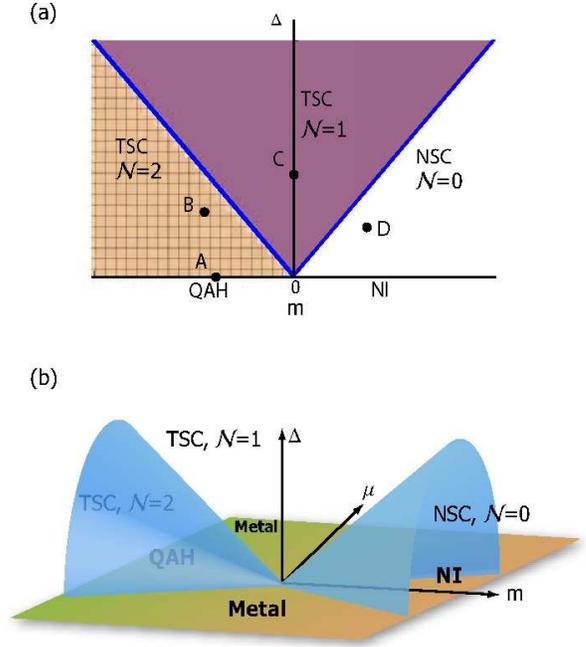}
\end{center}
\caption{(a) Phase diagram of the QAH-superconductor hybrid system for $\mu=0$. $m$ is the mass parameter, $\Delta$ is the magnitude of the superconducting gap, and ${\cal N}$ is the Chern number of the superconductor, which is equal to the number
of chiral Majorana edge modes. (b) Phase diagram for
finite $\mu$, shown only for $\Delta\geq 0$.
The QAH, normal insulator (NI) and metallic (Metal) phases are well-defined only for $\Delta=0$. From \onlinecite{qi2010}. } \label{fig:PD}
\end{figure}

More recently, a new approach to realize a topological
superconductor phase has been proposed~\cite{qi2010}, which is based on the proximity effect to a 2D QH or QAH insulator. Integer QH states are classified by an integer $N$ corresponding to the first Chern number in momentum space and equal to the Hall conductance in units of $e^2/h$. Consider a QH insulator with Hall conductance $Ne^2/h$ in close proximity to a superconductor. Even if the pairing strength inducing by the superconducting proximity effect is infinitesimally small, the resulting
state is topologically equivalent to a chiral topological superconductor with $\mathbb{Z}$ topological quantum number ${\cal N}=2N$. An intuitive way to understand such a relation between QH and topological superconducting phases is through the evolution of the
edge states. The edge state of a QH state with Chern
number $N=1$ is described by the effective 1D
Hamiltonian $H_\mathrm{edge}=\sum_{p_y}vp_y\eta_{p_y}^\dagger \eta_{p_y}$, where $\eta_{p_y}^\dag,\eta_{p_y}$ are creation/annihilation operators for a complex spinless fermion. We can
decompose $\eta_{p_y}$ into its real and imaginary parts,
$\eta_{p_y}=1/\sqrt{2}(\gamma_{p_y 1}+i\gamma_{p_y 2})$ and
$\eta^{\dagger}_{p_y}=1/\sqrt{2}(\gamma_{-p_y 1}-i\gamma_{-p_y 2})$,
where $\gamma_{p_y a}$ are Majorana fermion operators satisfying $\gamma^{\dagger}_{p_y a}=\gamma_{-p_y a}$ and
$\left\{\gamma_{-p_y
a},\gamma_{p^{'}_{y}b}\right\}=\delta_{ab}\delta_{p_{y}p^{'}_{y}}.$
The edge Hamiltonian becomes
\begin{equation}
H_\mathrm{edge}=\sum_{p_y\geq 0}p_{y}\left(\gamma_{-p_{y} 1}\gamma_{p_{y}
1}+\gamma_{-p_{y} 2}\gamma_{p_{y} 2}\right),
\end{equation} up to a
trivial shift of the energy. In comparison with the edge theory of
the chiral topological superconducting state, the QH edge state can be
considered as two identical copies of chiral Majorana fermions, so
that the QH phase with Chern number $N=1$ can be considered as a
chiral topological superconducting state with Chern number ${\cal N}=2$, {\em even for infinitesimal pairing amplitudes}.

An important consequence of such a relation between QH and topological superconducting phases is that the QH plateau transition from $N=1$ to $N=0$ will generically split into two transitions when superconducting
pairing is introduced. Between the two transitions, there will be a new
topological superconducting phase with odd winding number ${\cal
N}=1$ [Fig.~\ref{fig:PD}]. Compared to other approaches, the emergence of the topological superconducting phase at a QH plateau transition is determined topologically, so that this approach does not depend on any fine tuning or details of the theory.

A natural concern raised by this approach is that the strong magnetic field usually required for QH states can suppress superconductivity. The solution to this problem can be found in a special type of QH state --- the QAH state, which is a TR breaking gapped state with nonzero Hall conductance in the absence of an external orbital magnetic
field (Sec.~\ref{sec:qah}). There exist now two realistic proposals for realizing the QAH state experimentally, both of which make use of the TR invariant topological insulator materials Mn-doped HgTe QWs~\cite{liu2008}, and Cr- or Fe-doped Bi$_2$Se$_3$ thin
films~\cite{yu2010}. The latter material is proposed to be ferromagnetic, and can thus exhibit a quantized Hall conductance at zero magnetic
field. The former material is known to be paramagnetic for low Mn
concentrations, but only a small magnetic field is needed to polarize
the Mn spins and drive the system into a QAH phase. This requirement is not so prohibitive, because a nonzero magnetic field is already necessary to generate superconducting vortices and the associated Majorana zero modes.

\subsubsection{Detection of Majorana fermions}

\begin{figure}
          \includegraphics[width=0.3\textwidth]{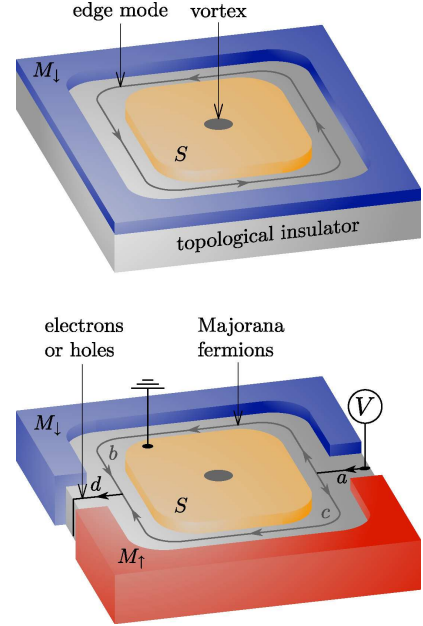}
\caption{ 3D topological insulator in proximity to ferromagnets with opposite polarization ($M_\uparrow$ and $M_\downarrow$) and to a superconductor ($S$). The top panel shows a single chiral Majorana mode along the edge between superconductor and ferromagnet. This mode is electrically neutral, and therefore cannot be detected electrically. The Mach-Zehnder interferometer in the bottom panel converts a charged current along the domain wall into a neutral current along the superconductor (and vice versa). This allows electrical detection of the parity of the number of enclosed vortices/flux quanta. From \onlinecite{akhmerov2009}. }\label{fig:elecTR invariantcMajorana}
  \end{figure}

The next obvious question is how to detect the Majorana fermion if such a proposal is experimentally realized. There exist two similar theoretical proposals of electrical transport
measurements to detect these Majorana
fermions~\cite{fu2009,akhmerov2009}. Consider the geometry shown in
Fig.~\ref{fig:elecTR invariantcMajorana}. This
device is a combination of the inhomogeneous structures on the
surface of a topological insulator discussed in the previous subsections. The input and output of the circuit consist of a chiral fermion coming from a domain wall between two ferromagnets. This chiral fermion is incident on a superconducting region where it splits into two chiral Majorana fermions. The chiral Majorana fermions then recombine into an outgoing electron or hole after traveling around the superconducting island. More explicitly, an electron incident from the source can be transmitted to the drain as an electron, or converted to a hole by an Andreev process in which charge $2e$ is absorbed into the superconducting condensate. To illustrate the idea we discuss the behavior
for a $E=0$ quasiparticle~\cite{fu2009}. A chiral fermion incident at point $a$ meets the superconductor and evolves from an
electron $c^{\dagger}_{a}$ into a fermion $\psi$ built from the
Majorana operators $\gamma_1$ and $\gamma_2$. The arbitrariness in the sign of $\gamma_{1,2}$ allows us to choose
$\psi=\gamma_1 + i \gamma_2$. After the quasiparticle winds
around the superconducting region, $\psi$ recombines into a
complex fermion at point $d$. This fermion must be {\it either} $c_d^\dagger$ or $c_d$, since a superposition of the two is not a fermion operator and is thus forbidden. To determine the correct operator we can use adiabatic continuity. When the size of the superconductor shrinks
continuously to zero, points $a$ and $d$ continuously tend to
each other. Adiabatic continuity implies that an incident $E=0$
electron is transmitted as an electron, $c_a^\dagger \rightarrow
c_d^\dagger$. However, if the ring encloses a quantized flux
$\Phi=nhc/2e$, this adiabatic argument must be reconsidered. When
$n$ is an odd integer, the two Majorana fermions acquire an additional relative phase of $\pi$, since each flux quantum $hc/2e$ is a $\pi$ flux for an electron, and thus $\pi$ for a Majorana fermion. Up to an overall sign, one can take $\gamma_1 \rightarrow -\gamma_1$ and $\gamma_2\rightarrow \gamma_2$. Thus, when
the ring encloses an odd number of flux quanta, $c_a^\dagger
\rightarrow c_d$, and an incident $E=0$ electron is converted to a
hole. The general consequences of this were calculated in detail~\cite{fu2009,akhmerov2009}, and it was shown that the output current
(through arm $d$ in the lower panel of Fig.~\ref{fig:elecTR invariantcMajorana}) changes sign when the number of flux quanta in the ring jumps between odd and even. This unique behavior of the current provides a way to electrically detect Majorana fermions.

Besides these two proposals reviewed above, several other theoretical proposals have also been made recently to observe the Majorana fermion state, which make use of the Coulomb charging energy \cite{fu2010} or a flux qubit \cite{hassler2010}. More indirectly, Majorana fermions can also be detected through their contribution to Josephson coupling \cite{fu2008,linder2010,linder2010a,tanaka2009,fu2009d,lutchyn2010}. For a topological superconductor ring with Majorana fermions at both ends, the period of Josephson current is doubled, independent from the physical realization \cite{kitaev2001,fu2009d,lutchyn2010}.

\section{Outlook}

The subject of topological insulators and topological superconductors is now one of the most active fields of research in condensed matter physics, developing at a rapid pace. Theorists have systematically classified topological states in all dimensions. Ref. ~\cite{qi2008b} initiated the classification program of all topological insulators according to discrete particle-hole symmetry and the TR symmetry, and noticed a periodic structure with period eight, which is known in mathematics as the Bott periodicity. More extended and systematic classification of all topological insulator and superconductor states are obtained according to TR, particle-hole and bipartite symmetries~\cite{qi2008b,schnyder2008,kitaev2009,ryu2010,stone2010}. Such classification scheme gives a ``periodic table" of topological states, which may play a similar role as the familiar period table of elements. For future progress on the theoretical side, the most important outstanding problems include interaction and disorder effects, realistic predictions for topological Mott insulator materials, a deeper understanding of fractional topological insulators and realistic predictions for materials realizations of such states, the effective field theory description of the topological superconducting state, and realistic materials predictions for topological superconductors. On the experimental side, the most important task is to grow materials with sufficient purity so that the bulk insulating behavior can be reached, and to tune the Fermi level close to the Dirac point of the surface state. Hybrid structures between topological insulators and magnetic and superconducting states will be intensively investigated, with a focus on detecting exotic emergent particles such as the image magnetic monopole, the axion and the Majorana fermion. The theoretical prediction of the QAH state is sufficiently realistic and its experimental discovery appears to be imminent. The topological quantization of the TME effect in 3D and the spin-charge separation effect in 2D could experimentally determine the topological order parameter of this novel state of matter.

Due to space limitations, we did not discuss in detail the potential for applications of topological insulators and superconductors. It would be interesting to explore the possibility of electronic devices with low power consumption based on the dissipationless edge channels of the QSH state, spintronics devices based on the unique current-spin relationship in the topological surface states, infrared detectors, and thermoelectric applications. Topological quantum computers based on Majorana fermions remain a great inspiration in the field.

Topological insulators and superconductors offer a platform to test many novel ideas in particle physics --- a ``baby universe" where the mysterious $\theta$ vacuum is realized, where exotic particles roam freely and where compactified extra dimension can be tested experimentally. In the introduction to this article we drew an analogy between the search for new states of matter and the discovery of elementary particles. Up to now, the most important states of quantum matter were first discovered empirically and often serendipitously. On the other hand, the Einstein-Dirac approach has been most successful in searching for the fundamental laws of nature: pure logical reasoning and beautiful mathematical equations guided and predicted subsequent experimental discoveries. The success of theoretical predictions in the field of topological insulators shows that this powerful approach works equally well in condensed matter physics, hopefully inspiring many more examples to come.

\section*{ACKNOWLEDGMENTS}
We are deeply grateful to Taylor L. Hughes, Chao-Xing Liu, Joseph Maciejko and Zhong Wang for their invaluable inputs which made the current manuscript possible. We would like to thank Andrei Bernevig, Hartmut Buhmann, Yulin Chen, Suk Bum Chung, Yi Cui, Xi Dai, Dennis Drew, Zhong Fang, Aharon Kapitulnik, Andreas Karch, Laurens Molenkamp, Naoto Nagaosa, S. Raghu, Zhi-Xun Shen, Cenke Xu, Qikun Xue, Haijun Zhang for their close collaboration and for their important contributions reviewed in this paper. We benefitted greatly from the discussions with colleagues Leon Balents, Mac Beasley, Carlo Beenakker, Marcel Franz, Liang Fu, David Goldhaber-Gordon, Zahid Hasan, Charlie Kane, Alexei Kitaev, Steve Kivelson, Andreas Ludwig, Joel Moore, Phuan Ong, Rahul Roy, Shinsei Ryu, Ali Yazdani and Jan Zaanen. This work is supported by the Department of Energy, Office of Basic Energy Sciences, Division of Materials Sciences and Engineering, under contract DE-AC02-76SF00515, the NSF under the grant number DMR-0904264, the ARO under the grant number W911NF-09-1-0508 and the Keck Foundation.

\bibliographystyle{apsrmp4-1}

\end{document}